\documentclass[useAMS,usenatbib]{mn2e}
\usepackage{subfigure}
\usepackage{natbib}
\usepackage{lscape}
\usepackage[dvips]{color}
\usepackage{amssymb}
\usepackage{graphicx}
\usepackage{url} 
\usepackage{ulem} 
\usepackage{setspace}
\usepackage{threeparttable}
\usepackage{multirow}

\title[Spectroscopic characterisation of SHK galaxy groups]{Characterising galaxy groups: spectroscopic observations of the Shakhbazyan sample}
\author[Capozzi D. et al. 2012] {Diego 
  Capozzi$^{1,2,3}$\thanks{E-mail: diego.capozzi@port.ac.uk}, Marilena
  Spavone$^{3,4,6}$, Silvio Barbati$^{3}$, Maurizio Paolillo$^{3,5}$,\and
  Elisabetta De Filippis$^{3,5}$, Giuseppe Longo$^{3,4,7}$ \\ \\ 
  1 - Institute of Cosmology and Gravitation, University of Portsmouth, Dennis Sciama Building, Burnaby Road,\\
  \quad \, Portsmouth, PO1 3FX, UK\\
  2 - Astrophysics Research Institute, Liverpool John Moores University, Twelve Quays House, Egerton Wharf,\\
  \quad \, Birkenhead, CH41 1LD, UK\\
  3 - Dipartimento di Scienze Fisiche, Universit\`{a} Federico II, via Cinthia 6, I-80126 Napoli, Italy\\ 
  4 - INAF-Astronomical Observatory of Naples, via Moiariello 16, I-80131 Napoli, Italy\\  
  5 - INFN - Napoli Unit, Dept. of Physical Sciences, via Cinthia 9, 80126, Napoli, Italy\\
  6 - Dipartimento di Fisica e Astronomia, Universit\`{a} di Padova, Vicolo dell'Osservatorio 2, I-35122 Padova, Italy\\
  7 - Visiting associate - California Institute of Technology, California Bvd., Pasadena, 90125 CA, USA}

\date{Accepted ;
  Received ; in original form }
\pagerange{\pageref{firstpage}--\pageref{lastpage}} \pubyear{2012}

\begin{document}

\maketitle

\label{firstpage}

\begin{abstract}
Groups of galaxies are the most common cosmic structures.
However, due to the poor statistics, projection effects and the lack of accurate distances, our understanding of 
their dynamical and evolutionary status is still limited. 
This is particularly true for the so called Shakhbazyan groups  (SHK) which are still largely unexplored 
due to the lack of systematic spectroscopic studies of both their member galaxies and the surrounding environment.
In our previous paper, we investigated the statistical properties of a large sample of SHK groups using Sloan Digital Sky Survey data and photometric redshifts. 
Here we present the follow-up of 5 SHK groups (SHK 10, 71, 75, 80, 259) observed within our spectroscopic campaign with the Telescopio Nazionale Galileo, aimed at confirming their physical reality and strengthening our photometric results.
For each of the selected groups we were able to identify between 6 and 13 spectroscopic members, thus 
confirming the robustness of the photometric redshift approach in identifying real galaxy over-densities.
Consistently with the finding of our previous paper, the structures studied here have properties spanning from those of compact and isolated groups to those of loose groups. 
For what the global physical properties are concerned (total mass, mass-to-light ratios, etc.), we find systematic differences with those reported in the literature by previous studies. Our analysis suggests
that previous results should be revisited; we show in fact that, if the literature data are re-analysed in a consistent and homogeneous way, the properties obtained are in agreement with those estimated for our sample.\\
\end{abstract}


\section{INTRODUCTION}
\label{sec:sec1}
The relevance of galaxy groups to the study of galaxy structure formation and evolution arises mainly from the fact
that they are the main building blocks in the framework of the hierarchical models for the formation of larger and more 
massive galaxy structures (e.g., \citealp{Ace2002, Don2005,Dia2010,Mend2011}). In spite of their importance, many aspects of the role they play in shaping the
observed structure of the Universe are still poorly understood.
For instance, compact groups have crossing times which are far too short to be reconciled with their observed number in the 
local Universe (e.g., \citealp{Dia1994, Gov1996, Mam2000}) unless initial conditions are correctly chosen (e.g. \citealp{Ath1997, Ace2002}) or mechanisms 
apt to prolong their estimated lifetimes are in place (e.g. \citealp{Dia1994, Gov1996}).  
Among those mechanisms, the so called ``secondary infall'' scenario predicts that the groups observed nowadays 
are being continuously rejuvinated by the infall of new galaxies captured from the field (\citealp{Gun1972, Mam2007} and references therein),
a hypothesis which is supported by both numerical simulations \citep{Gov1996,McC2008} and the observations that most compact groups 
are surrounded by larger and looser structures \citep{Ven1993, Pal1995, Bar1998, Rib1998, deC2000}.
This implies that even when an isolation criterion is adopted to identify compact groups (e.g., \citealp{Hic1982}), the surrounding environment should be explored.
For instance, the exclusion of members which lay in the outskirts of the groups leads to an underestimate of the group velocity dispersion 
and hence of the group total mass and dynamical lifetime.
For this reason, it is crucial to perform detailed studies also of less compact structures and of the environment surrounding 
groups of different multiplicity and compactness.

In the recent literature, one of the possible and most largely used scenarios for galaxy group formation and evolution describes  galaxy groups 
as accreting through different phases, starting from loose configurations (e.g., \citealp{deC2000}): 
i) [core+halo] (\citealp{deC2000,Cap2009}, hereafter Cap09); 
ii) compact (e.g., \citealp{Hic1982, Hic1992, deC2000}, Cap09); 
iii) fossil groups \citep{Vik1999, Jon2003, Men2006, Men2007}. 
However, the debate about such evolutionary path is far from settled as, for instance, several studies (e.g., \citealp{Jon2003, Kho2004, Don2005, Dar2007, Kho2007, von2008, Har2012}; see also \citealp{LaB2009}) suggested that fossil systems formed earlier than ``normal'' ones, within particularly over-dense environments, and possibly follow a different formation and evolutionary path compared to the latter.

Bearing in mind the importance of investigating the environment surrounding galaxy groups, in Cap09 we studied a sample of 58 Shakhbazyan (SHK) groups \citep{Sha1973,Sto1997} both in the projected space and photometric
redshift space, aiming at assessing its nature and properties.  Among other findings, we confirmed that not only are these groups real local galaxy over-densities (i.e. not just the result of projection effects) but they have heterogenous properties, encompassing compact groups, [core+halo] configurations and cluster cores in spite of the fact that they were 
originally defined as ``compact groups of compact galaxies''. 

This finding makes SHK group particularly interesting to characterise spectroscopically.
The sample of 376 SHK was in fact originally selected from visual inspection of the 
First Palomar Sky Survey (POSS) by means of the following (rather fuzzy) criteria: 
i) isolated structures with 5-15 compact members with POSS apparent magnitude $14<{\it R}<19$; 
ii) relative distances 3/5 times the characteristic diameter of a galaxy
and iii) extremely red galaxy colours. Because of these selection criteria, the SHK groups is as an interesting as a debated galaxy-group sample.

Despite SHK groups being so numerous and therefore potentially interesting to constrain the dynamical history of 
groups in general, spectroscopic redshifts are still missing for the great majority of them, thus preventing any robust
assessment of their properties. 
With the final goal of studying the properties of SHK groups in relation to their environment, we have recently carried out a
spectroscopic campaign at the Italian Telescopio Nazionale Galilei ({\it TNG}). In this paper we present the spectroscopic observations for the first 5 SHK groups and derive their global properties using dynamical and structural analyses.

Throughout this paper we make use of magnitudes in the Sloan Digital Sky Survey (SDSS) photometric
system and assume a standard cosmology with $H_{0}=74\ {\rm
km\ s^{-1}\ Mpc^{-1}}$, $\Omega_{\rm m}=0.3$ and $\Omega_{\rm
 \lambda}=0.7$.

\section{Sample Selection} 
\label{sec:sec2}
The candidate SHK groups for the spectroscopic observations were selected among those with no or poor spectroscopic coverage (less than five redshifts available in the literature or in the 
SDSS spectroscopic galaxy sample) and covered by the SDSS DR5\footnote{At the time of observing proposal submission, DR5 was the most recent release available. More information about the survey can be found at:\url{http://www.sdss.org/}.} \citep{Ade2007}. 
The advantage of focussing on the SHK groups within the SDSS is the availability of good quality photometric data and photometric redshifts \citep{Dab2007} coupled with a large sky coverage. 

In order to maximize the probability of dealing with physically bound structures rather than over-densities resulting from chance projections of galaxies on the sky, we pre-selected candidates for observations following the method developed by Cap09, based on the structural analysis of candidate groups in both the projected and photometric redshift space. We identify candidate group members as those within 500 {\rm kpc} from the group centroid reported in the original SHK catalog, and with photometric redshift  within $<z_{\rm phot}> \pm 3 \epsilon(z_{\rm phot})$, where  $\epsilon(z_{\rm phot})=0.02$ \citep{Dab2007}; we label a structure as real when: (i) it is characterised by an excess ($> 5$) of galaxies in the photometric redshift distribution, with respect to the local background; ii) $<z_{\rm phot}>$ is compatible with the available spectroscopic redshift of the (usually brightest) candidate members; iii) the background-corrected group richness, evaluated within a radial distance of $150\ {\rm kpc}$ as in Cap09, $N_{150}>3$.
We further decided to retain only groups where the Extension Index $EI=N_{500}/N_{150} > 1$, $N_{500}$ being the group richness evaluated within a radial distance of $500\ {\rm kpc}$ (see Cap09), in order to remove very poor groups or residual fortuitous alignments of galaxies, which would provide very few targets for spectroscopic observations. 

We finally selected 15 SHK groups among those falling within the observable sky region at {\it TNG} during the time slots we had been allocated. This final selection aimed at exploring the different types of structures identified in Cap09, i.e. compact groups, [core+halo] configurations and cluster cores.
In this paper we describe the results for the first 5 groups (namely: SHK 10, 71, 75, 80 \& 259), whose galaxies' main properties are
respectively summarized in Tables \ref{tab:TableA1}, \ref{tab:TableA2}, \ref{tab:TableA3}, \ref{tab:TableA4} and
\ref{tab:TableA5}. The study of the remaining 10 groups is currently under way and will be presented in a forthcoming paper.

\section{Observation and data reduction}
\label{sec:sec3}
Spectra were obtained with {\it DOLORES@TNG} (Device Optimized for the Low RESolution), in visitor mode during the
observing run AOT18-TAC38 on October 2008 (PI: Paolillo). 
{\it DOLORES} is installed at the Nasmyth B focus of the {\it TNG} and is
equipped with the E2V 4240 CCD with a scale of $0.252\ {\rm arcsec} \ {\rm pix^{-1}}$.

Using the photometric redshifts by \citet{Dab2007}, we again selected candidate targets for MOS (Multi-object spectroscopy) spectroscopy among galaxies with photometric redshifts within $<z_{\rm phot}> \pm 3 \epsilon(z_{\rm phot})$, in order to remove secure fore/background contaminants. 
By using the LR-B spectrograph, with a dispersion of 2.52 {\rm \AA/pix}, we obtained spectra for all galaxies
brighter than ${\it r} = 20$ within $\sim 5'$ (i.e., $\simeq 500\ {\rm kpc}$)\footnote{The distance constrain is due to the DOLORES field-of-view, which is limited to $6'\times8'$ for MOS spectroscopy and to the additional limitations imposed by the mask-design software.} from the group centre, subject to limitations due to slit positioning and spectra overlap, thus deriving velocities for $> 20$
objects per group.
The spectra covered the wavelength range 3000-8000 {\rm \AA}.
All slits were $1.6\ {\rm arcsec}$ wide. 
In the central regions of the groups, crowding did not allow to obtain spectra for all galaxies. This 
problem was partially solved by excluding those objects which already had a spectroscopic-redshift measure in the literature or the SDSS spectroscopic catalogue. However, few of these galaxies residing in less crowded regions were kept to check the quality of our measures. The total integration time for each mask was 1200 {\rm s}, with an average seeing of $1.5^{\prime\prime}$.

Data reduction was carried out using the CCDRED package in the IRAF\footnote{IRAF is distributed by the National Optical Astronomy Observatories, which is operated by the Associated Universities for Research in Astronomy, Inc. under cooperative agreement with the National Science Fundation.} ({\it Image Reduction and Analysis Facility}) environment. The main strategy adopted for each data-set included dark subtraction\footnote{Bias frame is included in the dark frame.}, flat-fielding correction, slits extraction, 
sky subtraction and rejection of bad pixels.\\
Wavelength calibration was achieved by
means of comparison spectra of Hg+Ne lamps acquired during each observing night, using the IRAF TWODSPEC.LONGSLIT package. 
Sky subtraction was performed by using the sky spectra obtained with three slits placed at the top, the middle and the bottom 
of each mask. 
The sky-subtracted frames were then co-added, by performing a $\sigma$-clipping in order to remove cosmic rays, 
obtaining median averaged 2D spectra for 86 galaxies. 

\begin{figure}
 \centering
 \includegraphics[width=8cm]{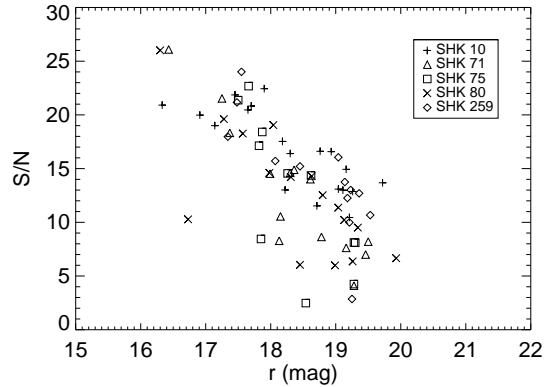}
 \caption{$S/N$ ratio per pixel vs. {\it r} magnitude for SHK 10 (plus signs), 71 (triangles), 75 (squares), 80 (crosses) and 259 (diamonds).}
 \label{fig:Fig1}
\end{figure}

In figure \ref{fig:Fig1} we show the signal-to-noise ($S/N$) 
ratios of our spectra as a function of their SDSS {\it r}-band magnitudes, empirically estimated from the average pixel-to-pixel variance over 100 \AA~  measured around 5000, 6000 and 7000 \AA~ where the spectra is devoid of strong spectral lines. 

\begin{figure}
 \centering
 \includegraphics[width=8cm]{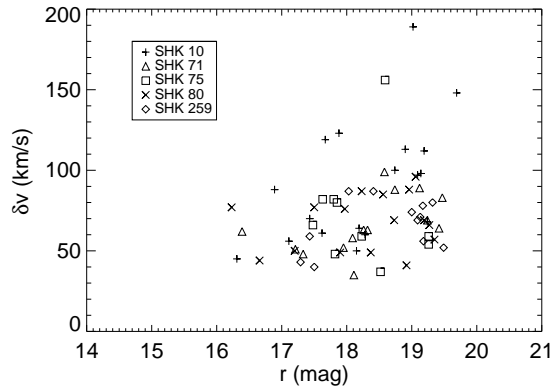}
 \caption{Error on galaxy velocities vs. {\it r}-band magnitude.}
 \label{fig:Fig2}
\end{figure}

\section{Data analysis}\label{sec:sec4}

\subsection{Dynamical analysis}
\label{subsec:subsec4.1}

Radial velocities and luminosities were estimated by using the {\it cross correlation} method described in
\citet{Ton1979}, which consists in deriving the cross-correlation function (CCF) between the spectra of the galaxies 
and that of a galaxy used as template. In particular, we used the IRAF task XCSAO to derive radial velocities 
and the spectrum of an early-type elliptical galaxy \citep{Cal1994}  as template\footnote{We verified that the precise choice of the adopted template does not affect our results, mainly
because the vast majority of our targets are early-type systems.}. 

In order to reduce the probability of a bad determination of the radial velocity due to the use of a ``false peak'' present in the CCF,
the statistical analysis was carried out only using those galaxies whose CCF satisfied all the following 
requirements:
{\it i)}  well defined and clearly visible peak and absence of ``secondary'' peaks comparable with the principal one (amplitude $<75$ per cent of main peak); 
{\it ii)} a confidence level \footnote{This parameter is used to quantify the reliability of the galaxy velocity as measured by means of the CCF and it is defined as the ratio 
between the height of the principal CCF peak and that averaged over the secondary peaks heights. The average error on the position of 
the main peak, so on galaxy redshift,  is proportional to $1/(1+R)$, as described in \citet{Ton1979}.}$R\geq\ 2.5$ \citep{Ton1979}.
Due to the low $S/N$ ratios of their spectra, 19 galaxies were rejected from further analysis either because it was not possible to carry out the CCF procedure (15 galaxies, listed in the tables in Appendix \ref{sec:appendix} with no velocity measurement) or because their CCFs did not satisfy the requirements described above (4 galaxies, listed in the tables in Appendix \ref{sec:appendix}), making it difficult to recognize the true peak of the CCF. Figure \ref{fig:Fig2} shows the relative error on galaxy 
velocities as a function of {\it r}-band magnitude for all the groups studied.
We then applied to the remaining 67 galaxies the criterion introduced by
\citet{Hic1992}, according to which galaxies are considered
``members'' of the respective groups if they have a radial velocity
$v_{\rm i}$ such that $\mid \Delta v \mid = \mid v_{\rm M} - v_{\rm i} \mid \leq
1000\ {\rm km/s}$, where $v_{\rm M}$ is the median radial velocity of the
group.

The mean radial velocity for each group ($\overline{v}$) was estimated by applying the {\it biweight} method \citep{Bee1990} 
to the concordant velocities (i.e., those satisfying the Hickson criterion explained above).
Velocity dispersions of each group were estimated as
the luminosity-weighted mean of their galaxies' peculiar velocities with respect to $\overline{v}$, using
the recipe provided by \citet{Fir2006}:
\begin{equation}
\sigma_{\rm r} = \left[\frac{\Sigma_{\rm i} L_{\rm i} (\Delta
    v_{\rm i})^{2}}{\Sigma_{\rm i} L_{\rm i}}\right]^{\frac{1}{2}}
\end{equation} where $\Delta v_{\rm i} = v_{\rm i} - \overline{v}$.
Finally, group masses were estimated by using the Virial theorem:
\begin{equation}
\label{vir_mass_eq}
M_{\rm vir} = \frac {3 \sigma^{2}_{\rm r}\ R_{\rm vir}} {G}
\end{equation} 
where the virial radius $R_{\rm vir}$ was calculated by using the following expression, given in \citet{Bin1987},
\begin{equation}
R_{\rm vir} = \frac {\pi}{2}\left[\frac{\Sigma_{\rm i}\Sigma_{\rm j>i}\frac{L_{\rm i}L_{\rm j}}{\mid R_{\rm i}-R_{\rm j}\mid}}{\Sigma_{\rm i}\Sigma_{\rm j>i}L_{\rm i}L_{\rm j}}\right]^{-1}
\end{equation} 
in which $\mid R_{\rm i}-R_{\rm j}\mid$ is the projected distance between each couple of galaxies belonging to the group.
Group total luminosities ($L_{\rm tot}$) were derived by summing
the {\it r}-band luminosities of the member galaxies. 

The virial masses,  virial radii and average radii and velocities were then used to estimate the mass-luminosity ratios $M/L =
M_{\rm vir}/L_{\rm tot}$ and the crossing times $\tau_{\rm cross} =\frac{\langle r \rangle}{\langle v \rangle}$ \citep{Fir2006}.
The properties obtained for each group are listed in Tables \ref{tab:Table1} and \ref{tab:Table2} and shown in Figures \ref{fig:Fig3}, \ref{fig:Fig4}, \ref{fig:Fig5}, \ref{fig:Fig6}, \ref{fig:Fig7} and \ref{fig:Fig8}.

As described later in more detail, our groups may not be fully virialized and we suffer from small number statistics; in such cases the dependence of our results from the exact mass estimator that we adopt can be a matter of concern. Several authors \citep{Bah1981,Hei1985,Per1990} explored the performances of different mass estimators as a function of the group richness, the level of contamination and the degree of anisotropy, finding that they are mainly consistent with each other within the uncertainties, especially for poor systems, yielding a factor of $\sim 2$ systematic difference even in severely contaminated systems 
\citep{Hei1985}. In particular, \citet{Per1990} pointed out that the mass-weighted virial estimator is the one least affected by the presence of anisotropies or substructure, although it is sensitive to interlopers. They also show that a good alternative is represented by the projected mass estimator ($M_{\rm PM}$), provided that mass weighting is applied as well. In Table \ref{tab:Table2} we show $M_{\rm PM}$ for our groups. We point out that since our groups are dominated by early type galaxies (with similar M/L ratios), our light-weighting approach yields a close approximation to a mass-weighted estimate.
We find that on average the projected masses are a factor $\sim 2.5$ smaller than the virial masses. Considering that both the uncertainties due to the possible presence of contaminants \citep[see][]{Hei1985}, and the systematics due to orbit anisotropies or presence of substructures, can each yield differences of a factor $\gtrsim 2$, we find the disagreement acceptable and we will adopt the virial estimate as our bona-fide mass measurement.

\begin{table}
\begin{center}
\caption{{\bf Observed groups' astrometric \& spectroscopic properties}. {\it Col 1}: group number; {\it col 2} \& {\it col 3}: luminosity-weighted centroid coordinates based on spectroscopically confirmed members only; {\it col 4}: average spectroscopic redshift; {\it col 5}: number of spectroscopically confirmed member galaxies (in brackets the new spectroscopic members confirmed by this study)}
\begin{tabular}{lcccc}

 \hline
  \multicolumn{1}{l}{\bf SHK} &
  \multicolumn{1}{c}{\bf RA} &
  \multicolumn{1}{c}{\bf Dec} &
  \multicolumn{1}{c}{$\mathbf{z_{\rm {\bf spec}}}$} &
  \multicolumn{1}{c}{$\mathbf{N_{\rm {\bf spec}}}$} \\
  \multicolumn{1}{c}{}&
  \multicolumn{1}{c}{\rm hh:mm:ss} &
  \multicolumn{1}{c}{\rm dd:mm:ss} &
  \multicolumn{1}{c}{} &
  \multicolumn{1}{c}{\rm {(gal)}}\\

\hline
  10   & 14:10:48.93 & +46:15:54.2   &  0.1339$\pm$0.0017 & 13 (8) \\
  71   & 13:02:06.58 & +38:03:20.1   &  0.1131$\pm$0.0020 &  \, 9 (6)  \\
  75   & 14:27:32.49 & +38:45:13.7   &  0.1651$\pm$0.0019 &  \, 7 (5)  \\
  80   & 15:37:08.20 & +41:38:57.5   &  0.1306$\pm$0.0014 &  \, 6 (5) \\
  259 & 15:39:31.00 & +37:50:43.5   &  0.1518$\pm$0.0015 &  \, 7 (6) \\
\hline
\end{tabular}
\label{tab:Table1}							    
\end{center}
\end{table}

\begin{table*}
\begin{center}
\caption{{\bf Dynamical properties}. {\it Col 1}: group number; {\it col 2}: radial velocity dispersion; {\it col 3}: virial radius; {\it col 4}: total SDSS {\it r}-band luminosity; {\it col 5}: virial mass;  {\it col 6}: projected mass; {\it col 7}: $M_{\rm vir}/L_{\rm tot}$ ratio; {\it col 8}: crossing time. The errors on $\sigma_{\rm r}$ were estimated via bootstrap as 68 per cent confidence intervals.}
\begin{tabular}{lccccccc}

 \hline
  \multicolumn{1}{l}{\bf SHK} &
  \multicolumn{1}{c}{$\mathbf{\sigma_{\rm r}}$} &
  \multicolumn{1}{c}{$\mathbf{R_{\rm {\bf vir}}}$} &
  \multicolumn{1}{c}{$\mathbf{L_{\rm {\bf tot}}}$} &
  \multicolumn{1}{c}{$\mathbf{M_{\rm {\bf vir}}}$} &
  \multicolumn{1}{c}{$\mathbf{M_{\rm {\bf PM}}}$} &
  \multicolumn{1}{c}{$\mathbf{M_{\rm {\bf vir}}/L_{\rm {\bf tot}}}$} &
  \multicolumn{1}{c}{$\mathbf{\tau_{\rm {\bf cross}}}$} \\
  \multicolumn{1}{c}{}&
  \multicolumn{1}{c}{$(\rm{km/s})$} &
  \multicolumn{1}{c}{$(\rm{kpc})$} &
  \multicolumn{1}{c}{$(10^{11}\ {\rm L_{\odot}})$} &
  \multicolumn{1}{c}{$(10^{13}\ {\rm M_{\odot}})$} &
 \multicolumn{1}{c}{$(10^{13}\ {\rm M_{\odot}})$} &
  \multicolumn{1}{c}{(${\rm M_{\odot}/L_{\odot}}$)} &
  \multicolumn{1}{c}{$(10^{6}\ {\rm yr})$} \\

\hline
  10   &  $440^{+70}_{-90}$ & 199 & 3.56 & 2.66 & 0.97 & 75   & 55\\
  71   &  $600^{+100}_{-100}$ & 173 & 1.32 & 4.46 & 1.99 & 337 & 28 \\
  75   &  $500^{+100}_{-100}$ & 351 & 2.22 & 6.07 & 2.32 & 273 & 70 \\
  80   &  $380^{+70}_{-100}$ & 74   & 1.56 & 0.76 & 0.30 & 49    & 35 \\
  259 &  $470^{+80}_{-80}$ &130 & 1.63 & 1.97 & 0.90 &121  & 29 \\
\hline
\end{tabular}
\label{tab:Table2}							    
\end{center}
\end{table*}
 
\subsection{Structural analysis}
\label{subsec:subsec4.2}
We studied the structural properties of the SHK groups presented here with two methods: i) following the approach of \citet{Rib1998}, limiting the analysis to spectroscopic confirmed members; ii) following the  method of Cap09 (introduced in Section \ref{sec:sec2}) using the deeper and more extended SDSS photometric data. 

\subsubsection{Spectroscopic sample diagnostics}
\label{subsubsec:subsubsec4.2.1}
This method \citep{Rib1998} is based on the analysis of group properties as a function of radial distance (as measured by the median of the inter-galactic distances) in comparison to those of the typical Hickson compact group (HCG), assumed as a representative template of compact, isolated structures. This approach requires spectroscopic redshifts in order to accurately distinguish between fore/background and member galaxies (and is thus limited in our case to ${\it r}\lesssim 20~{\rm mag}$) and the presence of at least 4 confirmed member galaxies per group. In particular the velocity dispersion, volume number and surface brightness densities of our groups are compared to those characterizing the typical HCG group, which are $\sigma_{\rm r}\sim 400\ {\rm km\ s^{-1}}$, $\rho\gtrsim 10^{3}\ {\rm gal\ Mpc^{-3}}$ (spatial density) and $\mu_{\rm r}\lesssim 26.3\ {\rm mag\ arcsec^{-2}}$ (surface brightness). 
Although Cap09 has shown that the large majority of the SHK groups do not satisfy Hickson criteria, in particular the compactness and isolation ones, this method is still useful to understand if our structures share some properties of compact groups or represent more dispersed structures.
The average properties derived four our 5 groups are listed in Table \ref{tab:Table3}. The dependence of such properties as a function of the median inter-galactic distance is shown instead in Appendix \ref{sec:appendix} for each group separately.

We also tested the group velocity distributions against gaussianity through Monte Carlo simulations, following the prescription of \citet{Bab2006}, finding that we cannot reject the gaussian hypothesis for any of our groups. However this is not surprising given the limited statistics. 

\begin{table}
\begin{center}
\caption{{\bf  Structural properties 1}. {\it Col 1}: group number; {\it col 2}: group spatial galaxy number density; {\it col 3}: group surface brightness; {\it col 4}: group projected radius, as determined by the median of inter-galactic distances. The error on $\rho$ is given by shot noise according to \citet{Geh1986} prescription while, as for the velocity dispersion, the error on $\mu$ was estimated as 68 per cent confidence intervals via bootstrap}
\begin{tabular}{lccc}
\hline
  {\bf SHK} &     {$\mathbf{\log(\rho)}$}     & {$\mathbf{\mu}$}                     & {$\mathbf{R_{\rm {\bf P}}}$}  \\
                   &     {$({\rm gal \ Mpc^{-3}})$} & {$({\rm mag\ arcesc^{-2}})$} & {$({\rm kpc})$}                         \\
\hline
10              & $2.0^{+0.3}_{-0.6}$                                         & $27.4^{+0.2}_{-0.2}$                                           & 310                                            \\
71              & $2.7^{+0.3}_{-0.6}$                                         & $26.9^{+0.2}_{-0.2}$                                           & 157                                           \\
75              & $1.8^{+0.3}_{-0.7}$                                         & $28.0^{+0.1}_{-0.1}$                                           & 308                                           \\
80              & $2.3^{+0.3}_{-0.7}$                                         & $27.1^{+0.6}_{-0.4}$                                           & 185                                           \\
259            & $3.0^{+0.3}_{-0.7}$                                         & $26.2^{+0.2}_{-0.2}$                                           & 120                                           \\
\hline
\end{tabular}
\label{tab:Table3}							    
\end{center}
\end{table}

\subsubsection{Photometric diagnostics}
\label{subsubsec:subsubsec4.2.2}
The photometric approach of Cap09 is more appropriate when the spectroscopic information is not available and allows us to derive the properties of each group down to a deeper magnitude limit (${\it r}\lesssim 21~{\rm mag}$, due by the use of the photometric redshifts from \citealp{Dab2007}) and over a larger region ($\gtrsim 500$ kpc); it further allows us to homogeneously compare the SHK groups' properties with those of the larger sample studied in Cap09. 
As in Cap09, we extracted all SDSS galaxies within 3 {\rm Mpc}\footnote{This radius was chosen to select a local background region which was reasonably far from the prospective structure under analysis, so to avoid over-subtraction when carrying out our  statistical background subtraction.} at the group spectroscopic redshift.
After the extraction, we measured individual group properties (richness, group mean photometric redshift, fraction of Red-Sequence galaxies, extension index, etc.); the main parameters relevant for the following analysis are listed in Table \ref{tab:Table4}. 

\begin{table*}
\begin{center}
\caption{{\bf Structural properties 2}. {\it Col 1}: group number; {\it col 2}: average photometric redshift; {\it col 3}: richness within $500\ {\rm kpc}$ from group centroid; {\it col 4}: richness within $150\ {\rm kpc}$ from group centroid; {\it col 5}: fraction of red-sequence galaxies measured by using the ({\it i-r}) vs. {\it r} diagram; {\it col 6}: group's brightest galaxy inside a radial distance of $150\ {\rm kpc}$; {\it col 7}: magnitude gap between the brightest and the faintest galaxies inside a radial distance of $150\ {\rm kpc}$; {\it col 8}: extension index of the group; {\it col 9}: structure typology. All these properties are measured as described in detail in Cap09. The asterisks highlight the richest compact structures ($N_{150}>7$) as in Cap09.}
\begin{tabular}{lcccccccc}
\hline
  \multicolumn{1}{l}{\bf SHK} &
  \multicolumn{1}{c}{$\mathbf{z_{\rm {\bf phot}}}$} &
  \multicolumn{1}{c}{$\mathbf{N_{\rm {\bf 500}}}$} &
  \multicolumn{1}{c}{$\mathbf{N_{\rm {\bf 150}}}$} &
  \multicolumn{1}{c}{$\mathbf{f(RS)_{150}}$} &
  \multicolumn{1}{c}{$\mathbf{r_{1}^{0}}$} &
   \multicolumn{1}{c}{$\mathbf{\Delta_{m}}$} & 
  \multicolumn{1}{c}{$\mathbf{EI}$} &
  \multicolumn{1}{c}{\bf Structure Typology}\\
  \multicolumn{1}{c}{}&
  \multicolumn{1}{c}{} &
  \multicolumn{1}{c}{$({\rm gal})$} &
  \multicolumn{1}{c}{$({\rm gal})$} &
  \multicolumn{1}{c}{} &
  \multicolumn{1}{c}{$({\rm mag})$} &
  \multicolumn{1}{c}{$({\rm mag})$} &
  \multicolumn{1}{c}{} &
  \multicolumn{1}{c}{} \\
\hline
  10   &  0.14$\pm$0.01 &  23$\pm$5  & 7 $\pm$3   & 0.70 & 16.31 & 2.95 & 3.0 & core+halo/loose\\
  71   &  0.14$\pm$0.01 &  15$\pm$4  & 11$\pm$3  & 0.90 & 16.39 & 2.85 & 1.4 & isolated-compact$^{\ast}$\\
  75   &  0.15$\pm$0.02 &  16$\pm$5  & 5 $\pm$2   & 0.80 & 19.26 & 1.68 & 3.4 & core+halo/loose\\
  80   &  0.15$\pm$0.02 &   9 $\pm$3  & 4 $\pm$2   & 0.75 &16.23 & 2.33 & 2.3 & core+halo/loose\\
  259 &  0.15$\pm$0.02 &   9 $\pm$4  & 9 $\pm$3   & 0.80 &17.29 & 2.73 & 1.0 & isolated-compact$^{\ast}$\\
\hline
\end{tabular}
\label{tab:Table4}							    
\end{center}
\end{table*}

\section{Observed groups' individual properties}
\label{subsubsec:subsubsec4.2.3}
{\it SHK 10} -- Formed by 13 spectroscopically confirmed member galaxies (Table \ref{tab:TableA1}), this group is characterized by a peaked velocity distribution (Fig. \ref{fig:figA1}, upper panel) and a $M/L=75\ {M_{\odot}\ L_{\odot}^{-1}}$, which suggests a structure close to virialisation. Despite this, its shape appears elongated in the sky plane (Fig. \ref{fig:figA1}, lower panel), which suggests the presence of other galaxies at larger radial distances, a possibility supported by our structural analysis. In fact, Ribeiro's diagnostic plots (Fig. \ref{fig:figA1b}, upper panel) show the presence of a denser and brighter central core at $R_{\rm P}\lesssim200\ {\rm kpc}$ (corresponding to the virial radius calculated for this group), with $\sigma_{\rm r}\sim 400\ {\rm km\ s^{-1}}$, surrounded by a few additional sparse galaxies with velocities consistent with the central members. In any case the characteristics of the group differ from those of compact groups,  since the group is neither as dense nor isolated.
This picture is also corroborated by the structural analysis with Capozzi's diagnostics, which shows the presence of an extended structure ($EI=3$) with numerous galaxies within $500\ {\rm kpc}$ of the group centroid ($N_{500}=23$) and by the presence of a well-defined colour-magnitude relation (Fig. \ref{fig:figA1b}, lower panel) which is populated also by galaxies with radial distances $R>150\ {\rm kpc}$.  
As found for other SHK groups, this group shows a high content of Red-Sequence galaxies (RSGs, galaxies on the RS), with a value of $f(RS)_{150}=0.7$ (the fraction of RSGs within $150\ {\rm kpc}$ of the group centroid, as defined in Cap09 by using the ({\it i-r}) vs. {\it r} colour-magnitude diagram). See Table \ref{tab:TableA1} and Figures \ref{fig:figA1} and \ref{fig:figA1b}.\\

\noindent {\it SHK 71} -- This is the second most massive ($M_{\rm vir}=4.46\times 10^{13}\ {\rm M_{\odot}}$) structure in our sample, characterised by a cluster-like velocity dispersion ($\sigma_{\rm r}=600\ {\rm km\ s^{-1}}$). Although we find 9 spectroscopically confirmed member galaxies, the abnormally large $M/L=337\ {M_{\odot}\ L_{\odot}^{-1}}$ ratio favours either a temporary alignment of galaxies (a scenario favoured by the value of $EI<1.5$) or an infalling structure (situation favoured by the possible presence of a secondary peak in the velocity distribution shown in the upper panel of Figure \ref{fig:figA2}, at $(v-\overline{v}) \sim -1000\ {\rm km\ s^{-1}}$, and of a narrow RS). 
The peak in the distribution of peculiar velocities corresponds to 
galaxies {\it b} and {\it f} showed in the lower panel of Figure \ref{fig:figA2}. We re-calculate all the properties after removing these two galaxies, finding slightly different values ($\sigma_{\rm r}=470^{+90}_{-200}\ {\rm km\ s^{-1}}$, $M_{\rm vir}=3.12\times 10^{13}\ {\rm M_{\odot}}$, $L_{\rm tot}=9.9\times 10^{10}\ {\rm L_{\odot}}$, $M/L=313\ {M_{\odot}\ L_{\odot}^{-1}}$, $R_{\rm vir}=204\ {\rm kpc}$, $\tau_{\rm cross}=49\times 10^{6}\ {\rm yr}$) which, however, do not significantly change the scenario described above. In particular, the value of the $M_{\rm vir}/L_{\rm tot}$ ratio remained abnormally large.
Ribeiro's diagnostics show that $\sigma_{\rm r}$, $\rho$ and $\mu$ remain constant within the errors with increasing $R_{\rm P}$ and that only the first two quantities may be consistent with those of the typical HCG in the group core. On the other hand the structure looks quite compact and isolated, with 9 spectroscopically confirmed member galaxies within $R_{\rm P}=157\ {\rm kpc}$, a result confirmed by the photometric diagnostics ($EI<1.5$, $N_{500}=15\pm 4$). 
With an $f(RS)_{150}=0.8$, this group is dominated by RS galaxies. Note that the galaxy identified as the brightest galaxy of the group (see Table \ref{tab:Table4}) by the structural analysis of Cap09, has been discarded by our spectroscopic measurements because of its velocity (see Table \ref{tab:TableA2}) not satisfying the spectroscopic Hickson criterion. 
See Table \ref{tab:TableA2} and Figures \ref{fig:figA2} and \ref{fig:figA2b}.\\

\noindent {\it SHK 75} -- As for SHK 71, the 7 spectroscopically confirmed member galaxies in this structure (the most massive in our sample with $M_{\rm vir}=6.07\times 10^{13}\ {\rm M_{\odot}}$) may be not virialised or constitute a temporary dense configuration within a larger structure. In fact, their velocity distribution appears flat, with no presence of a dominant peak (see upper panel of Fig. \ref{fig:figA3}). Ribeiro's structural plots (Fig. \ref{fig:figA3b}, upper panel) show the presence of a central triplet at $R_{\rm P}<180\ {\rm kpc}$ surrounded by the other four spectroscopic member galaxies at $R_{\rm P}\simeq300\ {\rm kpc}$. The spatial density and surface brightness of this group, as studied with these diagnostics, are always significantly lower than those of the typical compact groups.
The fairly large value of the $M/L=273\ {M_{\odot}\ L_{\odot}^{-1}}$ ratio probably favours the scenario of a temporary alignment within a larger structure, which is also supported by our structural analysis using photometric redshifts, whose results show the presence of additional galaxies outside the virial radius ($N_{500}=16\pm5$), constituting an extended structure ($EI=3.4$). Although our analysis suggests the presence of a fairly defined RS (Fig. \ref{fig:figA3b}, lower panel) within 500 {\rm kpc} from the group's centroid, most RS galaxies lie outside the central $R<150\ {\rm kpc}$. Despite this, we still see a relatively high value of $f(RS)_{150}= 0.8$. We also point out that, by comparing Tables \ref{tab:Table4} \& \ref{tab:TableA3}, it is possible to see that the brightest galaxy of the group does not reside within the first 150 {\rm kpc} of the group centre. See Table \ref{tab:TableA3} and Figures \ref{fig:figA3} and  \ref{fig:figA3b}.\\

\noindent {\it SHK 80} -- This is the least massive ($M_{\rm vir}=7.6\times 10^{12}\ {\rm M_{\odot}}$) group of our sample, with 6 spectroscopically confirmed member galaxies, whose velocity distribution presents an evident peak (see Fig. \ref{fig:figA4}, upper panel), with $\sigma_{\rm r}=380\ {\rm km\  s^{-1}}$. The radial profiles of $\sigma_{\rm r}$, $\log(\rho)$ and $\mu$ (Fig. \ref{fig:figA4b}, upper panel), show the presence of a probably virialised central core made of 5 spectroscopic member galaxies at $R_{\rm P}<70\ {\rm kpc}$ (which corresponds to the virial radius measured for this group), with properties similar to the typical HCG. Around this central core, at $R_{\rm P}\simeq190 \ {\rm kpc}$, the last spectroscopic member galaxy renders the properties of this group less consistent with those of the typical HCG and may be the sign of the presence of a larger galaxy halo. In fact,  our structural analysis indicates the presence of few more galaxies outside the virial region with an $EI=2.3$. However the S/N is too low to make a definitive conclusion. A fairly defined RS can be seen (Fig. \ref{fig:figA4b}, lower panel), which shows that also this group is dominated by red galaxies ($f(RS)=0.75$). See Table \ref{tab:TableA4} and Figures \ref{fig:figA4} and \ref{fig:figA4b}.\\

\noindent {\it SHK 259} -- This group has very similar properties to SHK 80, with 7 spectroscopically confirmed member galaxies and very similar mass ($M_{\rm vir}=1.97\times 10^{13}\ {\rm M_{\odot}}$), velocity dispersion ($\sigma_{\rm r}=470\ {\rm km\ s^{-1}}$) and total luminosity ($L_{\rm tot}=1.63\times 10^{11}\ {\rm L_{\odot}}$). However, it appears more compact than SHK 80. In fact, Ribeiro's diagnostics (Fig. \ref{fig:figA5b}, upper panel) show a structure with properties always consistent with those of the typical HCG and almost constant with increasing projected distance. We note that, as SHK 75, this group presents a central triplet (at $R_{\rm P}\sim75 \ {\rm kpc}$), surrounded by a compact galaxy halo (made of other 4 galaxies) displaced at $R_{\rm P}\sim120\ {\rm kpc}$ ($\sim$ the $R_{\rm vir}$ of measured for this group). However, despite SHK 259 and 75 have a very similar shape of their radial profiles, SHK 259 has all its spectroscopic member galaxies enclosed in a radial distance about half of the corresponding one of SHK 75. Because of this, SHK 259 can be regarded more as a compact structure rather than, as for SHK 75, a core+halo one. In addition and in accordance to this, Capozzi's structural analysis indicates a compact and isolated structure ($EI=1$) even going deeper in magnitude. According to these diagnostics, this group also has a visible RS and is mostly made of RSGs ($f(RS)_{150}=0.8$).  See Table \ref{tab:TableA5} and Figures \ref{fig:figA5} and \ref{fig:figA5b}.

\section{SHK groups in the literature}
\label{sec:sec5}
In order to understand the general properties of SHK groups we integrate our spectroscopic sample with the additional groups available from the literature, although we have to keep in mind that the two samples are not selected with the same criteria and the data properties are not homogeneous. 
The literature data were collected from the work of Tovmassian, Tiersch and collaborators (\citealp{Tov2008} and references therein; hereafter Tovmassian sample), after applying the corrections needed to match the cosmological model used here. To mitigate the biases involved in using different samples, we also create a \textit{restricted} Tovmassian subsample by applying the same selection criteria ($N_{150}>3$ and $EI>1$) used to select our candidates for spectroscopic follow-up, as described in Section \ref{sec:sec2}. 

\begin{figure}
 \centering
 \includegraphics[width=0.485\textwidth]{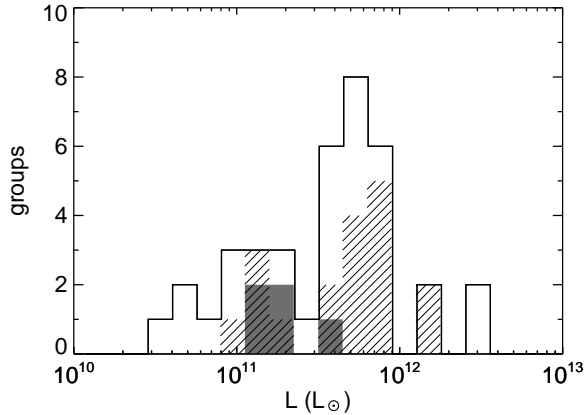}
 \caption{Distributions of r-band total luminosities: our sample (full-colour shaded histogram), Tovmassian sample (empty histogram), Tovmassian restricted sample (line-shaded histogram, see text).}
 \label{fig:Fig3}
\end{figure}

\begin{figure}
 \centering
 \includegraphics[width=0.485\textwidth]{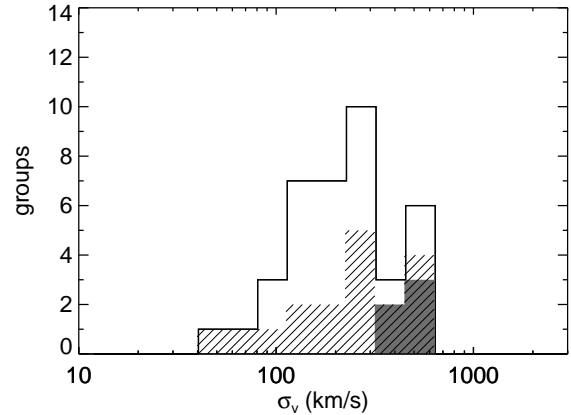}
 \caption{Distributions of radial velocity dispersions: our sample (full-colour shaded histogram), Tovmassian sample (empty histogram), Tovmassian restricted sample (line-shaded histogram, see text).}
 \label{fig:Fig4}
\end{figure}

\begin{figure}
 \centering
 \includegraphics[width=0.485\textwidth]{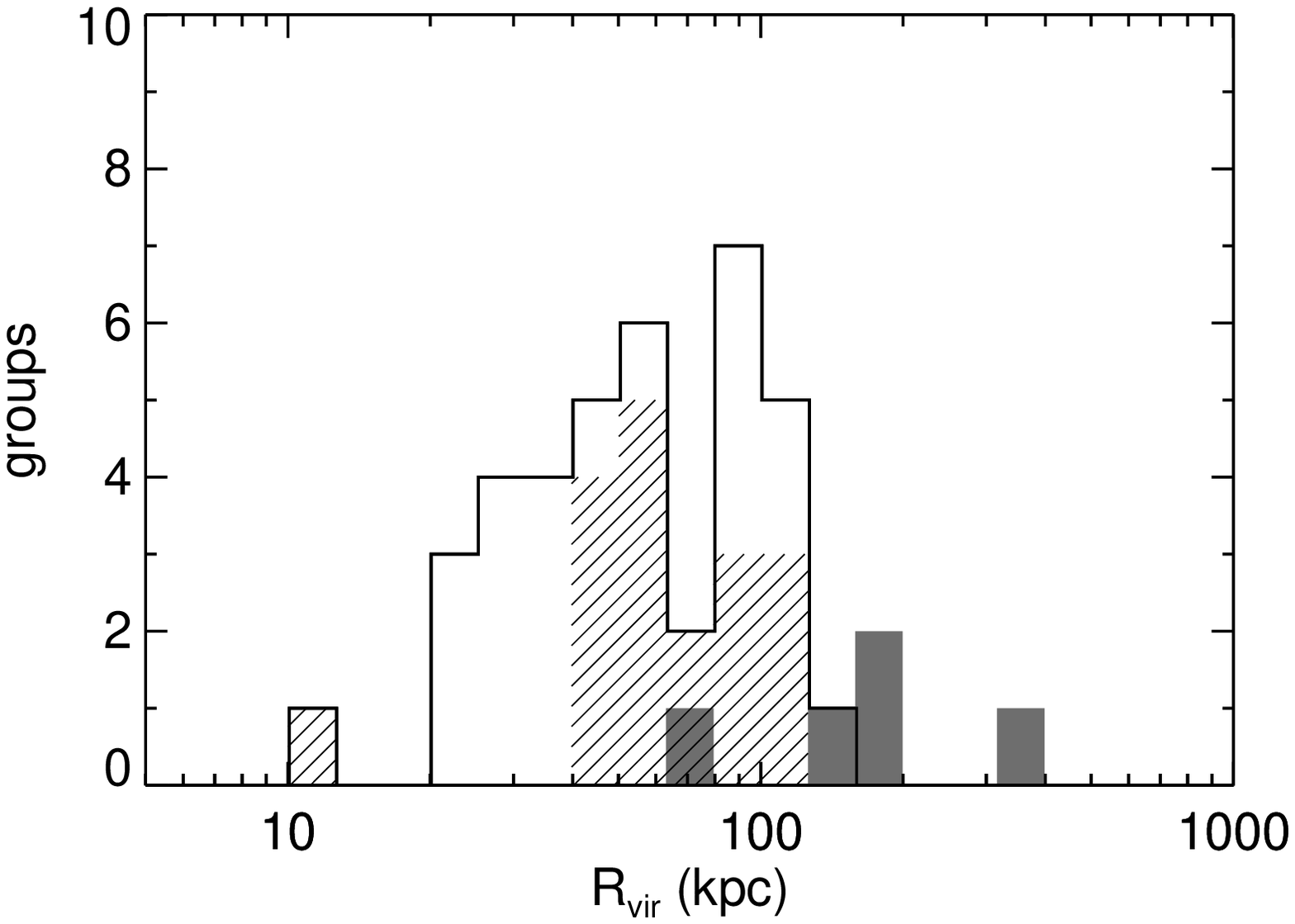}
 \caption{Distributions of virial radius: our sample (full-colour shaded histogram), Tovmassian sample (empty histogram), Tovmassian restricted sample (line-shaded histogram, see text).}
 \label{fig:Fig5}
\end{figure}

\begin{figure}
 \centering
 \includegraphics[width=0.485\textwidth]{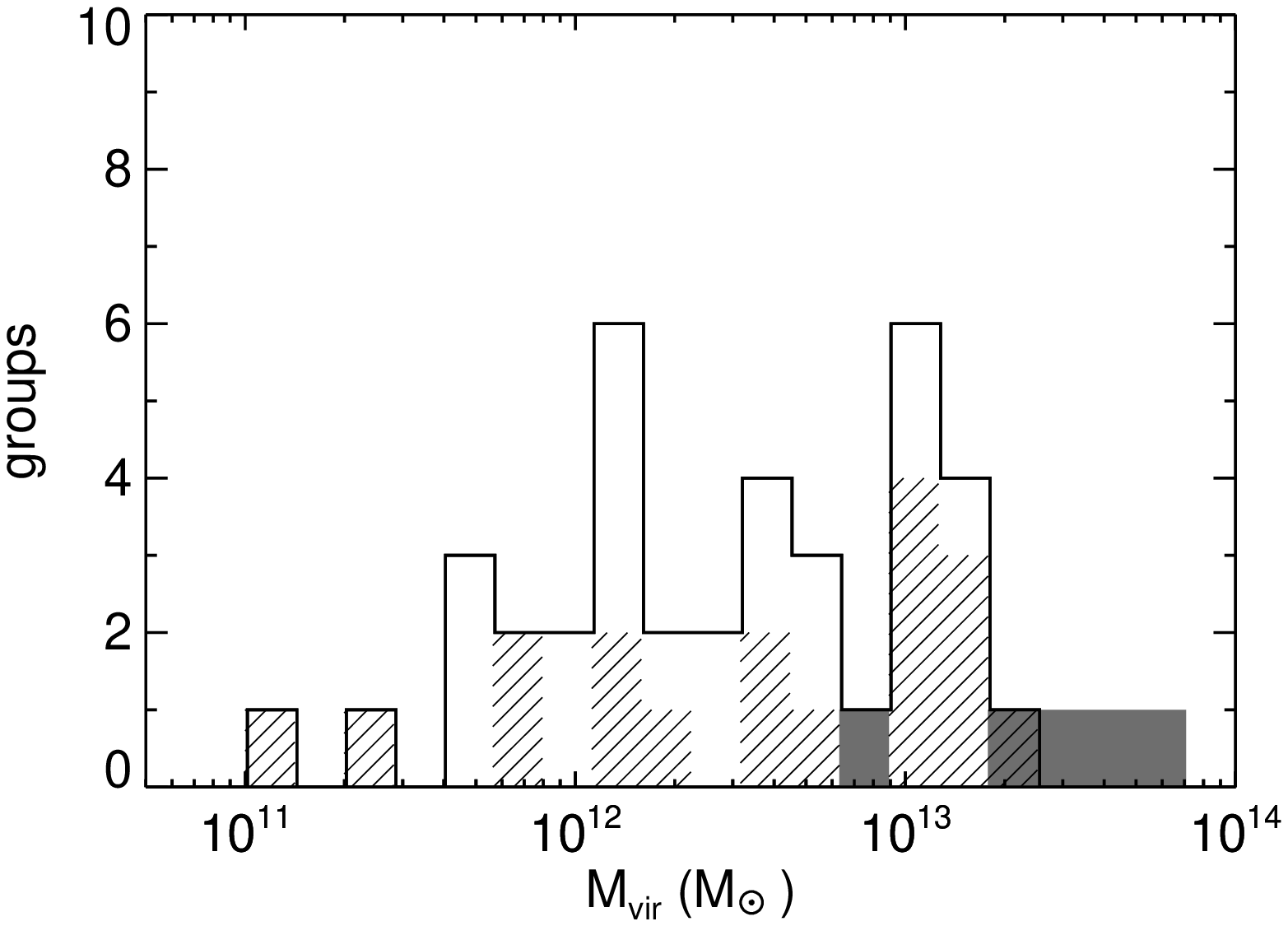}
 \caption{Distribution of virial masses: our sample (full-colour shaded histogram), Tovmassian sample (empty histogram), Tovmassian restricted sample (line-shaded histogram, see text).}
 \label{fig:Fig6}
\end{figure}

\begin{figure}
 \centering
 \includegraphics[width=0.485\textwidth]{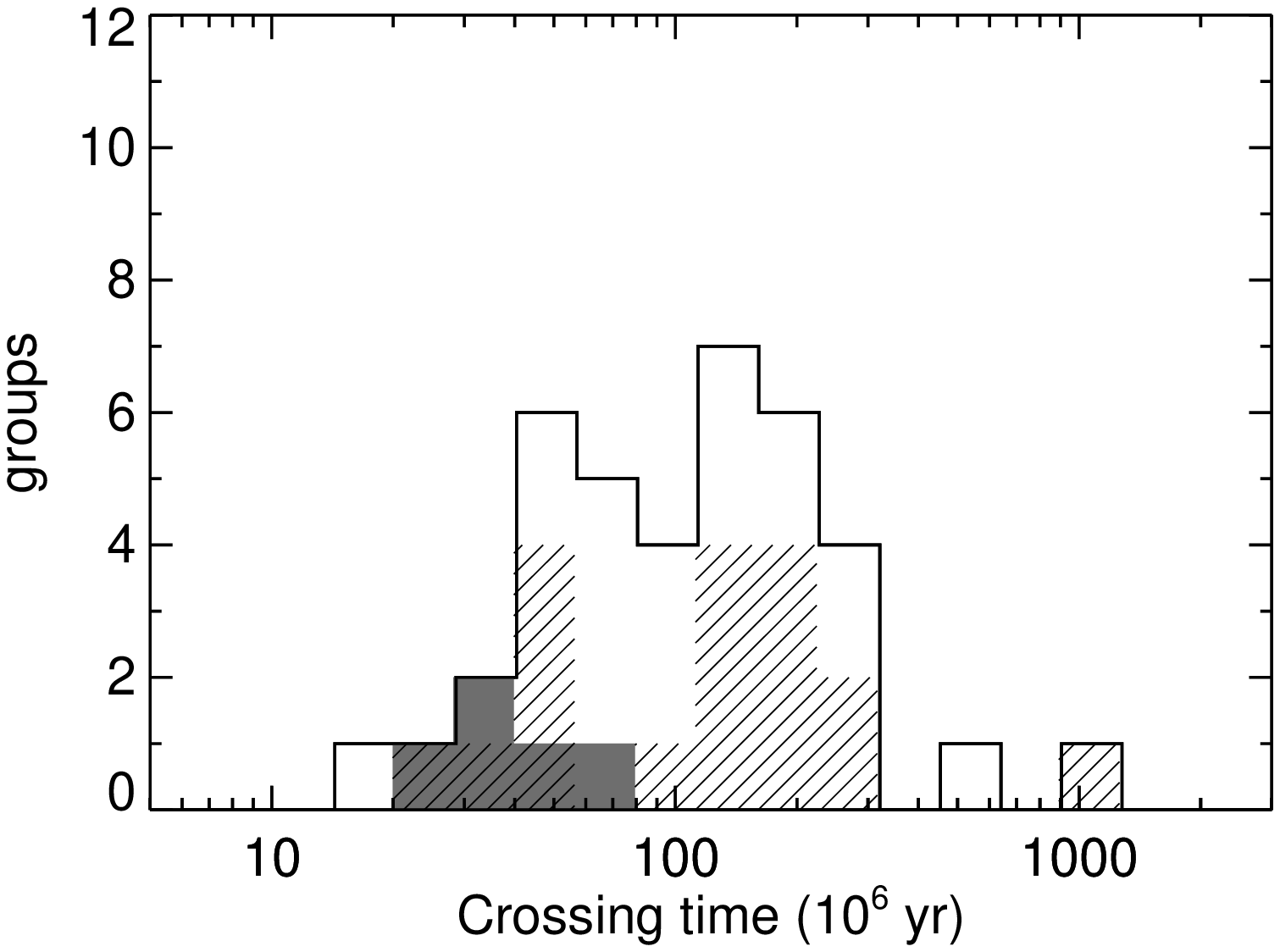}
 \caption{Distributions of crossing times: our sample (full-colour shaded histogram), Tovmassian sample (empty histogram), Tovmassian restricted sample (line-shaded histogram, see text).}
 \label{fig:Fig7}
\end{figure}

\begin{figure}
 \centering
 \includegraphics[width=0.485\textwidth]{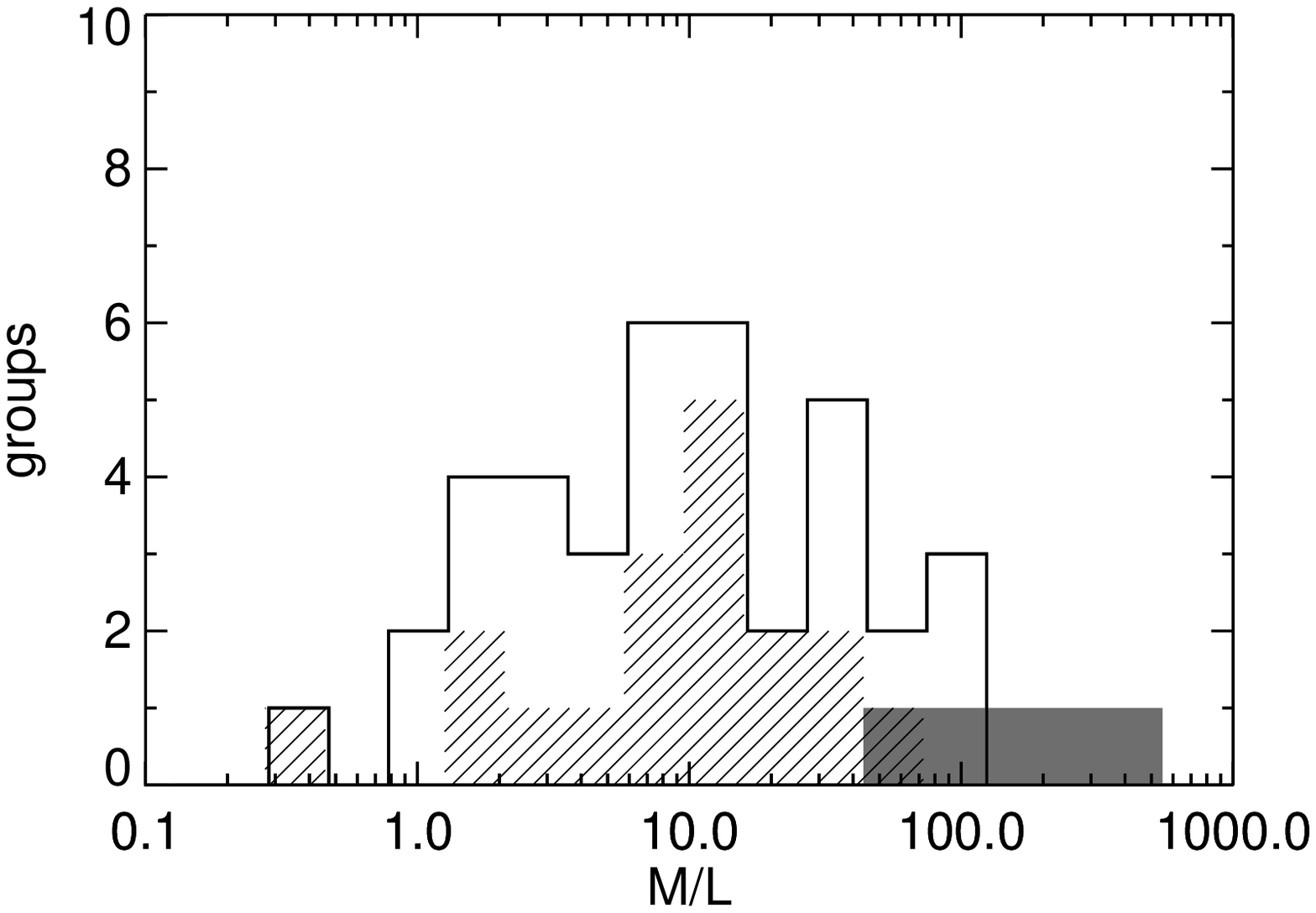}
 \caption{Distributions of $M_{\rm vir}/L_{\rm tot}$ ratios: our sample (full-colour shaded histogram), Tovmassian sample (empty histogram), Tovmassian restricted sample (line-shaded histogram, see text).}
 \label{fig:Fig8}
\end{figure}

\begin{table*}
\begin{center}
\caption{{\bf Tovmassian restricted group sample dynamical properties}. {\it Col 1}: group number; {\it col 2} \& {\it col 3}: luminosity-weighted centroid coordinates based on spectroscopically confirmed members only; {\it col 4}: average spectroscopic redshift; {\it col 5}: number of spectroscopically confirmed member galaxies; {\it col 6}: radial velocity dispersion; {\it col 7}: virial radius; {\it col 8}: total SDSS {\it r}-band luminosity; {\it col 9}: virial mass; {\it col 10}: $M_{\rm vir}/L_{\rm tot}$ ratio; {\it col 11}: crossing time. The errors on $\sigma_{\rm r}$ were estimated via bootstrap as 68 per cent confidence intervals.}
\begin{threeparttable}
\begin{tabular}{lcccccccccc}
\hline
  \multicolumn{1}{l}{\bf SHK} &
  \multicolumn{1}{c}{\bf RA} &
  \multicolumn{1}{c}{\bf Dec} &
  \multicolumn{1}{c}{$\mathbf{z_{\rm {\bf spec}}}$} &
  \multicolumn{1}{c}{$\mathbf{N_{\rm {\bf spec}}}$} &
  \multicolumn{1}{c}{$\mathbf{\sigma_{\rm r}}$} &
  \multicolumn{1}{c}{$\mathbf{R_{\rm {\bf vir}}}$} &
  \multicolumn{1}{c}{$\mathbf{L_{\rm {\bf tot}}}$} &
  \multicolumn{1}{c}{$\mathbf{M_{\rm {\bf vir}}}$} &
  \multicolumn{1}{c}{$\mathbf{M_{\rm {\bf vir}}/L_{\rm {\bf tot}}}$} &
  \multicolumn{1}{c}{$\mathbf{\tau_{\rm {\bf cross}}}$} \\
  \multicolumn{1}{c}{}&
  \multicolumn{1}{c}{\rm hh:mm:ss} &
  \multicolumn{1}{c}{\rm dd:mm:ss} &
  \multicolumn{1}{c}{} &
  \multicolumn{1}{c}{\rm {(gal)}}&
  \multicolumn{1}{c}{$(\rm{km/s})$} &
  \multicolumn{1}{c}{$(\rm{kpc})$} &
  \multicolumn{1}{c}{$(10^{11}\ {\rm L_{\odot}})$} &
  \multicolumn{1}{c}{$(10^{13}\ {\rm M_{\odot}})$} &
  \multicolumn{1}{c}{(${\rm M_{\odot}/L_{\odot}}$)} &
  \multicolumn{1}{c}{$(10^{6}\ {\rm yr})$} \\
\hline
  14  & 14:25:19.20  & +47:15:00.0  & 0.0745$\pm$0.0018     &\, 7	& $500^{+200}_{-100}$  & 43   & 1.20  & 0.89  & 74  & 8    \\
  19  & 13:28:30.05  & +15:50:21.5  & 0.0688$\pm$0.0018     &\, 4	& $490^{+50}_{-300}$  & 18   & 0.50  & 0.29  & 59   & 2     \\
  31  & 00:58:18.22  & +13:54:45.0  & 0.1877$\pm$0.0016     &\, 5	& $500^{+100}_{-200}$  & 114  & 3.90  & 1.85  & 47   & 16    \\
  74  & 14:21:10.18  & +43:03:49.3  & 0.1038$\pm$0.0014     &\, 8	& $290^{+80}_{-70}$  & 131  & 2.41  & 0.75  & 31   & 53   \\
  120 & 11:04:29.59  & +35:52:40.1  & 0.0701$\pm$0.0015     &\, 7	& $400^{+100}_{-200}$  & 61   & 0.44  & 0.79  & 182  & 12    \\
  154 & 11:22:53.57  & +01:06:59.0  & 0.0725$\pm$0.0016     &\, 5	& $400^{+70}_{-200}$  & 123  & 2.04  & 1.34  & 66   & 19    \\
  181\tnote{1} & 08:28:01.08  & +28:16:03.7  & 0.0934$\pm$0.0017     &\, 7	& $340^{+70}_{-80}$  & 61   & 2.06  & 0.48  & 23   & 16    \\
  188\tnote{2} & 09:56:59.62  & +26:10:18.1  & 0.0823$\pm$0.0020     &\, 6	& $500^{+100}_{-300}$  & 172  & 2.44  & 3.28  & 134  & 26    \\
  191 & 10:48:08.47  & +31:28:45.1  & 0.1151$\pm$0.0020     & 11	& $500^{+100}_{-100}$  & 99   & 3.88  & 2.05  & 53   & 24   \\
  223 & 15:49:46.63  & +29:10:10.6  & 0.0835$\pm$0.0017     & 10        & $500^{+100}_{-200}$  & 70   & 2.64  & 1.24  & 47   & 20    \\
  245 & 12:24:43.99  & +31:56:48.8  & 0.0616$\pm$0.0020     &\, 5	& $500^{+100}_{-200}$  & 79   & 1.48  & 1.65  & 112  & 13    \\
  254 & 13:56:23.57  & +35:11:24.4  & 0.0652$\pm$0.0009     &\, 4	& $220^{+61}_{-100}$  & 190  & 0.42  & 0.64  & 154  & 55   \\
  344 & 08:47:31.73  & +03:42:03.2  & 0.0775$\pm$0.0002     &\, 4	& $50^{+10}_{-10}$   & 99   & 1.01  & 0.02  & 2    & 96  \\
  346 & 09:15:10.82  & +05:14:21.8  & 0.1349$\pm$0.0005     &\, 5	& $140^{+50}_{-50}$  & 56   & 2.09  & 0.08  & 4    & 34   \\
  348 & 09:26:34.94  & +03:26:56.4  & 0.0886$\pm$0.0024     &\, 7	& $600^{+100}_{-200}$  & 126  & 2.29  & 3.24  & 141  & 21    \\
  351 & 11:10:20.86  & +04:48:16.2  & 0.0295$\pm$0.0019     &\, 7	& $400^{+100}_{-100}$  & 91   & 0.90  & 0.90  & 99   & 14    \\
  360 & 15:41:26.38  & +04:44:04.2  & 0.1062$\pm$0.0022     &\, 8	& $900^{+200}_{-100}$  & 74   & 2.95  & 4.06  & 138  & 9    \\
  376 & 13:56:35.33  & +23:21:31.3  & 0.0667$\pm$0.0017     &\, 8	& $300^{+100}_{-100}$  & 106  & 1.63  & 0.59  & 36   & 19    \\
\hline							     
\end{tabular}
\begin{tablenotes}\footnotesize 
\item[1] One of the 8 objects considered as member galaxies of this group in the literature is classified as a star in the SDSS catalogue. In order to be conservative, we have excluded it when carrying out our analysis. However there is not a significant change in this group's properties when the mentioned dubious object is included in our calculations. 
\item[2] As with SHK 181 for the 7 objects considered as member galaxies in the literature. 
\end{tablenotes}
\end{threeparttable}
\label{tab:Table5}							    
\end{center}
\end{table*}

In Fig. \ref{fig:Fig3} we compare the distribution of the total optical luminosities: our groups' total luminosities overlap with those quoted in the literature 
for the Tovmassian total and restricted samples. Despite this overlap, our groups' luminosities are on average lower than those of the Tovmassian groups; even limiting the comparison to the restricted sample, we see that more than half of its members have total luminosities greater than that of our brightest group. This could be due to the small size of our sample, but we point out that assuming an average luminosity $L_{\rm r}^{\star}\simeq 10^{10.3}\ L_{\odot}$ (e.g. \citealp{Ber2003a, Bla2003}), a total luminosity close to $10^{12}\ L_{\odot}$ would imply a group with $>50\ L^{\star}$ galaxies or 10 giant galaxies of $M_{\rm r}<-23\ {\rm mag}$. As will be discussed in more detail below, it is likely that the total luminosities quoted in the literature (\citealp{Tov2008} and references therein) were incorrectly estimated.

In Figures  \ref{fig:Fig4}, \ref{fig:Fig5}, \ref{fig:Fig6}, \ref{fig:Fig7} and \ref{fig:Fig8} we compare the velocity dispersion, virial radius, virial mass, crossing time and $M_{\rm vir}/L_{\rm tot}$ ratio. 
Our groups always tend to occupy the upper/lower-end tails of the distributions derived from the literature, with 
some of our values being completely out of the range spanned by those measured by Tovmassian and collaborators, even when applying the same
selection criteria. This is further surprising since the targets chosen for MOS spectroscopy were selected in order to span the whole range of richness and extensions of SHK groups.

It is difficult to understand the origin of these discrepancies. It is true that the formulae used to obtain the derived properties for the SHK groups reported in the literature are slightly different from those used here (i.e., we estimate the dynamical quantities following \citealp{Fir2006}, while the values reported 
in the literature are more similar to those described in \citealp{Per1990}). However, we did test the influence of using different formulations of the virial mass and find no significant change in the picture just described above. 
The lower values of $M_{\rm vir}/L_{\rm tot}$ ratio found for SHK groups within Tovmassian sample and restricted sample may thus be explained if we take into account
that their total luminosities may have been over-estimated, possibly explaining the observed differences with our groups' luminosities. In fact, summing the published luminosities of all member galaxies in the catalogues for some of these groups, we obtain total luminosities which are a factor 2 lower than those quoted in the same papers, partially reconciling the observed differences.  In any case, several of the total luminosities reported in the literature appear unrealistically large for galaxy groups, as discussed in the previous section.

\begin{figure*}
 \centering
 \includegraphics[width=0.42\textwidth]{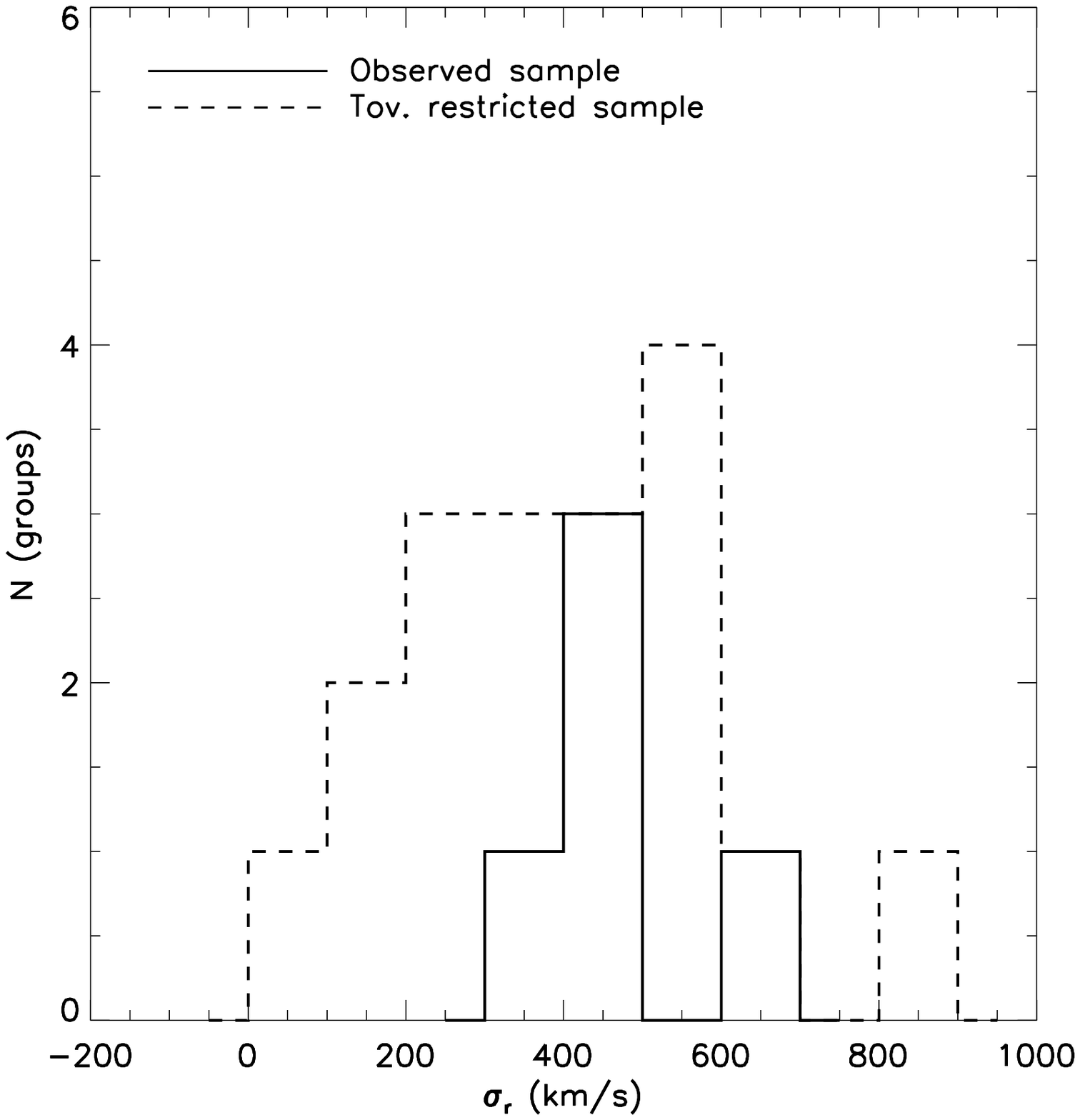}
 \includegraphics[width=0.42\textwidth]{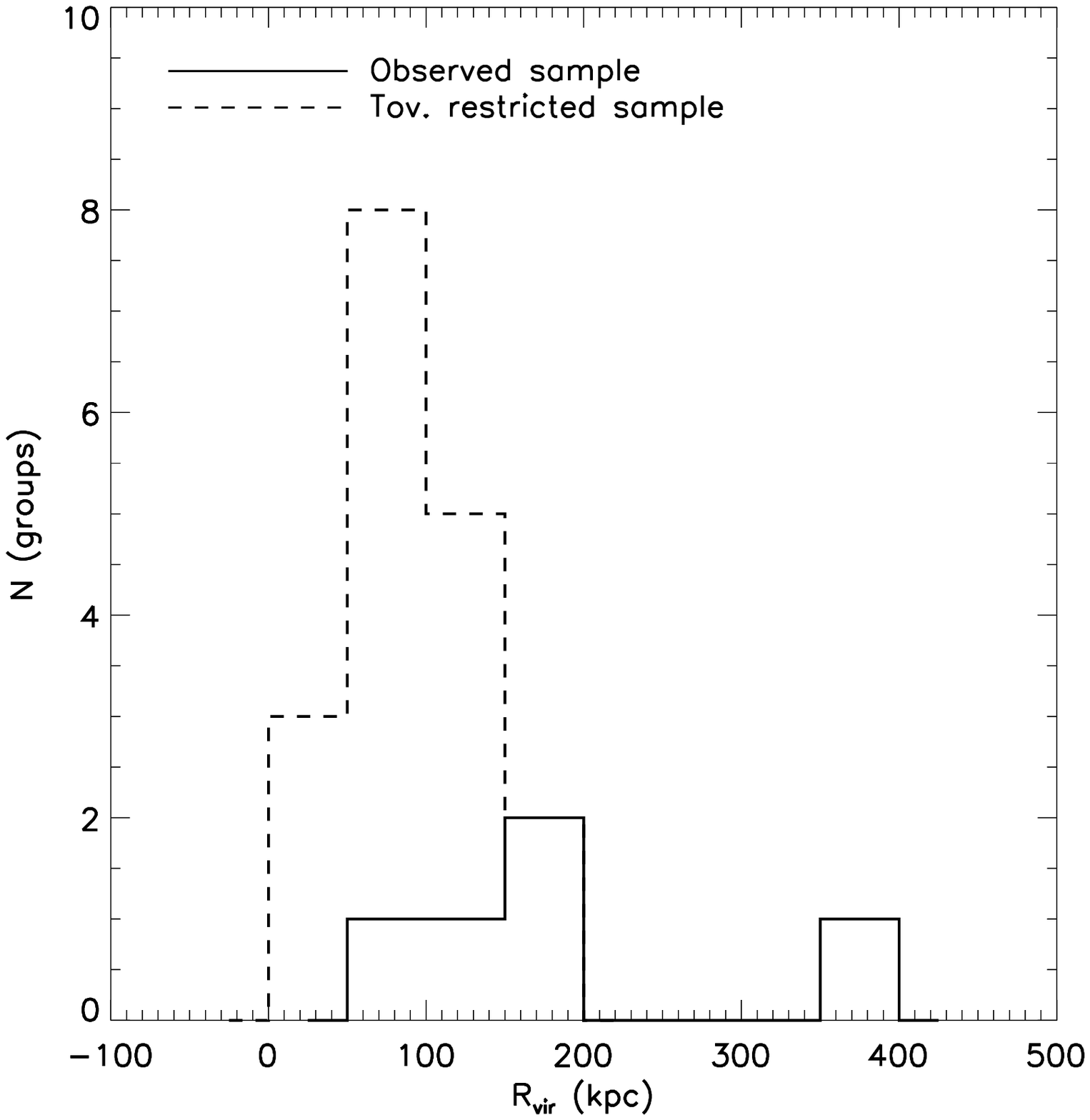}
 \includegraphics[width=0.42\textwidth]{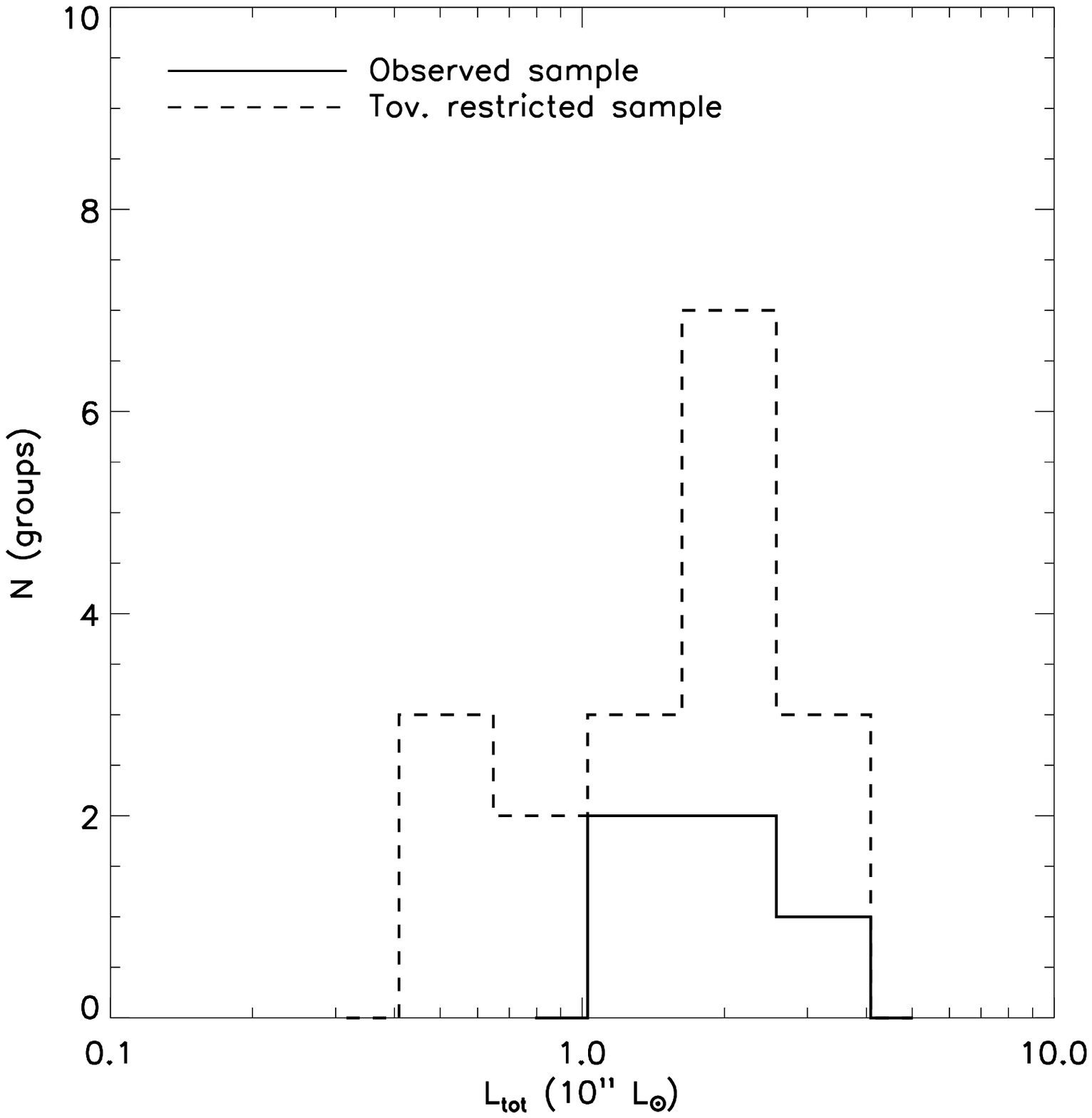}
 \includegraphics[width=0.42\textwidth]{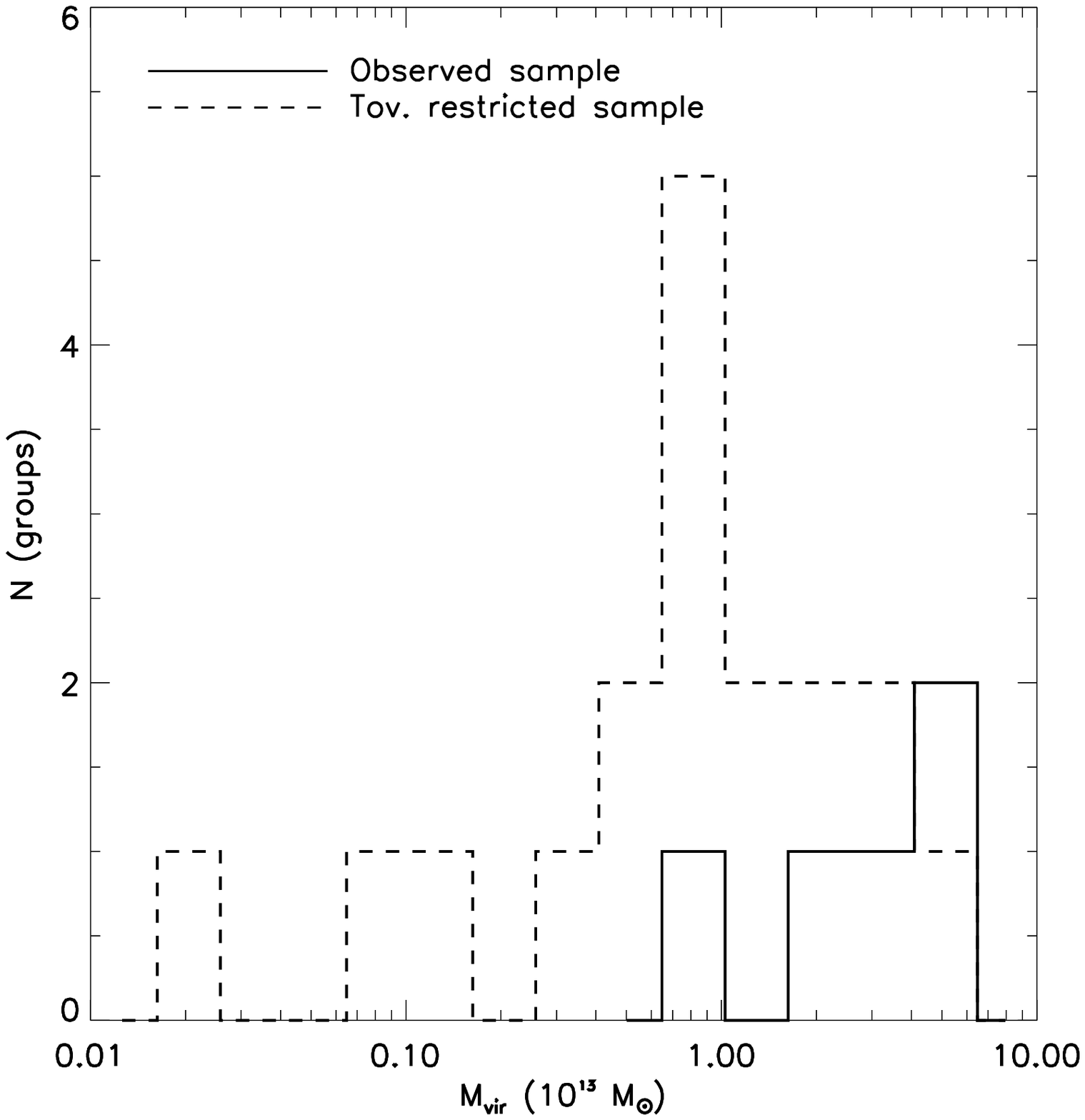}
 \includegraphics[width=0.42\textwidth]{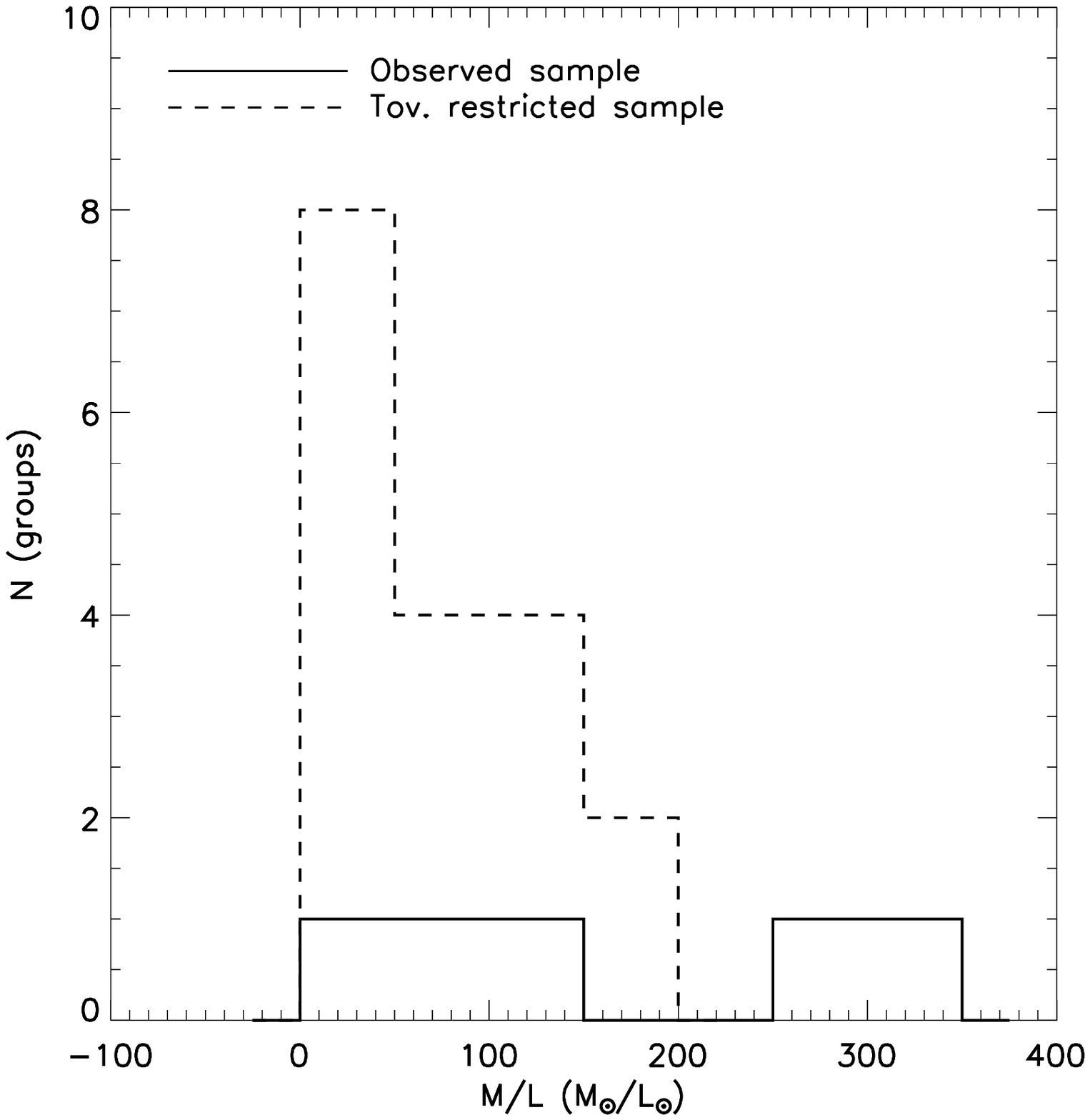}
 \includegraphics[width=0.42\textwidth]{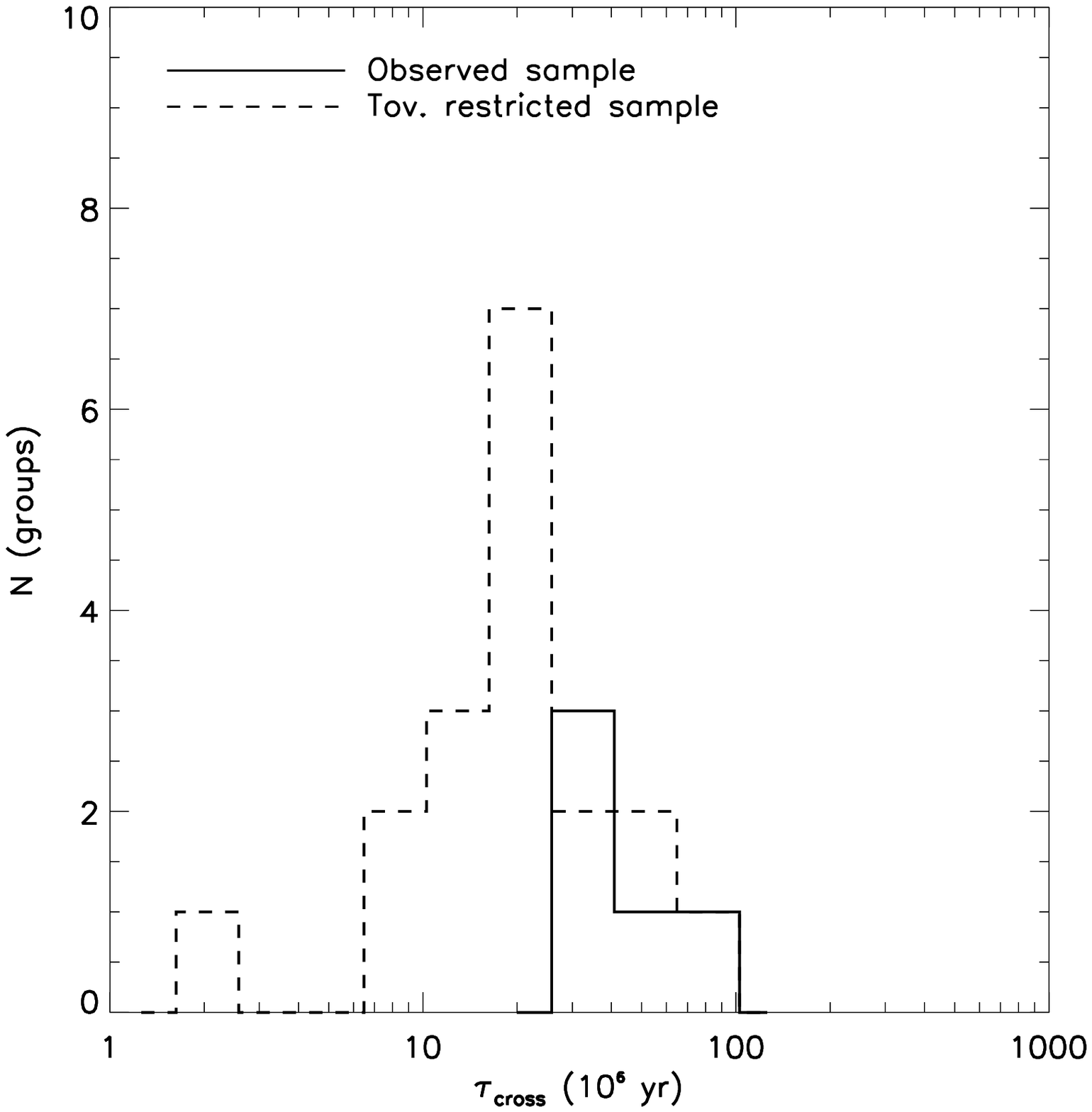}
 \caption{Property distributions of our observed SHK sample (full line) and of Tovmassian restricted SHK sample (dashed lines): $\sigma_{\rm r}$ (upper left panel), $R_{\rm vir}$ (upper right panel), $L_{\rm tot}$ (middle left panel), $M_{\rm vir}$ (middle right panel), $M_{\rm vir}/L_{\rm tot}$ ratio (lower left panel) and $\tau_{\rm cross}$ (lower right panel). We point out that all the properties of the latter sample have been re-determined by us.}
 \label{fig:Fig9}
\end{figure*}

To further investigate these discrepancies, we re-estimate the published properties of the SHK groups in these works, 
applying their formulae to the tabulated data of individual galaxies, and compare our estimates with those quoted there. 
We find discrepancies among the quoted values of $R_{\rm vir}$, $M_{\rm vir}$, $L_{\rm tot}$, $M_{\rm vir}/L_{\rm tot}$ and $\tau_{\rm cross}$ and those we re-measure using the very same 
method. 
For instance, if we repeat the analysis of SHK 245 described in \citet{Tov2007}, we find that the values for $M_{\rm vir}$, $R_{\rm vir}$, 
$\tau_{\rm cross}$, $M_{\rm vir}/L_{\rm tot}$ and $L_{\rm tot}$ quoted in their study are respectively a factor 0.23, 0.49, 3.9, 0.1 and 2.3 of those we find using their data and formulae. 
For these reasons, we re-determined the dynamical properties of the SHK groups contained in the Tovmassian restricted sample defined above, by applying our formulae to the published radial velocities, positions and using SDSS\footnote{The Sloan Digital Sky Survey web site is http://www.sdss.org} magnitudes. The re-determined dynamical properties of these groups are listed in Table \ref{tab:Table5}. In addition we performed the same structural analysis by using Ribeiro's diagnostics (the analysis on these groups with Capozzi's diagnostics was already carried out in Cap09). Results of the dynamical and structural analysis of SHK 14 and SHK 181 groups are shown in {\sc Appendix} \ref{sec:appendix} as explanatory cases. The re-determined properties of the Tovmassian restricted sample will be compared to our observed groups in next section.We simply point out now, as an explanatory case study,  that by applying Tovmassian and collaborators formula to their published data for SHK 154  \citep{Tie2002}, we find a value of $\sigma_{\rm r}=430\ {\rm km/s}$, which is  a factor 2 larger than what quoted in Table 8 of their paper ($\sigma_{\rm r}=215\ {\rm km/s}$), but in much better agreement (better than 1 $\sigma$ level) with the value estimated by us and quoted in Table \ref{tab:Table5}.

\section{Results and Discussion}
\label{sec:sec6}
\begin{figure}
 \centering
 \includegraphics[width=0.485\textwidth]{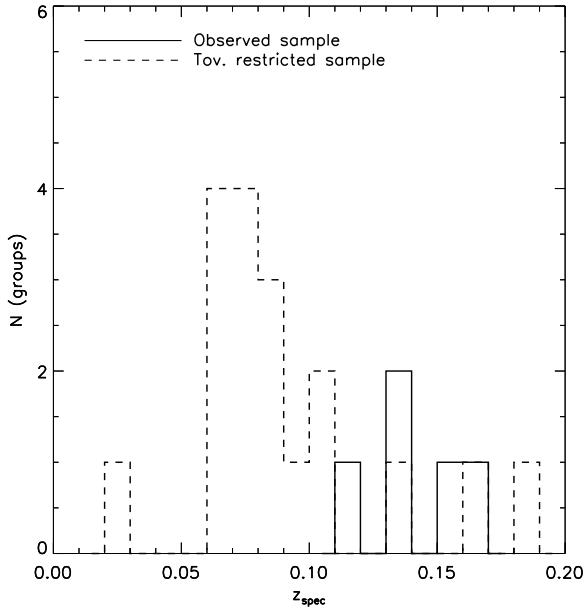}
 \caption{Spectroscopic redshift distribution of our observed sample (full line) and of Tovmassian restricted sample (dashed line).}
 \label{fig:Fig10}
\end{figure}

In Figure \ref{fig:Fig9} we now compare the properties of our spectroscopic targets (full lines) with those we re-estimated for the Tovmassian restricted sample (dashed lines), i.e. $\sigma_{\rm r}$, $R_{\rm vir}$, $L_{\rm tot}$, $M_{\rm vir}$, $M_{\rm vir}/L_{\rm tot}$ ratio and $\tau_{\rm cross}$. On average we notice that now the two samples are in much better agreement: in particular the total luminosities and the velocity dispersions are largely consistent. On the other hand our 5 SHK groups still seem to have larger than average virial radii, crossing times and masses. 
As it can be seen in Figure \ref{fig:Fig10}, our SHK groups tend to lie at higher redshift than those in Tovmassian restricted sample; this could explain part of the difference observed in the virial radius and consequently in the crossing times, as both of these quantities depend on the field-of-view (FOV) covered by the observations. Being at higher redshift (cf. Tables \ref{tab:Table1} \& \ref{tab:Table5}), our DOLORES observations tend to sample most of the group extension, as suggested by the fact that there are no confirmed group members among the target galaxies for SHK 71, 80 and 259 at the edge of the FOV, and only one in SHK 10 and 75. This may not be the case for the literature data which cover a smaller linear FOV due to their lower redshifts and sometimes due to the instrument used (for instance, when the {\it DSAZ Calar Alto} spectrtograph was used, the available FOV was $\sim 4'\times4'$.) In addition, even when the instrument used was able to cover a larger FOV (like when the {\it Landessternwarte Faint Object Spectrograph} on the {\it Guillermo Haro Observatory} was used, with a FOV of $10'\times6'$), the fact that the authors restricted their analysis only to galaxies included in the catalogue by \citet[and references therein]{Sto1997} could be equivalent to covering a smaller FOV. We also note that the outliers in our sample distributions are represented by SHK 71 and 75, which our analysis suggests to be non-relaxed or contaminated structures.

\begin{figure*}
 \centering
 \includegraphics[width=0.485\textwidth]{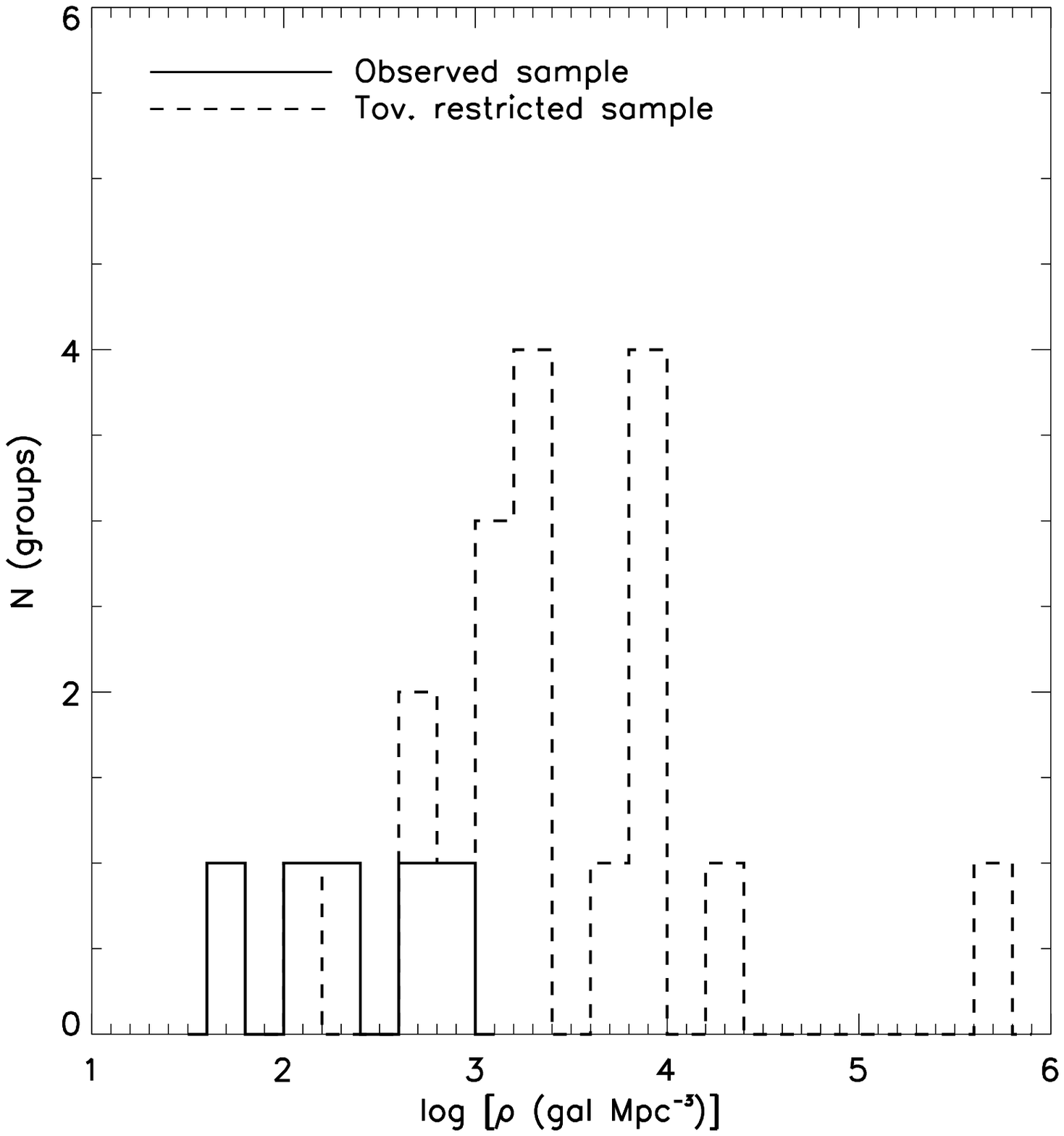}
  \includegraphics[width=0.485\textwidth]{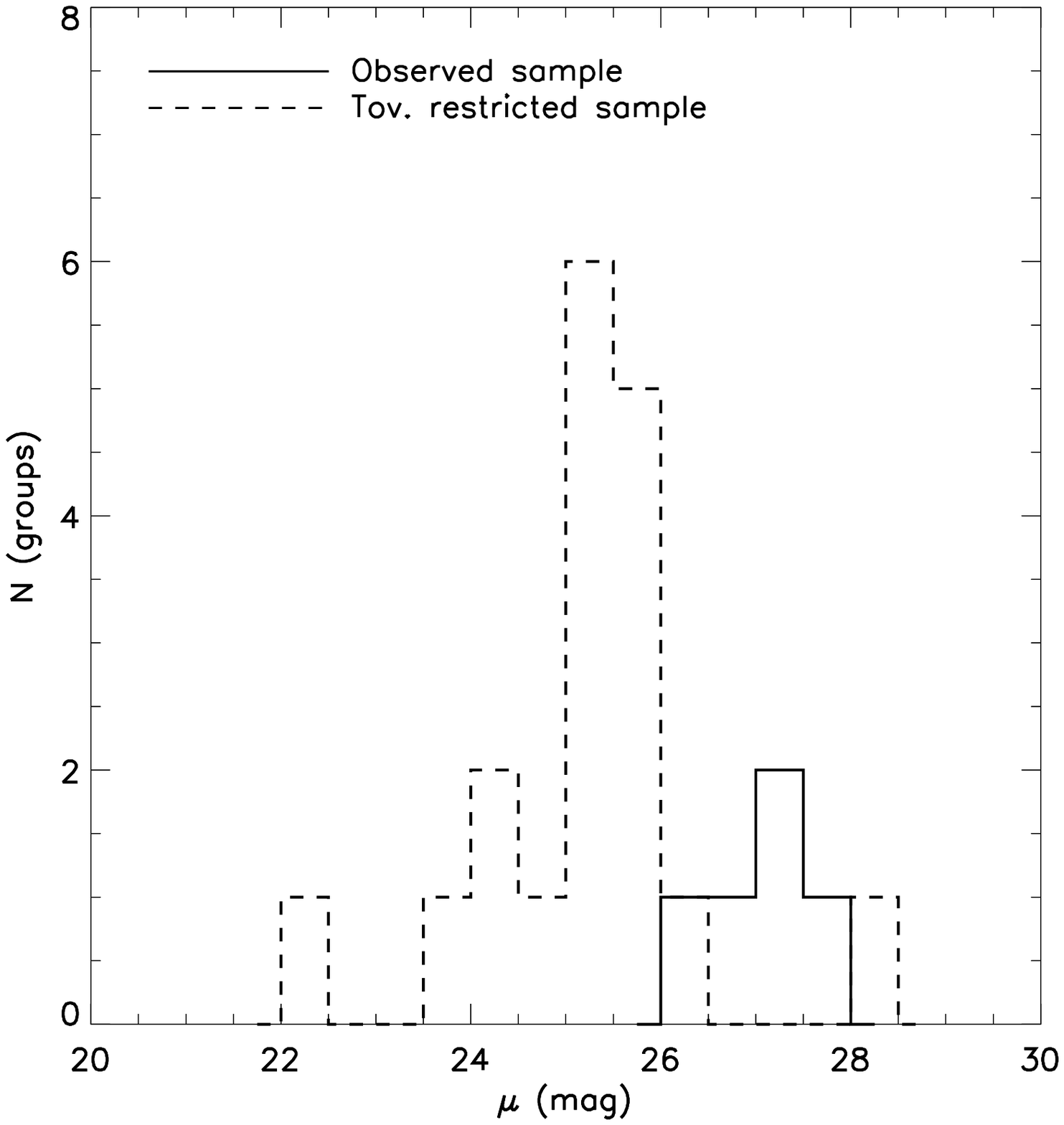}
 \caption{Distributions of $\log(\rho)$ (left panel) and $\mu$ (right panel)  for our observed groups (full line) and Tovmassian restricted sample (dashed line).}
 \label{fig:Fig11}
\end{figure*}

Turning to the structural properties of the groups, as derived with Ribeiro's diagnostics, we note that the observed sample tends to be constituted by different types of structures, only SHK 259 being similar to a HCG group. In addition, they tend to be less dense and less bright groups compared to the Tovmassian restricted sample, as Figure \ref{fig:Fig11} highlights. This may again be due to the selection effects explained above. 
We also point out that the groups contained in Tovmassian restricted sample have properties closer to those of the typical HCG, when using Ribeiro's diagnostics (apart from SHK 254, 348 and 360), than our observed groups. The latter, in fact, appear more heterogeneous, including among them structures ranging from HCG-like groups (like SHK 259) to systems formed by a core with HCG-like properties and a surrounding halo, as for SHK 10. 

It is also worth noticing that, in addition to the selection effects mentioned above,Tovmassian observations were $\sim 1\ {\rm mag}$ shallower than ours. As a consequence of all these effects, SHK 181 (whose diagnostics are shown in Appendix \ref{sec:appendix}), for instance, would be classified as an HCG-like group based on Ribeiro's diagnostics only. However, this structure is definitely more extended: it is less than 6 {\rm arcmin} distant from an Abell or Zwicky cluster (Cap09) and has been classified as a maxBCG cluster by \citet{Koe2007}, with 26 member galaxies within $R_{200}$ (the radius containing an over-density 200 times critical). In accordance to this picture, Cap09 diagnostics, which can probe galaxies within a magnitude limit $\sim 2\ {\rm mag}$ deeper than that of Tovmassian galaxy sample thanks to the use of photometric redshifts, are able to correctly detect SHK 181 as an extended structure, classifying it as a cluster core within a $500\ {\rm kpc}$ scale. 

Another example is SHK 19, which satisfies Hickson selection criteria (Cap09) and is found to be an HCG-like group after carrying out Ribeiro's diagnostics on Tovmassian's data, with 4 spectroscopic member galaxies within $R_{\rm P}\sim12\ {\rm kpc}$, $\sigma_{\rm r}=490^{+70}_{-200}$, $\log(\rho)=5.7^{+0.2}_{-0.8}$ and $\mu=22.2^{+0.4}_{-0.3}$. However, when analyzing the same group down to a fainter magnitude limit and out to a larger radial distance, making use of photometric diagnostics based on SDSS data, one realises that SHK 19 is actually an extended structure with many more galaxies surrounding the central structure. In fact, Cap09 found $N_{500}=17\pm5$ galaxies within $500\ {\rm kpc}$ from the group centroid, with an Extension Index of $EI=3.6$. 

\begin{figure}
 \centering
 \includegraphics[width=0.485\textwidth]{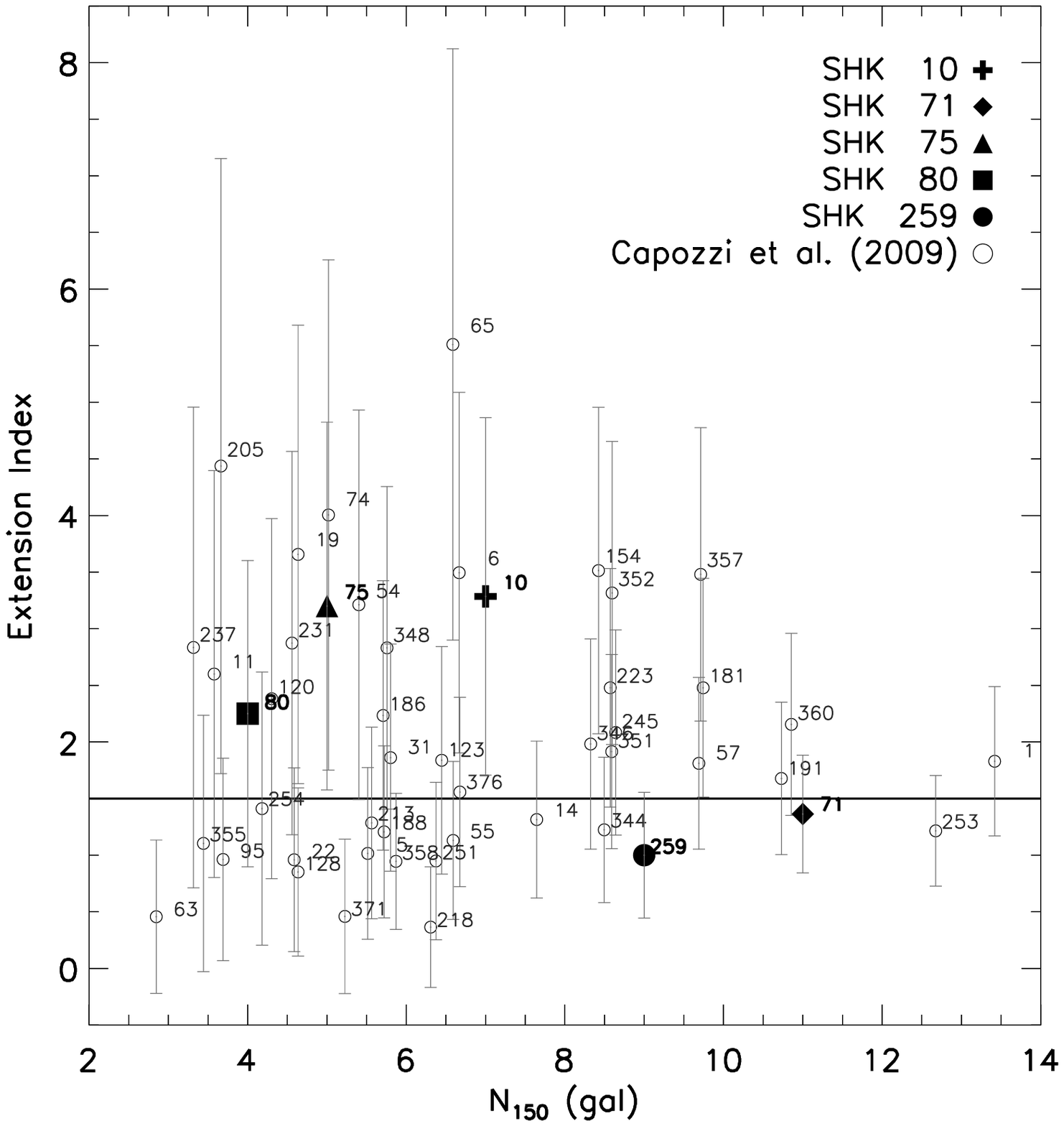}
 \caption{Extension index vs. richness within $150\ {\rm kpc}$ from group centroid. These quantities are measured as described in Cap09 (see text).}
 \label{fig:Fig12}
\end{figure}

\begin{figure}
 \centering
 \includegraphics[width=0.485\textwidth]{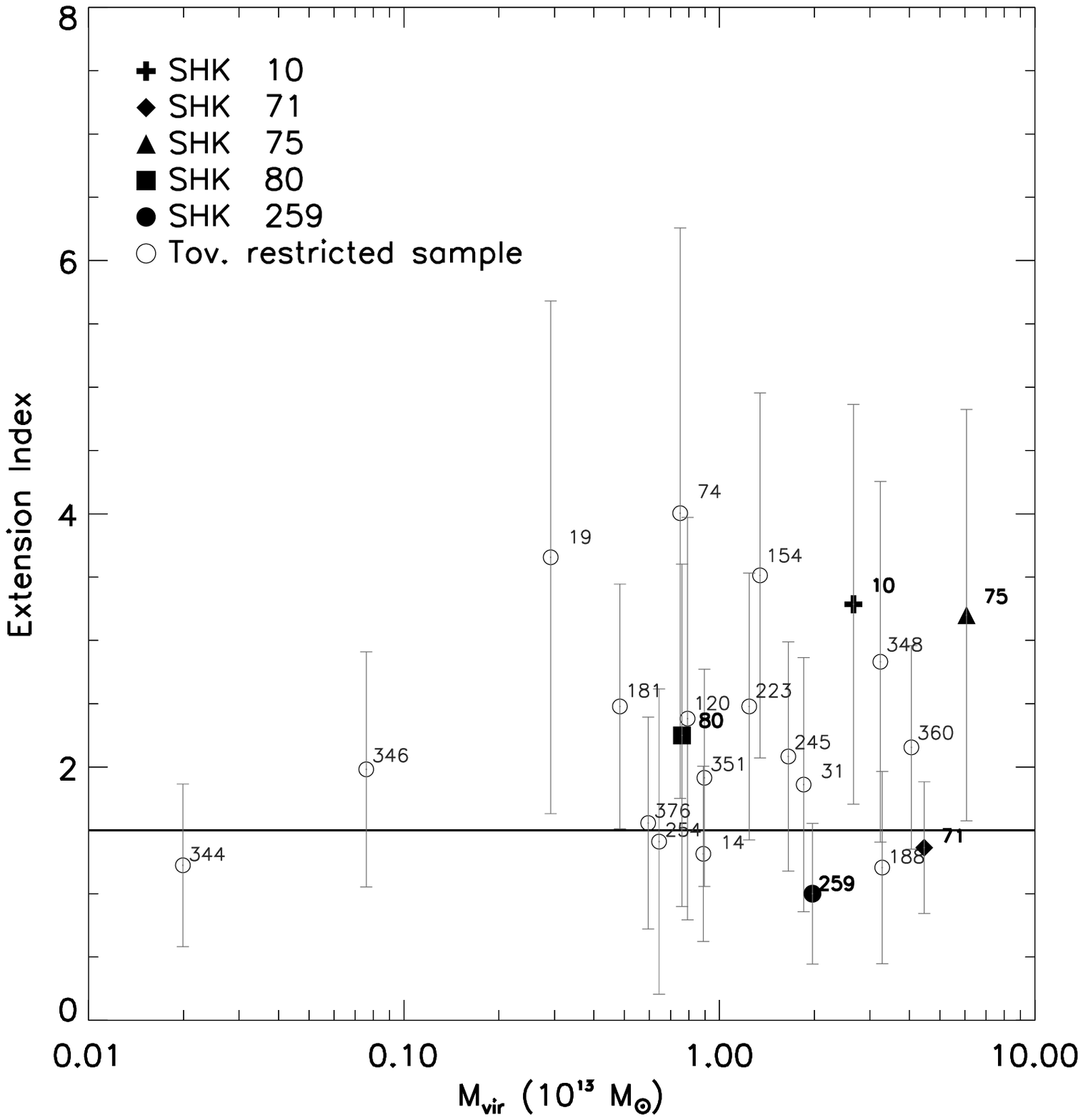}
 \caption{Extension index (as measured in Cap09) vs. virial mass.}
 \label{fig:Fig13}
\end{figure}

The photometric diagnostics thus allow a more homogeneous comparison with our observed sample, with respect to what can be done by using spectroscopic data assembled from the literature. Capozzi's diagnostics in fact have the advantage of relying on the same data and, due to the large SDSS coverage, are less affected by selection biases for all the SHK groups studied in this work.
Figures  \ref{fig:Fig12} and \ref{fig:Fig13} and Table \ref{tab:Table4} show that the SHK groups in our sample 
present characteristics similar to those of the groups studied in Cap09, when the same structural analysis is carried out. 
They cover the entire $EI$ vs. $N_{150}$ plot (Fig. \ref{fig:Fig12}), thus indicating the presence of systems in 
different configurations. However, no cluster-core systems  ($N_{150}>7$ and $EI>1.5$) are found among the 5 structures studied here. Considering that SHK 71 is possibly an un-bound system, only SHK 10 has properties very close to those of a cluster-core system as classified in Cap09. 
When inspecting the dependence of $EI$ with $M_{\rm vir}$ (Fig. \ref{fig:Fig13}), despite our 5 groups occupying the massive end of the plot (as also shown in the distribution middle right panel of Figure \ref{fig:Fig9})
we do not see a significant trend between these two quantities, as also indicated by a Spearmann's correlation test.   

Finally, as shown in Table \ref{tab:Table4}, our groups present a high content of RSGs ($f(RS)_{150}$ is always $>0.7$), in accordance with the findings of Cap09 and as expected, given the original selection criteria used to identify SHK groups, set up to select red galaxies. In fact the majority of the studied groups seems to have a RS in place (only SHK 75 shows a weakly defined RS when considering only galaxies within $R=150\ {\rm kpc}$).

The results of our analysis confirm those found in our previous paper (based on photometric redshifts) indicating that most SHK groups are real spatial over-densities of galaxies and not the result of projection effects. In fact we can now confirm that, considering the 45 confirmed groups (out of 58) studied in our previous paper and the 5 studied here, $\sim 80$ per cent (49 groups, SHK 10 is included in both these samples) of the 62 SHK groups studied in total are likely to be real structures, with their spectroscopic redshift being consistent with the photometric estimate.  
The method is also effective in identifying member galaxies: out of 67 galaxies with good S/N ratio and with $|z_{\rm phot}-\langle z_{\rm phot}\rangle|\leq3\epsilon(z_{\rm phot})$ (see Section \ref{sec:sec3}), 87 per cent (58 galaxies) were confirmed to have $z_{\rm spec}$ such that $|z_{\rm spec}-\langle z_{\rm phot}\rangle|\leq3\epsilon(z_{\rm phot})$, and 50 per cent were confirmed group members.
 
As proposed in Cap09 for the final sample of 45 confirmed SHK groups, the measured properties identify heterogeneous structures, with characteristics ranging from those of compact and isolated systems to those of loose groups 
(Figs. \ref{fig:Fig12} \& \ref{fig:Fig13}), and mostly populated by red galaxies.
In particular, SHK 71 and 75 could be unbound structures (Section \ref{sec:sec6}), given their high $M_{\rm vir}/L_{\rm tot}$ ratios. This situations holds independently of being 
an extended or a more compact and isolated structure (see Fig. \ref{fig:Fig12}), since SHK 71 and 75 show values of $EI=1.4$ and $EI=3.4$, respectively. 
As for the remaining groups studied, while SHK 10 and 80 can be considered as [core+halo] or loose configurations (for SHK 80 such a classification is more uncertain 
given the high noise due to its low multiplicity), SHK 259 can be included among those few rich ($N_{150}>7$) and compact ($EI<1.5$) SHK groups which, despite rare, 
were present also in the sample studied by Cap09. It is worth pointing out that none of the studied groups could be classified as cluster core according to Cap09 criteria 
($N_{150}>7$ and $EI>1.5$), a fact which will be further investigated when studying the remaining 10 SHK groups of our spectroscopic follow-up.\\ 

\section{Conclusions}
\label{sec:sec8}
In this work we aimed at confirming and extending the investigation on the properties of SHK groups of galaxies, undertaken by Cap09.
We pre-selected candidate groups within the SHK sample using the method described in Cap09, which makes use 
of diagnostics in both the projected and photometric redshift spaces, to carry out a spectroscopic follow-up. We present here the 
results for the first five groups (SHK 10, 71, 75, 80 \& 259), derive their dynamical and structural 
properties and compare them with those of other SHK groups in the literature. Our main conclusions are as follows:\\

\noindent (i) We find between 6 and 13 spectroscopically confirmed members per group (i.e., with $|v_{\rm M}-v_{\rm i}|<1000\ {\rm km\ s^{-1}}$ 
out to {\it r}=20) thus confirming that our pre-selection of group candidates based on photometric redshifts works well in 
selecting physically bound structures of low multiplicity. In particular, 87 per cent of observed galaxies with reliable velocity measures (67 out of 86) have photometric redshifts within the expected $3\epsilon(z_{\rm phot})$ range from group spectroscopic one and 50 per cent are found to be spatially associated to their groups. Individually, 100, 86, 70, 73 and 100 per cent of the photometrically-selected and observed galaxies of SHK 10, 71, 75, 80 and 259, respectively, found within the expected $3\epsilon(z_{\rm phot})$ scatter, are confirmed spectroscopic members. This, together with the results of our previous paper, supports the conclusion that most SHK groups are real spatial over-densities of galaxies and are not due to chance projection or contamination effects, although some of them (in our sample SHK 71 and 75) may be not virialised structures.\\

\noindent (ii) When compared to other studies in the literature our analysis strongly suggests that the properties of the samples published in the literature so far have been incorrectly determined, indicating that the properties of the SHK sample should be revisited with a more robust and homogeneous approach.  In fact, when literature data are re-analysed by us, we find that the newly derived  properties are much more consistent with those measured for our spectroscopic sample. The remaining differences are most likely due to selection effects, caused by different FOV, magnitude depth, redshift range and  the observation of catalogued galaxies only. \\

\noindent (iii) The studied group sample is characterised by heterogeneous structures, corroborating the claim of Cap09 according to which SHK
groups are an heterogeneous collection of galaxy structures, ranging from compact and isolated systems to loose groups. However, the studied sample does not show the presence of cluster-core systems. The 
results from the forthcoming study of the remaining 10 SHK groups of our spectroscopic campaign will allow us to more thoroughly investigate whether 
the lack of such systems is due to the poorness of the sample studied here. \\

\noindent (iv) Investigating the environment surrounding galaxy groups when carrying out spectroscopic observations revealed to be crucial. In fact, when this is not done, systems found to be poor, compact and isolated may actually be richer, more dispersed and/or dwelling within over-dense environments, as shown, for instance, for SHK 181 and SHK 19 and as already found by other authors (e.g. \citealp{Rib1998}) for the more studied HCGs.\\

Because of the significant discrepancies found in the estimates of the SHK groups' physical properties found in the literature, 
it will be fundamental estimating those of our entire spectroscopic sample (15 SHK groups). In fact, given the homogeneous approach we adopt in both our dynamical and structural analyses, this will allow us to more robustly assess the characteristics of the SHK group sample as a whole, confirm their heterogeneous nature and estimate the fraction of likely bound structures. We plan to address all these matters in detail in a future paper, once properties for all our groups are measured.\\

\subsection*{Acknowledgments}

Based on observations made under the programme ID AOT18-TAC38 with the Italian Telescopio Nazionale Galileo (TNG)
operated on the island of La Palma by the Fundaci\'{o}n Galileo Galilei of the INAF
(Istituto Nazionale di Astrofisica) at the Spanish Observatorio del Roque de los
Muchachos of the Instituto de Astrofisica de Canarias.
The authors thank the anonymous referee for his valuable comments. They also thank Ilaria Formicola for the help provided with the photometric pre-selection.
DC acknowledges financial support from the University of Portsmouth and thanks Claudia Maraston and Lado Samushia for valuable discussions.
MS acknowledges financial contribution from the  ``Fondi di Ateneo 2011'' (ex 60 $\%$) of Padua University.
MP acknowledges support from Miur grant PRIN 2009 and grant FARO 2011 from the Federico II University.
GL wishes to thank the Department of Astronomy at the California Institute of Technology for support and hospitality.

\bibliographystyle{mn2e}

\bibliography{Reference}
\appendix

\section{Notes on individual systems}
\label{sec:appendix}
Here we show the tables and the images summarizing the spectroscopic membership of the galaxies within the SHK groups studied individually. In addition, Ribeiro's diagnostic plots and the colour-magnitude diagrams
produced by our structural analysis are also shown for each studied SHK group.\\

\begin{table*}
\begin{center}
\begin{large}
\caption{{\bf Spectroscopic membership of SHK 10}. {\it Col 1}: galaxy number; {\it col 2} \& {\it col 3}: galaxy coordinates; {\it col 4}: SDSS dereddened {\it r}-band magnitude; {\it col 5}: radial velocity; {\it col 6}: spectroscopic membership. Note that when a spectroscopic member galaxy is within both our and the SDSS spectroscopic catalogues, this galaxy is only counted once for the determination of the group spectro-photometric properties. In particular only the galaxy contained in our spectroscopic galaxy catalogue is considered. Galaxies with no velocity measurement are those whose spectra had too a low S/N ratio to produce significant CCFs (see Section \ref{sec:sec4}).}  
\begin{threeparttable} 
\begin{tabular}{l c c c c l}        
\hline\hline
 & & &{\bf SHK 10} & & \\
\hline                 
Galaxy\tnote{1} & R.A.        & Dec.        & {\it r}                            & v                                 & Membership \\    
             &({\rm hh:mm:ss})& ({\rm dd:mm:ss})&  ({\rm mag})           & ({\rm km/s})               &                        \\
\hline                       
01 (a) &14:10:48.19    &+46:15:57.7	&16.31  		 &$40396\pm45$      &yes\\		     
02 (b) &14:10:33.27    &+46:15:30.4	&16.89  		 &$40292\pm60$      &yes (SDSS)\\	     
03     &14:10:32.89    &+46:14:57.3	&16.89  		 &$41789\pm88$      &no\\		     
04 (c) &14:11:01.43    &+46:15:38.6	&16.98  		 &$40532\pm60$      &yes (SDSS)\\	     
05     &14:10:56.38    &+46:15:55.4	&17.11  		 &$33698\pm56$      &no\\		     
06     &14:10:56.38    &+46:15:55.4	&17.11  		 &$33397\pm30$      &no$\:$ (SDSS same as 05)\\  
07 (d) &14:10:51.44    &+46:16:34.6	&17.44  		 &$40082\pm60$      &yes (SDSS)\\	     
08 (e) &14:10:50.10    &+46:16:46.2	&17.43  		 &$39637\pm70$      &yes\\		     
09 (f) &14:10:43.46    &+46:16:57.8	&17.53  		 &$40262\pm60$      &yes (SDSS)\\	     
10 (g) &14:10:50.77    &+46:16:08.0	&17.62  		 &$39375\pm61$      &yes\\		     
11 (h) &14:10:33.41    &+46:15:27.0	&17.67  		 &$40854\pm119      $&yes\\		     
12 (i) &14:10:27.24    &+46:18:14.3	&17.70  		 &$39303\pm30$      &yes (SDSS)\\	     
13     &14:10:27.16    &+46:16:12.1	&17.88  		 &$42091\pm123      $&no \\		     
14     &14:10:58.15    &+46:14:34.0	&18.15  		 &$42199\pm50$      &no\\		     
15 (j) &14:10:58.53    &+46:16:01.3	&18.19  		 &$40794\pm64$      &yes\\		     
16     &14:10:50.48    &+46:16:14.9	&18.28  		 &$36557\pm60$      &no\\		     
17     &14:10:40.95    &+46:18:52.7	&18.68  		 &--		    &--\\		     
18 (k) &14:10:58.18    &+46:15:47.6	&18.74  		 &$40196.5\pm100$   &yes\\		     
19 (l) &14:10:47.56    &+46:17:02.7	&18.90  		 &$39737\pm113$     &yes \\		     
20     &14:10:54.12    &+46:16:20.5	&19.02  		 &$36618\pm189$     &no\tnote{2} \\		     
21     &14:10:37.62    &+46:17:02.0	&19.09  		 &$50233\pm97$      &no \\		     
22 (m) &14:10:44.38    &+46:17:02.1	&19.14  		 &$39961\pm98$      &yes \\		     
23     &14:11:02.27    &+46:14:55.0	&19.19  		 &$40771\pm112$     &no\tnote{3} \\		     
24     &14:10:26.39    &+46:17:45.4	&19.24  		 &$58885\pm68$      &no\\		     
25     &14:10:59.51    &+46:15:12.6	&19.69  		 &$37971\pm148$     &no \\		     
\hline  				   		    						
\end{tabular}
\begin{tablenotes}\footnotesize 
\item[1] Letters are associated to spectroscopically confirmed member galaxies only, ordered from the brightest to the faintest. They do not indicate the original nominal scheme used by Shakhbazyan and collaborators.
\item[2] This galaxy has been excluded because it does not satisfy the criteria described in Section \ref{subsec:subsec4.1}, due to the presence of multiple and comparable peaks in the CCF. However, no one of these peaks corresponds to a velocity satisfying Hickson criterion. 
\item[3] As for 2, however, one of the secondary peaks in the CCF corresponds to a velocity satisfying Hickson criterion. The inclusion or exclusion of this galaxy does not affect our results.
\end{tablenotes}
\end{threeparttable}
\label{tab:TableA1}
\end{large}      
\end{center}
\end{table*}

\vspace{5cm} 
\clearpage

\begin{figure*}
\begin{center}
\includegraphics[width=0.62\textwidth]{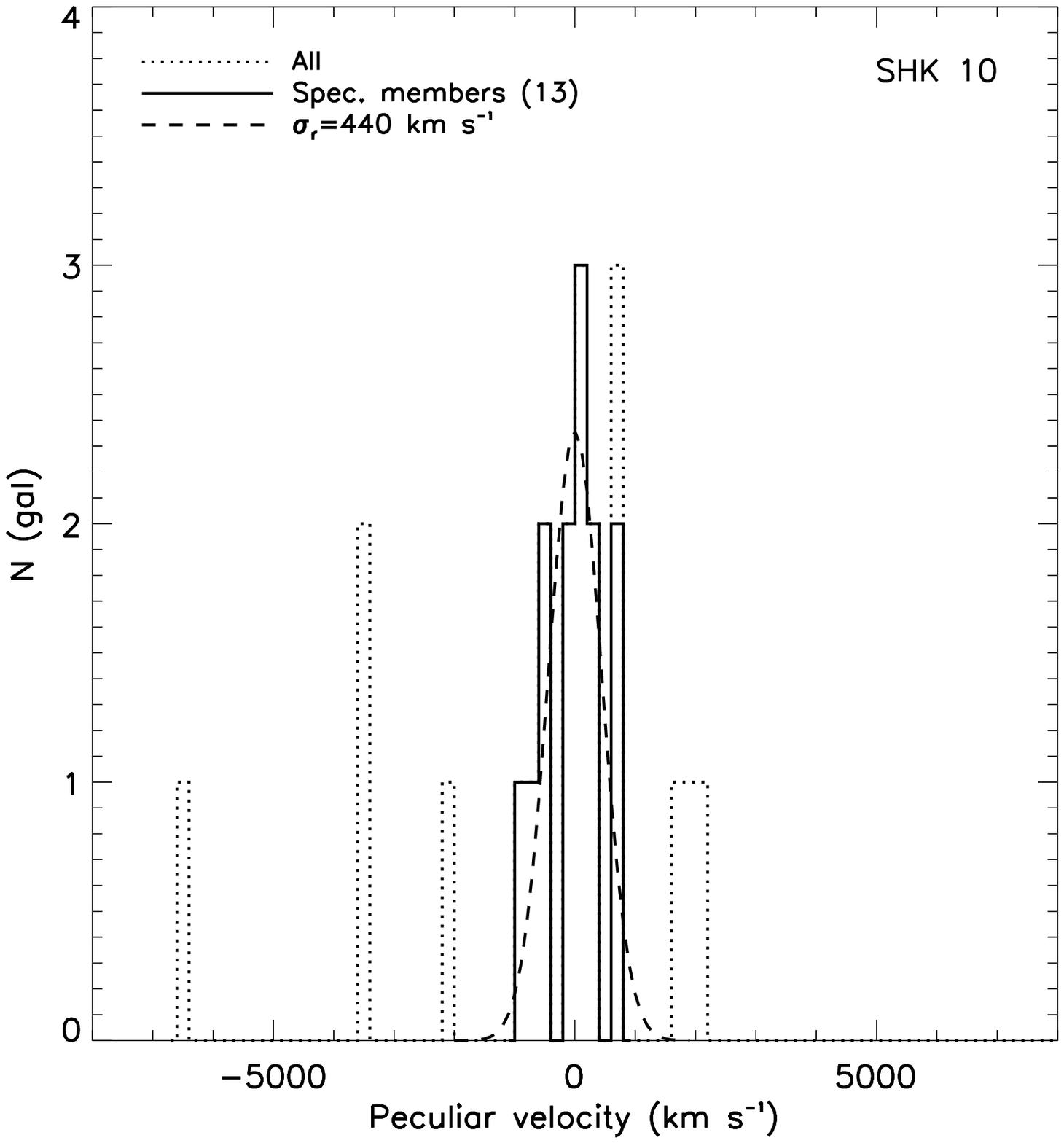}\\
\qquad  \includegraphics[width=0.55\textwidth]{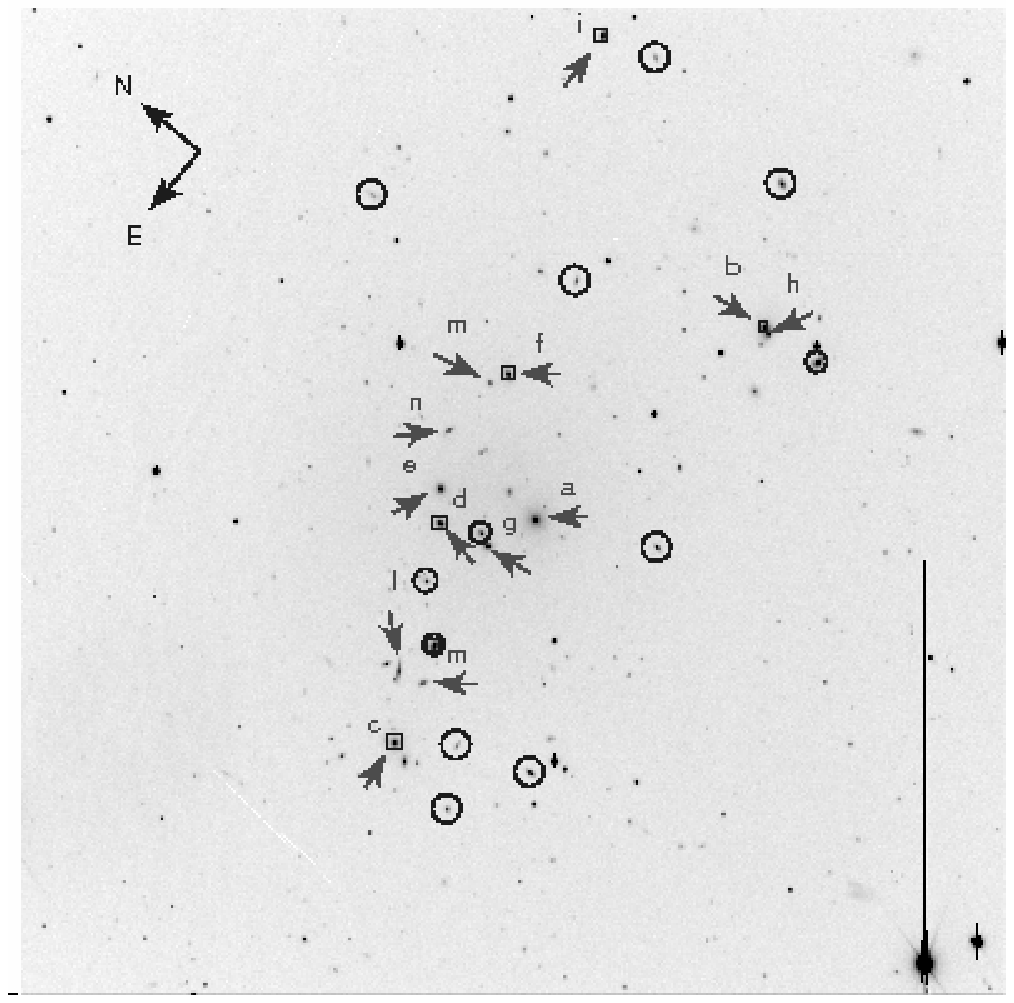}
\caption{{\bf SHK 10.} {\it Upper panel}: velocity distribution obtained by using spectroscopically confirmed member galaxies. A Gaussian distribution centred on zero and defined by the value of the estimated $\sigma_{\rm r}$ is over-plotted (dashed line). {\it Lower panel}: {\it r}-band image showing observed and SDSS spectroscopically confirmed group member galaxies (arrows), observed non-member galaxies (circles) and galaxies from the SDSS spectroscopic catalogue (squares). The size of the field is $8.6\ {\rm arcmin}\times 8.6\ {\rm arcmin}$ (DOLORES imaging mode).}
\label{fig:figA1}
\end{center}
\end{figure*}

\clearpage

\begin{figure*}
\begin{center}
\includegraphics[width=0.62\textwidth]{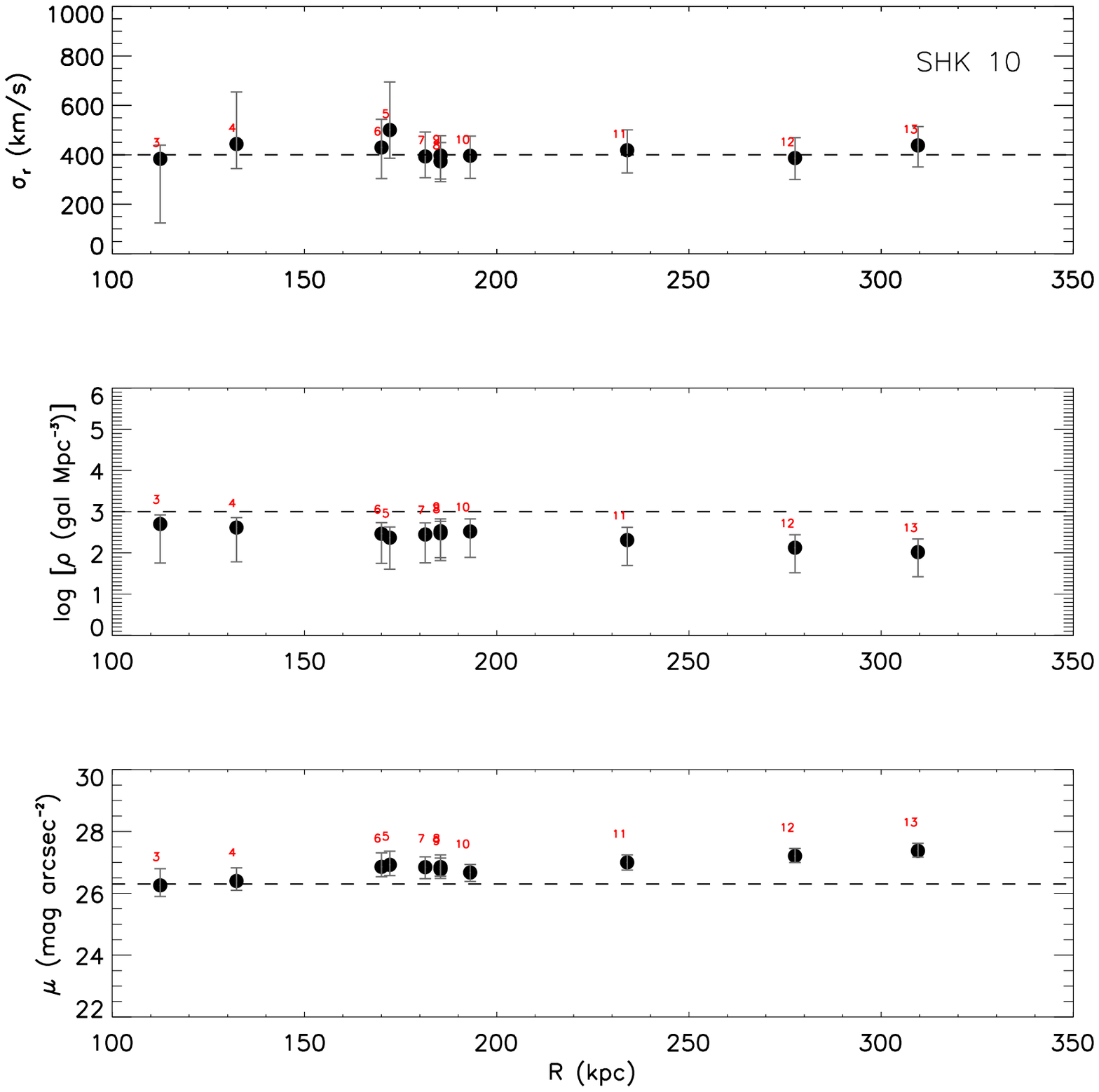}
\includegraphics[width=0.55\textwidth]{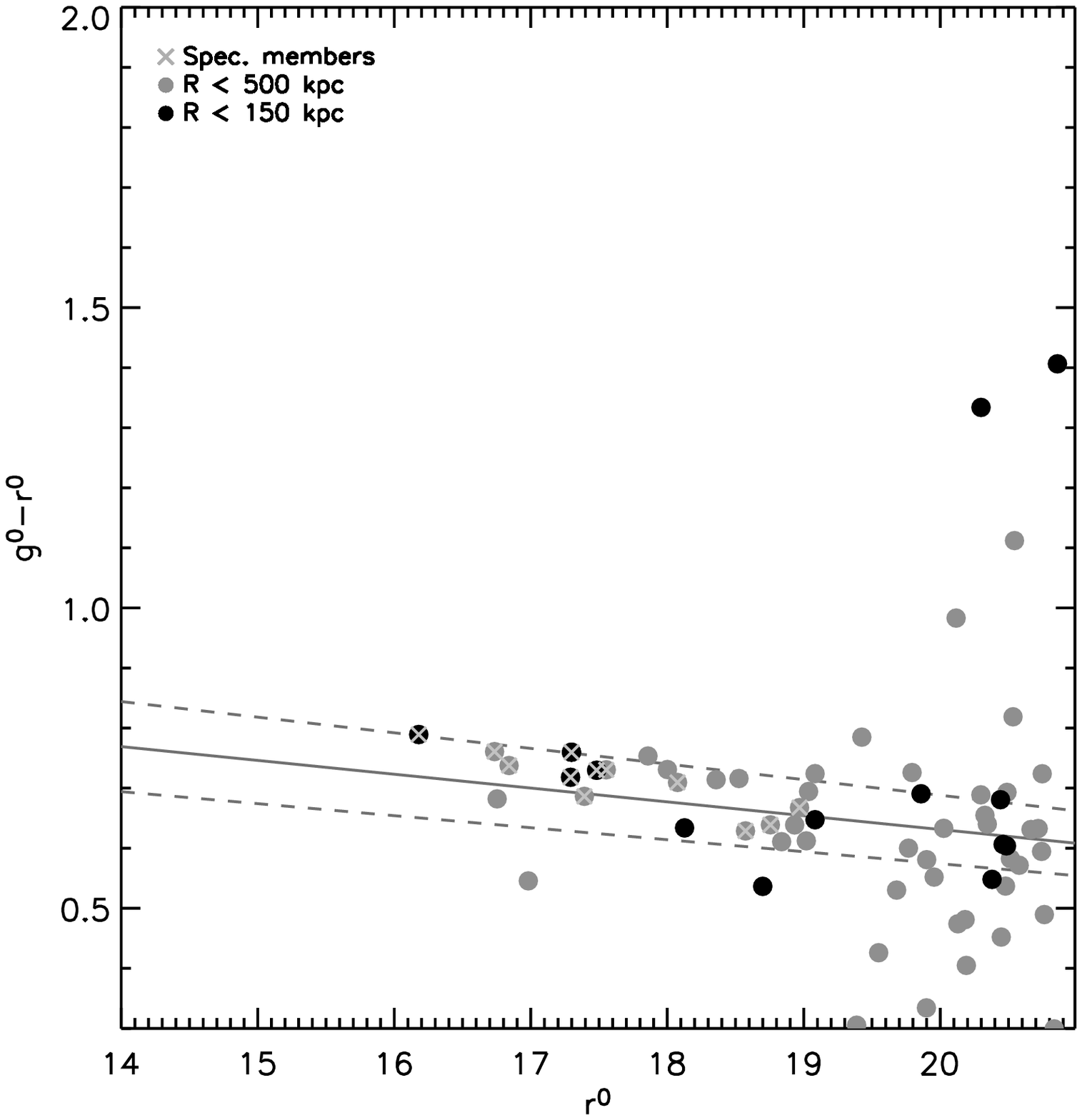}
\caption{{\bf SHK 10.} {\it Upper panel}: radial profiles of $\sigma_{\rm r}$, $\log(\rho)$ and $\mu$ (top to bottom) starting from the central galaxy triplet. Numbers on top of data points indicate the number of galaxies considered. {\it Lower Panel}: colour-magnitude diagram for all the galaxies contained within $500\ {\rm kpc}$ from group centroid and satisfying the criteria 2 and 3 defined in Section 3 of Cap09. Galaxies within the inner region ($R<150\ {\rm kpc}$) are represented by black dots. In addition, confirmed spectroscopic member galaxies are indicated with crosses. The solid line represents the colour-magnitude relation obtained by \citet{Ber2003b} at the groupÕs redshift, while the dashed ones indicate the scatter in its slope reported in the mentioned study.}
\label{fig:figA1b}
\end{center}
\end{figure*}

\clearpage

\begin{table*}
\begin{center}
\begin{large}
\caption{{\bf Spectroscopic membership of SHK 71}. {\it Col 1}: galaxy number; {\it col 2} \& {\it col 3}: galaxy coordinates; {\it col 4}: SDSS dereddened {\it r}-band magnitude; {\it col 5}: radial velocity; {\it col 6}: spectroscopic membership. Note that when a spectroscopic member galaxy is within both our and the SDSS spectroscopic catalogues, this galaxy is only counted once for the determination of the group spectro-photometric properties. In particular only the galaxy contained in our spectroscopic galaxy catalogue is considered. Galaxies with no velocity measurement are those whose spectra had too a low S/N ratio to produce significant CCFs (see Section \ref{sec:sec4}).}             
\label{tab:TableA2}                             
\begin{threeparttable}
\begin{tabular}{l c c c c l}      
\hline\hline
 & & &{\bf SHK 71} & & \\
\hline                 
Galaxy\tnote{1} & R.A.         & Dec.           & {\it r}                         & v                                 & Membership \\    
             &({\rm hh:mm:ss})& ({\rm dd:mm:ss})&  ({\rm mag})           & ({\rm km/s})               &                        \\
\hline                        
01     &13:01:47.77    &+38:02:17.7  &16.17		      &$10553\pm60$    &no$\:$ (SDSS)\\
02     &13:02:07.03    &+38:03:17.4  &16.39		      &$32337\pm62$    &no\\  
03 (a) &13:02:10.15    &+38:05:22.5  &17.12		      &$33787\pm30$    &yes (SDSS)\\
04 (b) &13:02:07.94    &+38:03:17.0  &17.14		      &$33067\pm30$    &yes (SDSS)\\
05     &13:02:11.76    &+38:00:29.5  &17.21		      &$26192\pm51$    &no\\
06 (c) &13:02:07.12    &+38:04:03.4  &17.33		      &$34418\pm48$    &yes\\
07 (d) &13:02:04.76    &+38:02:29.1  &17.33		      &$34806\pm30$    &yes (SDSS)\\
08 (e) &13:02:08.30    &+38:01:58.9  &17.95		      &$33729\pm52$    &yes\\
09     &13:02:22.96    &+38:59:43.8  &18.09		      &$47363\pm58$    &no\\
10     &13:02:09.70    &+38:00:03.7  &18.11		      &$26145\pm35$    &no\\
11 (f) &13:02:07.15    &+38:02:46.4  &18.26		      &$33041\pm63$    &yes\\
12     &13:02:07.30    &+38:02:21.4  &18.32		      &$59975\pm63$    &no\\
13     &13:02:04.83    &+38:03:44.2  &18.53		      &--	       &--\\
14 (g) &13:02:12.62    &+38:03:39.2  &18.58		      &$34102\pm99$    &yes\\
15 (h) &13:02:07.74    &+38:04:27.6  &18.74		      &$33955\pm88$    &yes\\
16     &13:02:17.62    &+38:02:15.0  &19.12		      &$20678\pm89$    &no \\
17     &13:02:08.92    &+38:02:27.3  &19.14		      &--	       &--\\
18 (i) &13:02:03.60    &+38:03:07.7  &19.24		      &$33941\pm69$    &yes\\
19     &13:02:00.34    &+38:01:50.2  &19.40		      &--	       &--\\
20     &13:02:16.35    &+38:06:00.5  &19.42		      &$28072\pm64$    &no\\
21     &13:02:12.89    &+38:04:49.8  &19.47		      &$35542\pm83$    &no\\
\hline                                   
\end{tabular}
\begin{tablenotes}\footnotesize 
\item[1] Letters are associated to spectroscopically confirmed member galaxies only, ordered from the brightest to the faintest. They do not indicate the original nominal scheme used by Shakhbazyan and collaborators.
\end{tablenotes}
\end{threeparttable}
\end{large}  
\end{center}
\end{table*}

\vspace{5cm} 
\clearpage

\begin{figure*}
\begin{center}
\includegraphics[width=0.62\textwidth]{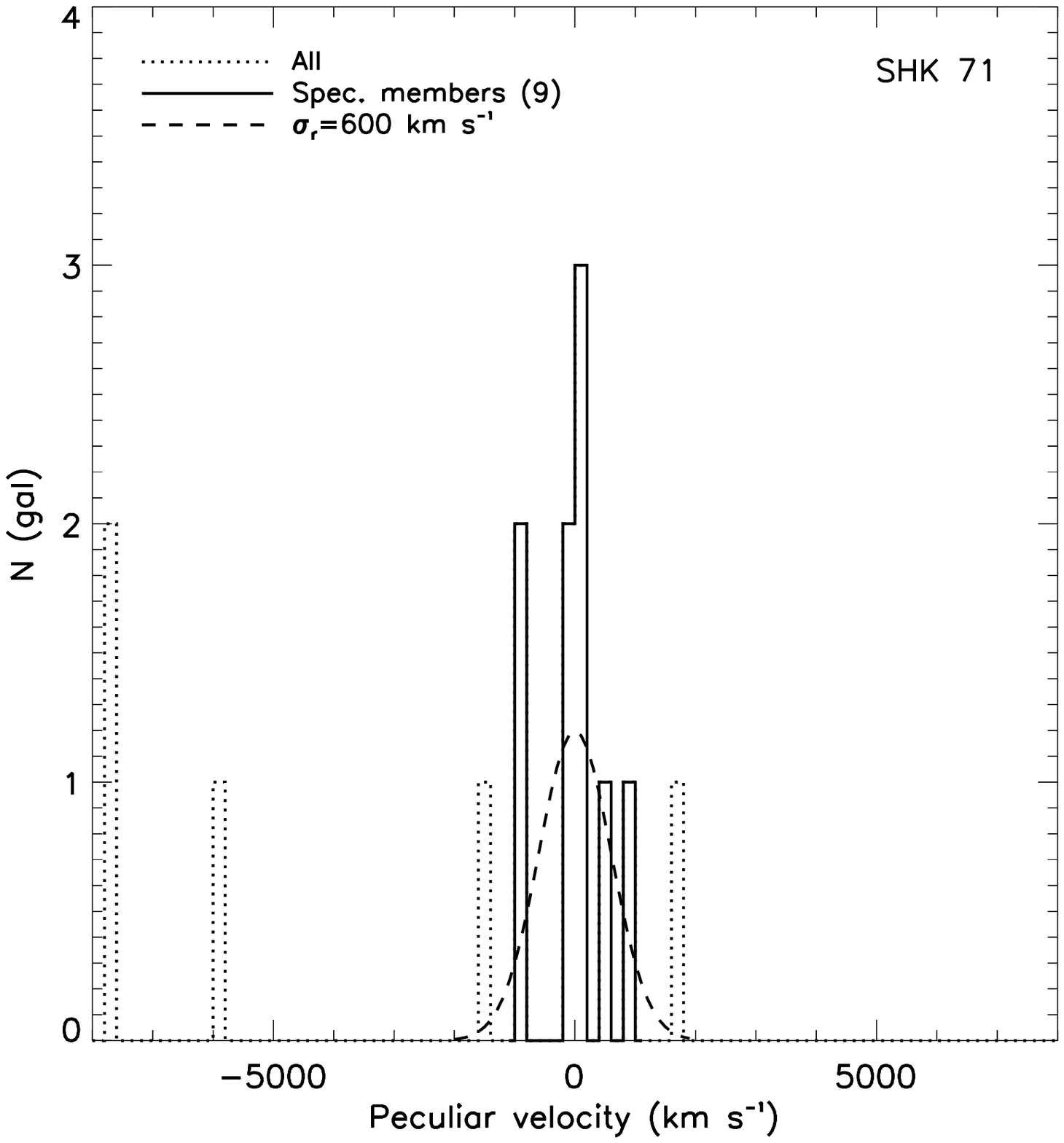}\\
\qquad \includegraphics[width=0.55\textwidth]{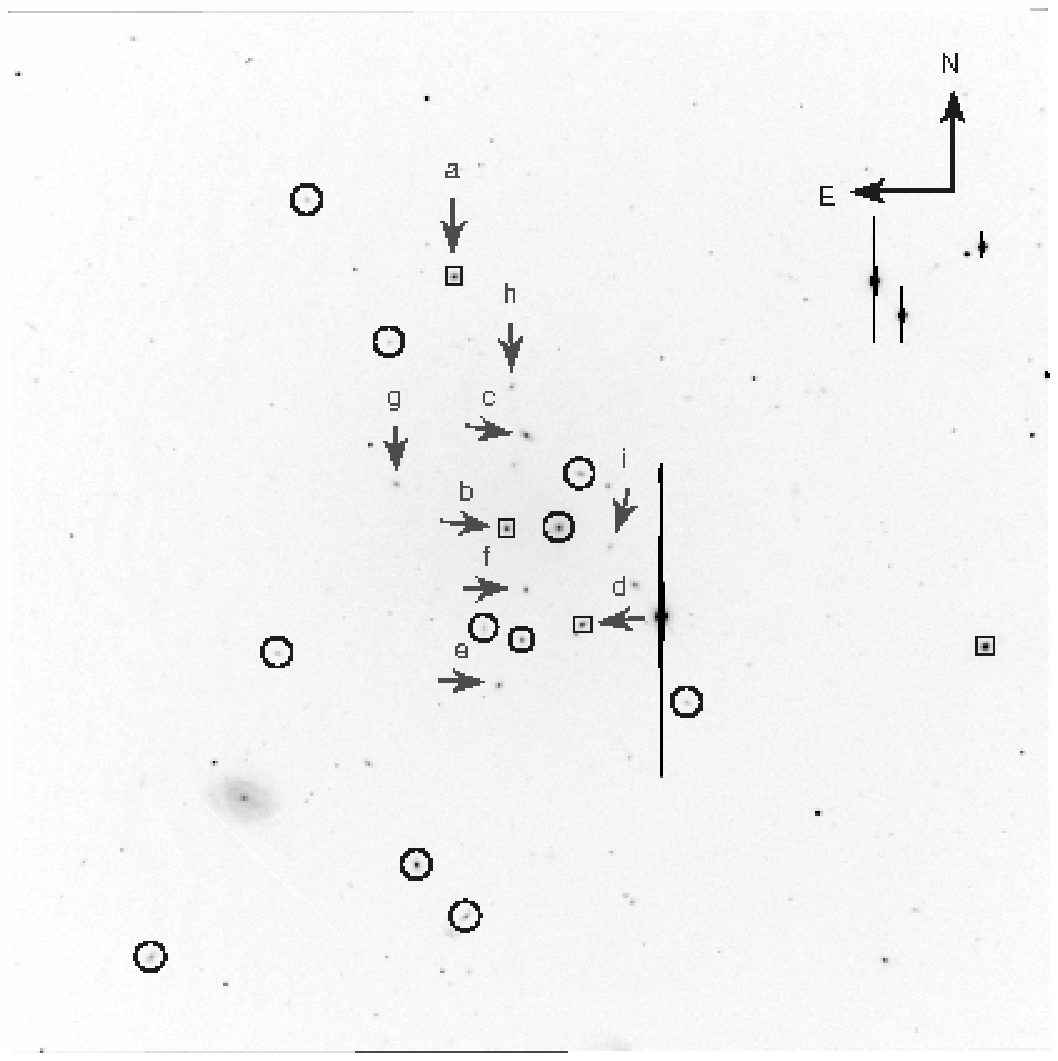}
\caption{{\bf SHK 71.} {\it Upper panel}: velocity distribution obtained by using spectroscopically confirmed member galaxies. A Gaussian distribution centred on zero and defined by the value of the estimated $\sigma_{\rm r}$ is over-plotted (dashed line). {\it Lower panel}: {\it r}-band image showing observed and SDSS spectroscopically confirmed group member galaxies (arrows), observed non-member galaxies (circles) and galaxies from the SDSS spectroscopic catalogue (squares). The size of the field is $8.6\ {\rm arcmin}\times 8.6\ {\rm arcmin}$ (DOLORES imaging mode).}
\label{fig:figA2}
\end{center}
\end{figure*}

\clearpage

\begin{figure*}
\begin{center}
\includegraphics[width=0.62\textwidth]{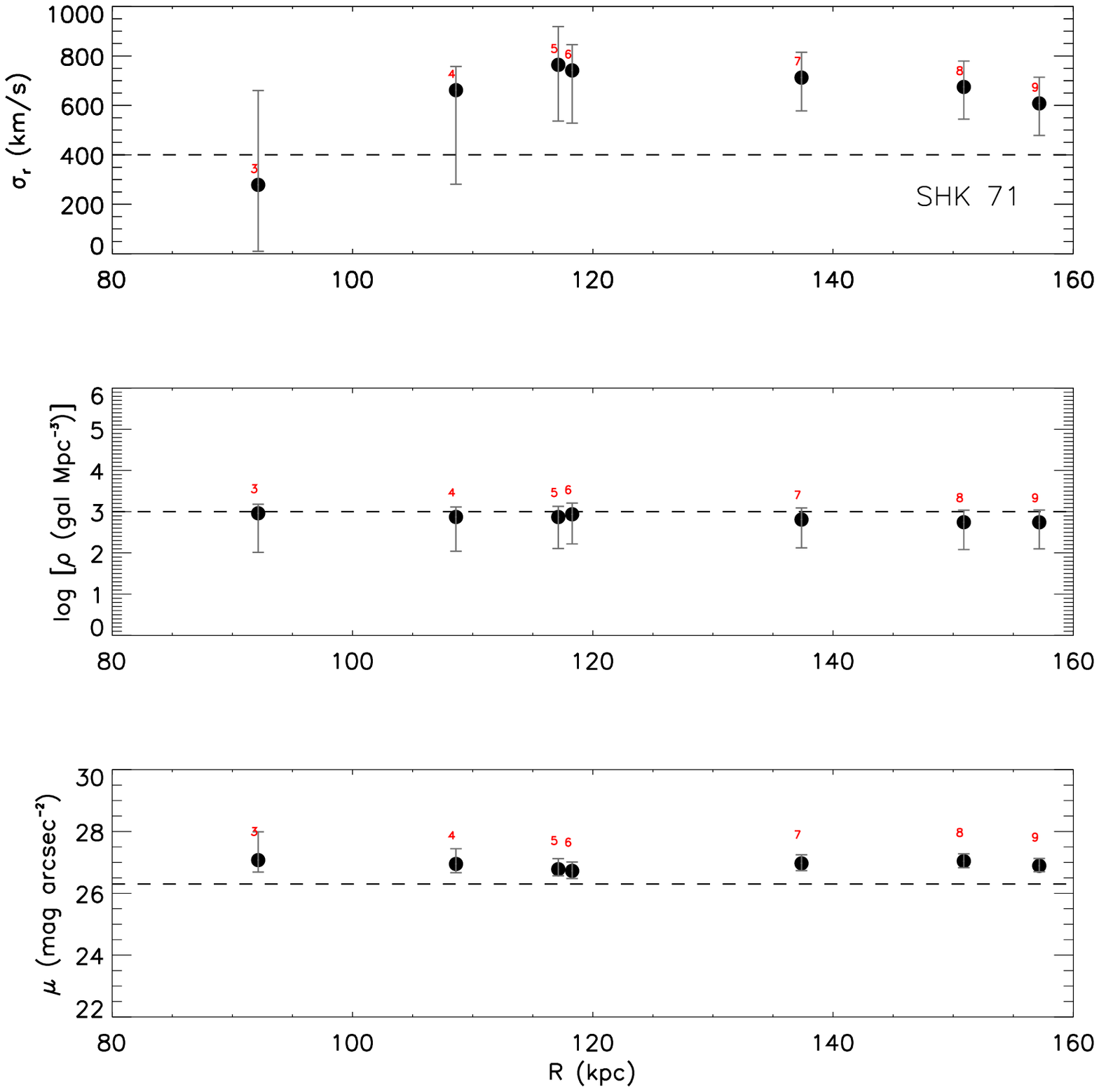}
\includegraphics[width=0.55\textwidth]{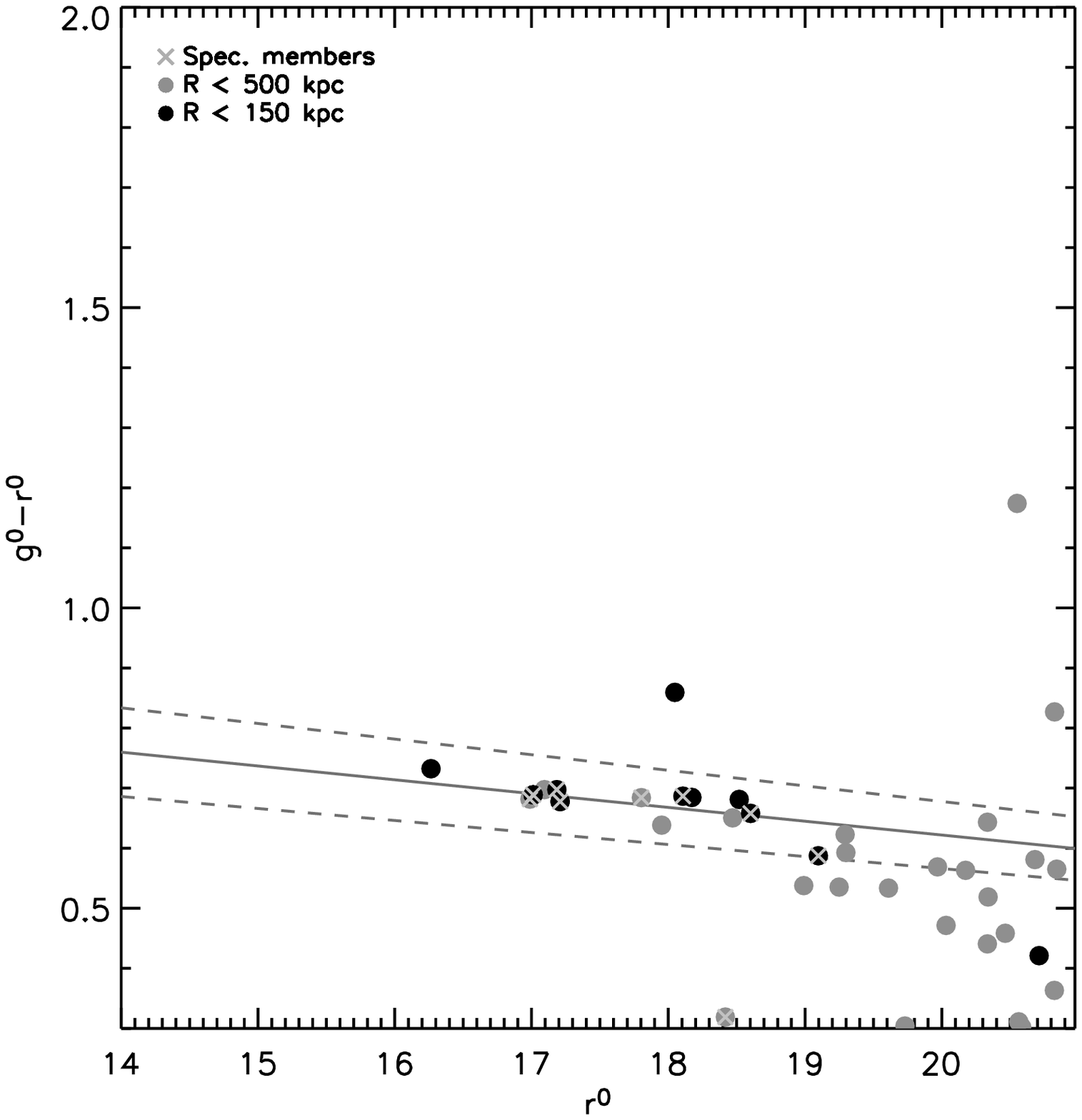}
\caption{{\bf SHK 71.} {\it Upper panel}: radial profiles of $\sigma_{\rm r}$, $\log(\rho)$ and $\mu$ (top to bottom) starting from the central galaxy triplet. Numbers on top of data points indicate the number of galaxies considered. {\it Lower Panel}: colour-magnitude diagram for all the galaxies contained within $500\ {\rm kpc}$ from group centroid and satisfying the criteria 2 and 3 defined in Section 3 of Cap09. Galaxies within the inner region ($R<150\ {\rm kpc}$) are represented by black dots. In addition, confirmed spectroscopic member galaxies are indicated with crosses. The solid line represents the colour-magnitude relation obtained by \citet{Ber2003b} at the groupÕs redshift, while the dashed ones indicate the scatter in its slope reported in the mentioned study.}
\label{fig:figA2b}
\end{center}
\end{figure*}

\clearpage

\begin{table*}
\begin{center}
\begin{large}
\caption{{\bf Spectroscopic membership of SHK 75}. {\it Col 1}: galaxy number; {\it col 2} \& {\it col 3}: galaxy coordinates; {\it col 4}: SDSS dereddened {\it r}-band magnitude; {\it col 5}: radial velocity; {\it col 6}: spectroscopic membership. Note that when a spectroscopic member galaxy is within both our and the SDSS spectroscopic catalogues, this galaxy is only counted once for the determination of the group spectro-photometric properties. In particular only the galaxy contained in our spectroscopic galaxy catalogue is considered. Galaxies with no velocity measurement are those whose spectra had too a low S/N ratio to produce significant CCFs (see Section \ref{sec:sec4}).}             
\label{tab:TableA3}                          
\begin{threeparttable} 
\begin{tabular}{l c c c c l}       
\hline\hline
 & & &{\bf SHK 75} & & \\
\hline                 
Galaxy\tnote{1} & R.A.        & Dec.                   & {\it r}                   & v                               & Membership \\    
             &({\rm hh:mm:ss})& ({\rm dd:mm:ss})&  ({\rm mag})           & ({\rm km/s})               &                        \\
\hline   
01     &14:27:45.42    &+38:47:44.6		   &17.07		    &$50665\pm60$    &no$\:$ (SDSS)\\
02     &14:27:27.26    &+38:41:42.7		   &17.42		    &$24493\pm30$    &no$\:$ (SDSS)\\
03 (a) &14:27:42.11    &+38:46:59.9		   &17.46		    &$49436\pm60$    &yes (SDSS)\\	    
04 (b) &14:27:33.13    &+38:46:34.6		   &17.48		    &$48804\pm66$    &yes\\  
05 (c) &14:27:36.07    &+38:45:58.4		   &17.63		    &$50175\pm82$    &yes\\
06 (c) &14:27:36.07    &+38:45:58.4		   &17.63		    &$50425\pm30$    &yes (SDSS same as 05)\\
07 (d) &14:27:51.52    &+38:46:25.8		   &17.80		    &$49713\pm82$    &yes\\
08     &14:27:23.64    &+38:42:36.8		   &17.82		    &$23881\pm48$    &no\\
09 (e) &14:27:38.27    &+38:46:54.2		   &17.85		    &$48979\pm80$    &yes\\
10 (f) &14:27:43.45    &+38:47:28.3		   &18.23		    &$50069\pm59$    &yes\\
11     &14:27:23.93    &+38:43:43.5		   &18.52		    &$23679\pm37$    &no\\
12     &14:27:12.48    &+38:43:16.6		   &18.54		    &-- 	     &--\\
13     &14:27:14.86    &+38:48:07.9		   &18.57		    &-- 	     &--\\
14 (g) &14:27:37.39    &+38:44:24.8		   &18.59		    &$49256\pm156$   &yes \\
15     &14:27:28.51    &+38:43:52.8		   &19.26		    &$18434\pm59$    &no \\
16     &14:27:30.57    &+38:44:59.0		   &19.26		    &$18705\pm54$    &no\tnote{2} \\
17     &14:27:33.50    &+38:45:17.8		   &19.27		    &-- 	     &--\\
18     &14:27:36.02    &+38:47:04.2		   &19.68		    &-- 	     &--\\
19     &14:27:38.11    &+38:46:08.0		   &22.45		    &$50784\pm130$   &no\\
\hline                                   
\end{tabular}
\begin{tablenotes}\footnotesize 
\item[1] Letters are associated to spectroscopically confirmed member galaxies only, ordered from the brightest to the faintest. They do not indicate the original nominal scheme used by Shakhbazyan and collaborators.
\item[2] This galaxy has been excluded because it does not satisfy the criteria described in Section \ref{subsec:subsec4.1}, due to the presence of multiple and comparable peaks in the CCF. One of these peaks corresponds to a velocity satisfying Hickson criterion. The inclusion or exclusion of this galaxy does not affect our results.
\end{tablenotes}
\end{threeparttable}
\end{large}  
\end{center}
\end{table*}

\vspace{5cm} 
\clearpage

\begin{figure*}
\begin{center}
\includegraphics[width=0.62\textwidth]{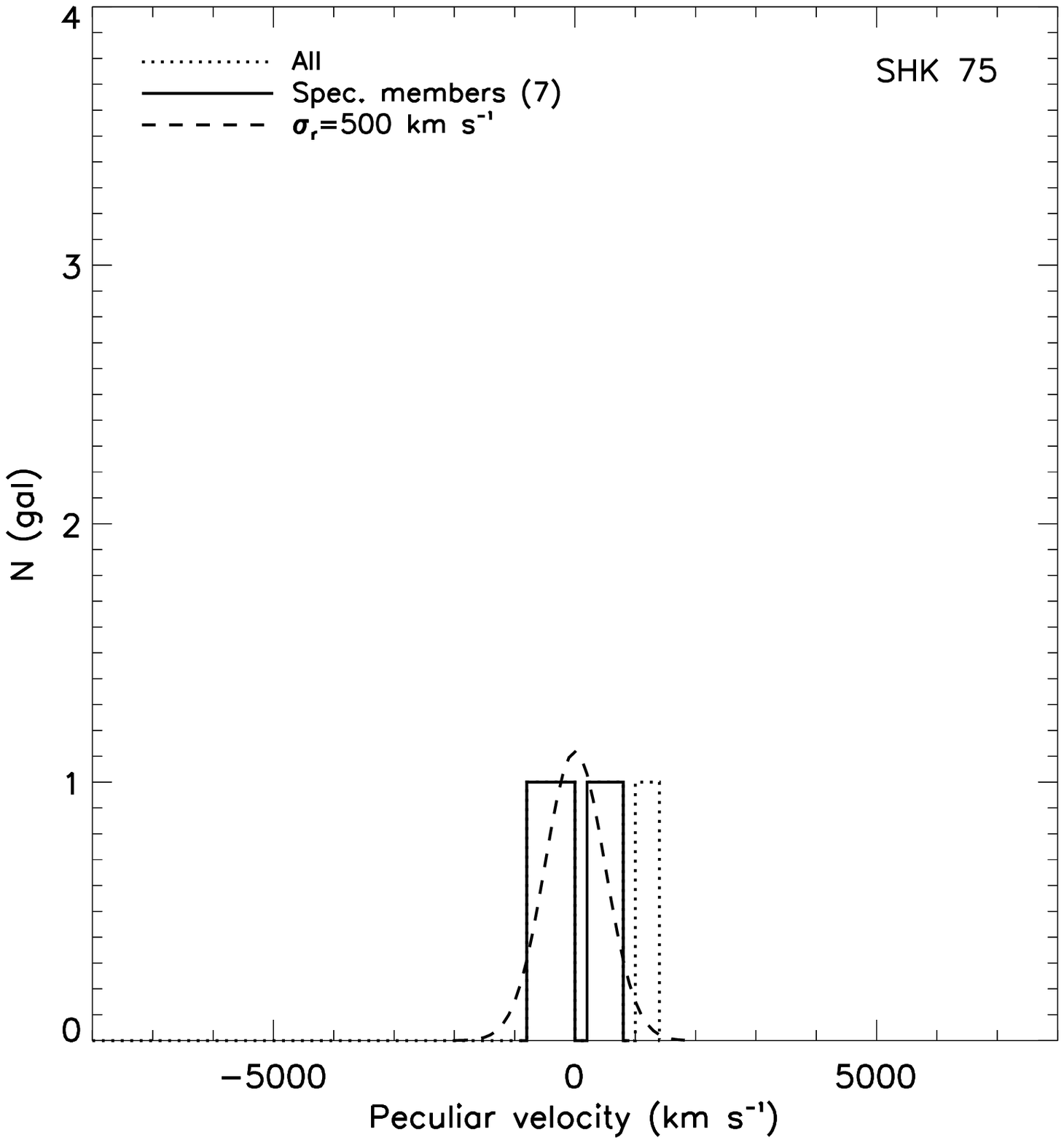}\\
\qquad \includegraphics[width=0.55\textwidth]{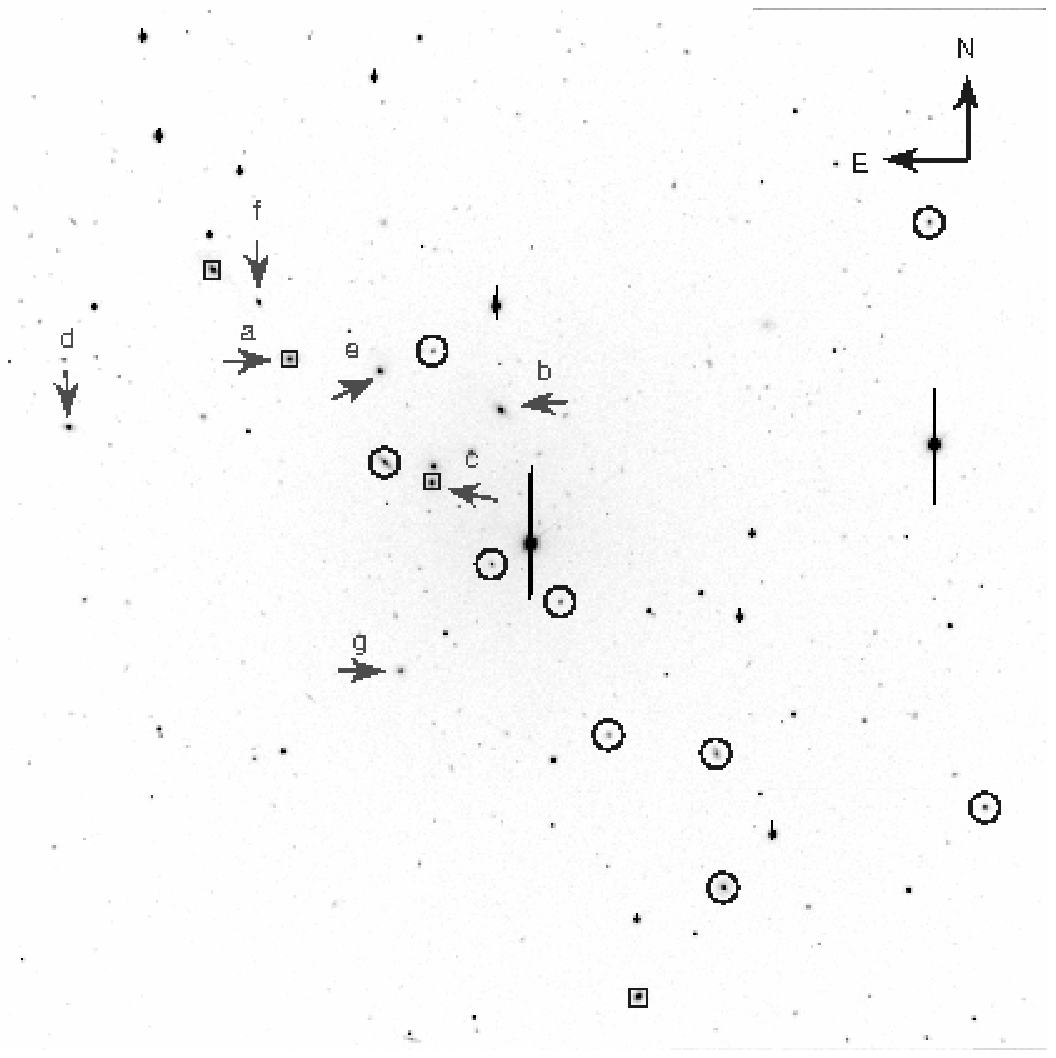}
\caption{{\bf SHK 75.} {\it Upper panel}: velocity distribution obtained by using spectroscopically confirmed member galaxies. A Gaussian distribution centred on zero and defined by the value of the estimated $\sigma_{\rm r}$ is over-plotted (dashed line). {\it Lower panel}: {\it r}-band image showing observed and SDSS spectroscopically confirmed group member galaxies (arrows), observed non-member galaxies (circles) and galaxies from the SDSS spectroscopic catalogue (squares). The size of the field is $8.6\ {\rm arcmin}\times 8.6\ {\rm arcmin}$ (DOLORES imaging mode).}
\label{fig:figA3}
\end{center}
\end{figure*}

\clearpage

\begin{figure*}
\begin{center}
\includegraphics[width=0.62\textwidth]{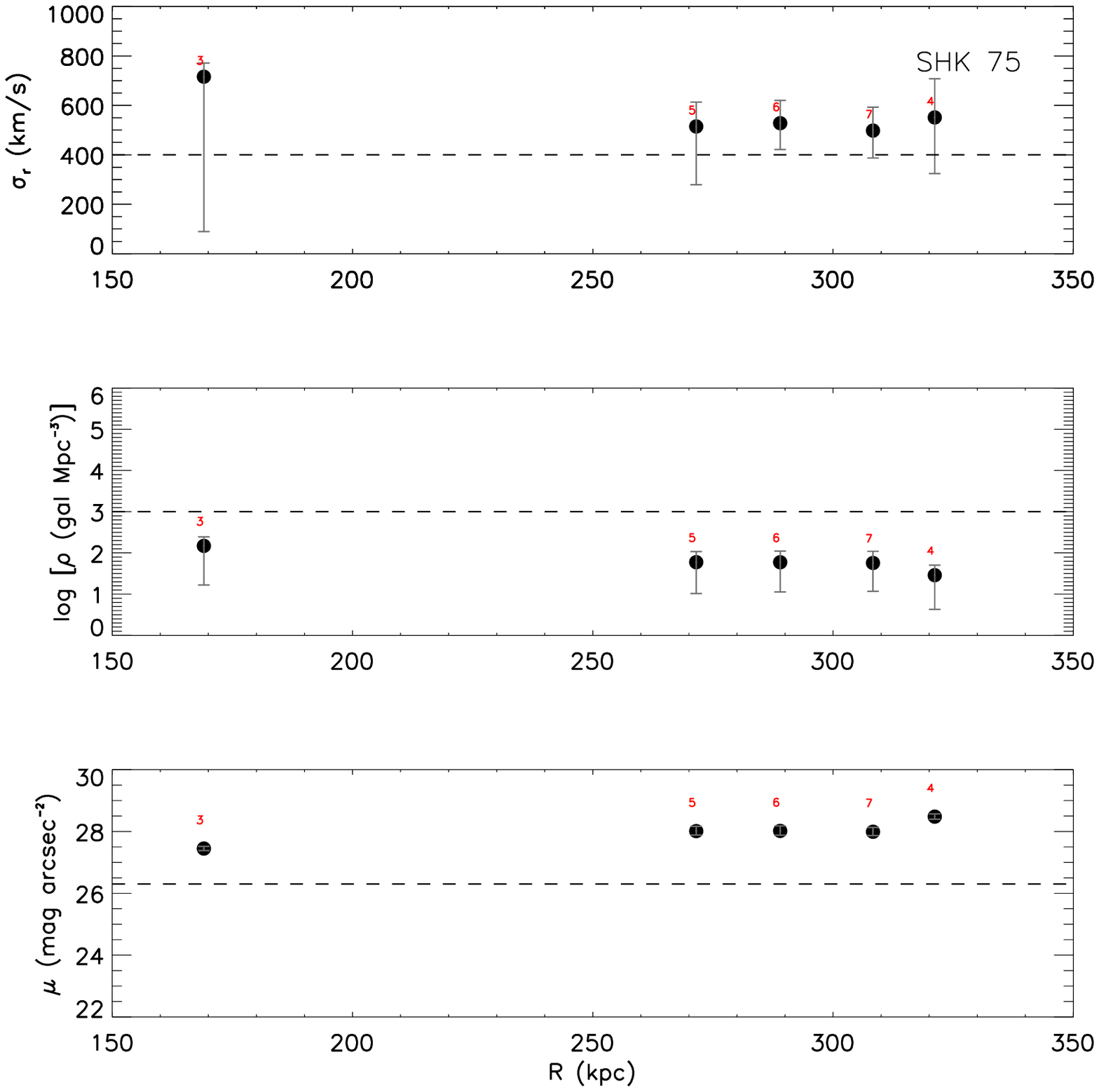}
\includegraphics[width=0.55\textwidth]{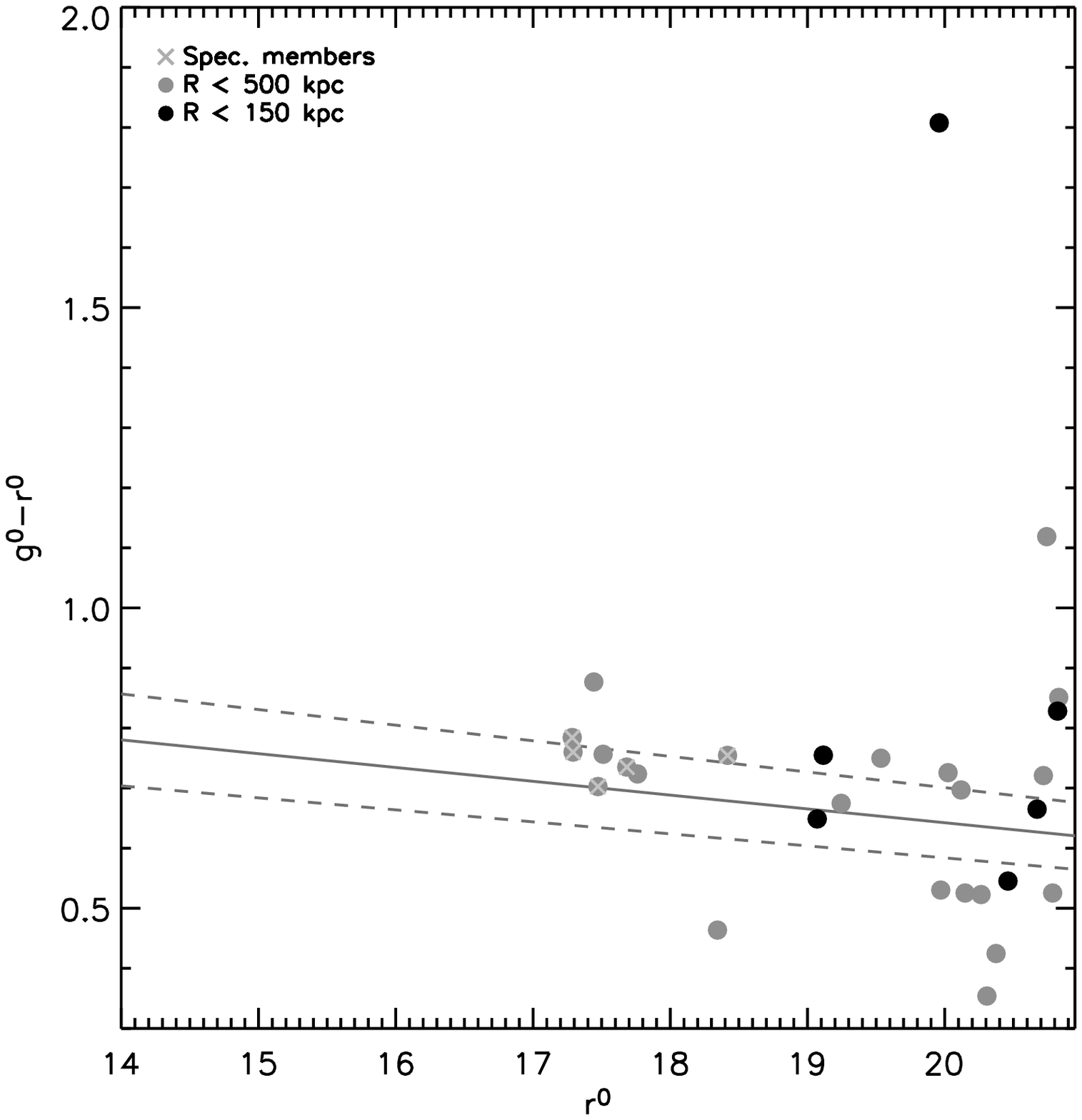}
\caption{{\bf SHK 75.} {\it Upper panel}: radial profiles of $\sigma_{\rm r}$, $\log(\rho)$ and $\mu$ (top to bottom) starting from the central galaxy triplet. Numbers on top of data points indicate the number of galaxies considered. {\it Lower Panel}: colour-magnitude diagram for all the galaxies contained within $500\ {\rm kpc}$ from group centroid and satisfying the criteria 2 and 3 defined in Section 3 of Cap09. Galaxies within the inner region ($R<150\ {\rm kpc}$) are represented by black dots. In addition, confirmed spectroscopic member galaxies are indicated with crosses. The solid line represents the colour-magnitude relation obtained by \citet{Ber2003b} at the groupÕs redshift, while the dashed ones indicate the scatter in its slope reported in the mentioned study.}
\label{fig:figA3b}
\end{center}
\end{figure*}

\clearpage

\begin{table*}
\begin{center}
\begin{large}
\caption{{\bf Spectroscopic membership of SHK 80}. {\it Col 1}: galaxy number; {\it col 2} \& {\it col 3}: galaxy coordinates; {\it col 4}: SDSS dereddened {\it r}-band magnitude; {\it col 5}: radial velocity; {\it col 6}: spectroscopic membership. Note that when a spectroscopic member galaxy is within both our and the SDSS spectroscopic catalogues, this galaxy is only counted once for the determination of the group spectro-photometric properties. In particular only the galaxy contained in our spectroscopic galaxy catalogue is considered. Galaxies with no velocity measurement are those whose spectra had too a low S/N ratio to produce significant CCFs (see Section \ref{sec:sec4}).}            
\label{tab:TableA4}             
\begin{threeparttable}              
\begin{tabular}{l c c c c l}        
\hline\hline
 & & &{\bf SHK 80} & & \\
\hline                
Galaxy\tnote{1} & R.A.        & Dec.      & {\it r}                              & v                                 & Membership \\    
             &({\rm hh:mm:ss})& ({\rm dd:mm:ss})&  ({\rm mag})           & ({\rm km/s})               &                        \\
\hline                        
01 (a) &15:37:07.33    &+41:39:07.5   &16.23		       &$39570\pm77$	&yes\\  
02     &15:36:46.11    &+41:38:04.6   &16.66		       &$16403\pm44$	&no\\
03     &15:36:46.11    &+41:38:04.6   &16.66		       &$16219\pm60$	&no$\:$ (SDSS same as 02)\\
04     &15:37:03.36    &+41:42:23.7   &17.20		       &$35539\pm50$	&no\\
05     &15:37:10.94    &+41:36:28.9   &17.50		       &$37769\pm77$	&no\\
06     &15:37:10.94    &+41:36:28.9   &17.50		       &$37654\pm30$	&no$\:$ (SDSS same as 05)\\
07     &15:37:08.26    &+41:38:35.3   &17.60		       &$39063\pm30$	&yes (SDSS)\\
08 (b) &15:37:08.20    &+41:39:01.0   &17.90		       &$39086\pm49$	&yes\\
09 (c) &15:37:16.27    &+41:38:23.7   &17.97		       &$39260\pm76$	&yes\\
10     &15:37:11.06    &+41:42:48.0   &18.23		       &$50422\pm87$	&no\\
11     &15:36:44.79    &+41:39:41.3   &18.37		       &$24042\pm49$	&no\\
12 (d) &15:37:09.50    &+41:38:59.7   &18.56		       &$38300\pm85$	&yes\\
13 (e) &15:37:16.99    &+41:38:24.4   &18.73		       &$39069\pm69$	&yes\\
14     &15:36:53.14    &+41:38:37.9   &18.92		       &$23926\pm41$	&no\\
15     &15:37:13.94    &+41:41:47.0   &18.96		       &$27218\pm88$	&no\tnote{2} \\
16     &15:37:04.95    &+41:41:28.3   &19.06		       &$36282\pm96$	&no \\
17     &15:37:19.99    &+41:36:07.4   &19.19		       &$20666\pm69$	&no\\
18     &15:37:09.62    &+41:38:48.5   &19.27		       &$33947\pm66$	&no\\
19     &15:37:16.72    &+41:35:46.1   &19.35		       &$51583\pm57$	&no\\
20     &15:36:42.39    &+41:39:18.8   &19.44		       &--		&--\\
21     &15:37:20.23    &+41:40:33.6   &19.86		       &--		&--\\
\hline                                   
\end{tabular}
\begin{tablenotes}\footnotesize 
\item[1] Letters are associated to spectroscopically confirmed member galaxies only, ordered from the brightest to the faintest. They do not indicate the original nominal scheme used by Shakhbazyan and collaborators.
\item[2] This galaxy has been excluded because it does not satisfy the criteria described in Section \ref{subsec:subsec4.1}, due to the presence of multiple and comparable peaks in the CCF. However, no one of these peaks corresponds to a velocity satisfying Hickson criterion. 
\end{tablenotes}
\end{threeparttable}
\end{large}  
\end{center}
\end{table*}

\vspace{5cm} 
\clearpage

\begin{figure*}
\begin{center}
\includegraphics[width=0.62\textwidth]{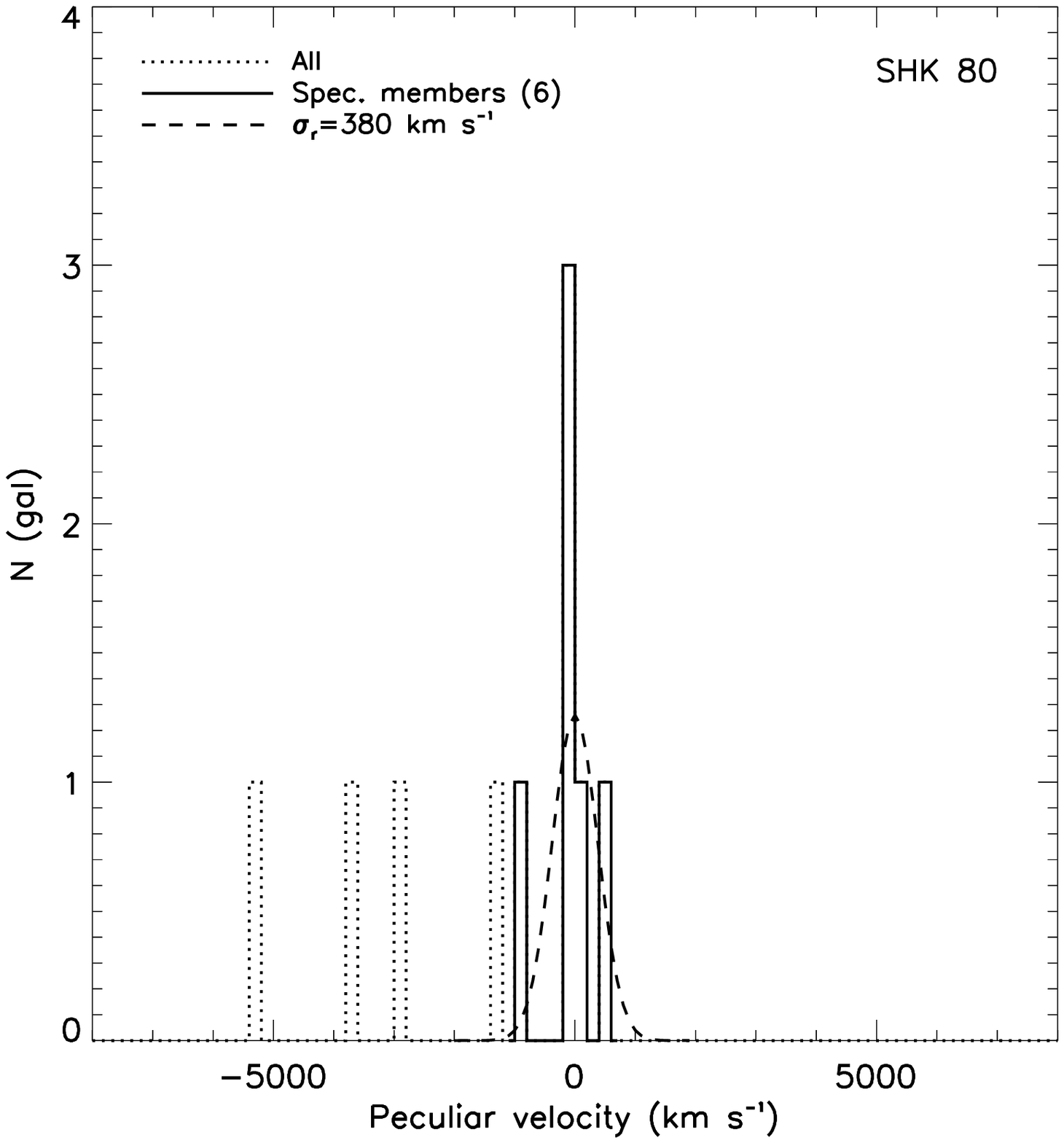}\\
\qquad \includegraphics[width=0.55\textwidth]{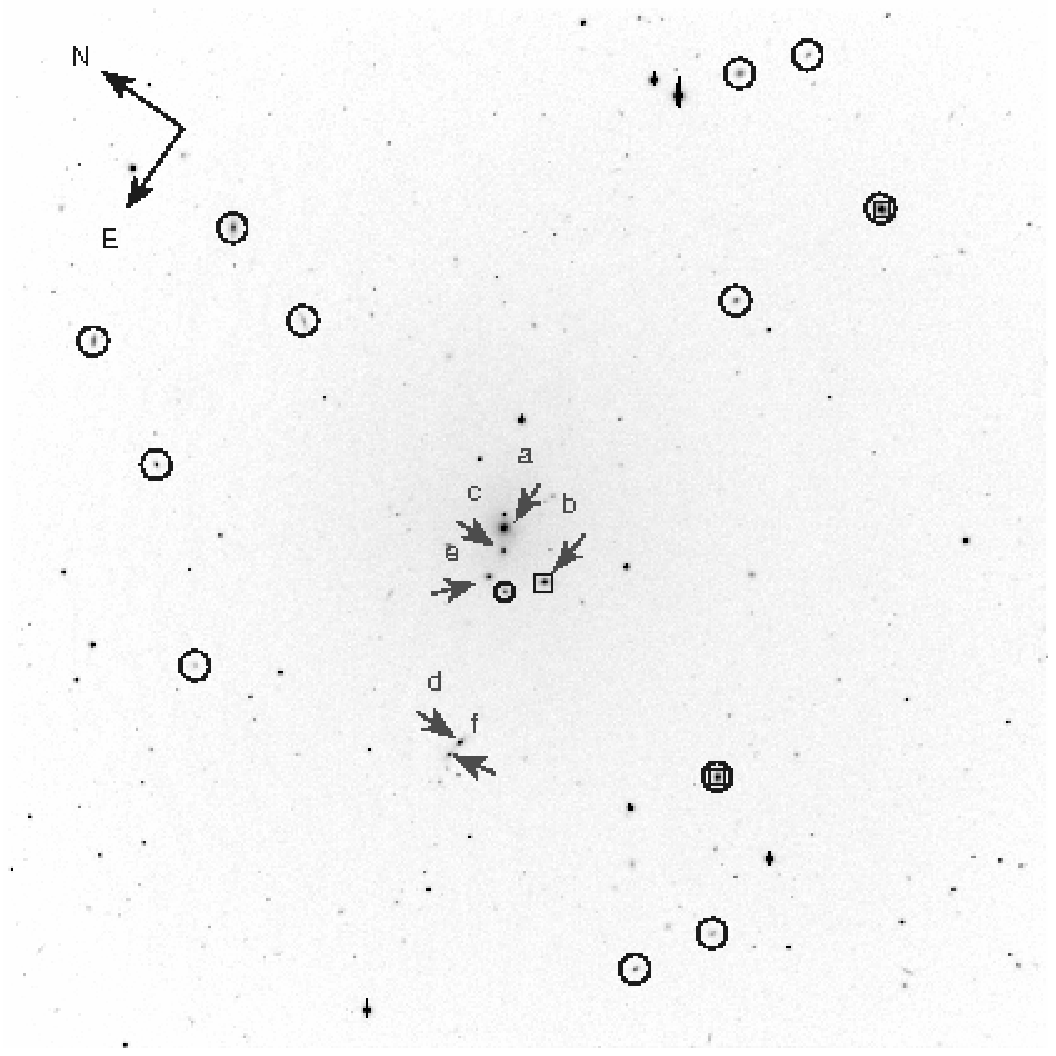}
\caption{{\bf SHK 80.} {\it Upper panel}: velocity distribution obtained by using spectroscopically confirmed member galaxies. A Gaussian distribution centred on zero and defined by the value of the estimated $\sigma_{\rm r}$ is over-plotted (dashed line). {\it Lower panel}: {\it r}-band image showing observed and SDSS spectroscopically confirmed group member galaxies (arrows), observed non-member galaxies (circles) and galaxies from the SDSS spectroscopic catalogue (squares). The size of the field is $8.6\ {\rm arcmin}\times 8.6\ {\rm arcmin}$ (DOLORES imaging mode).}
\label{fig:figA4}
\end{center}
\end{figure*}

\clearpage

\begin{figure*}
\begin{center}
\includegraphics[width=0.62\textwidth]{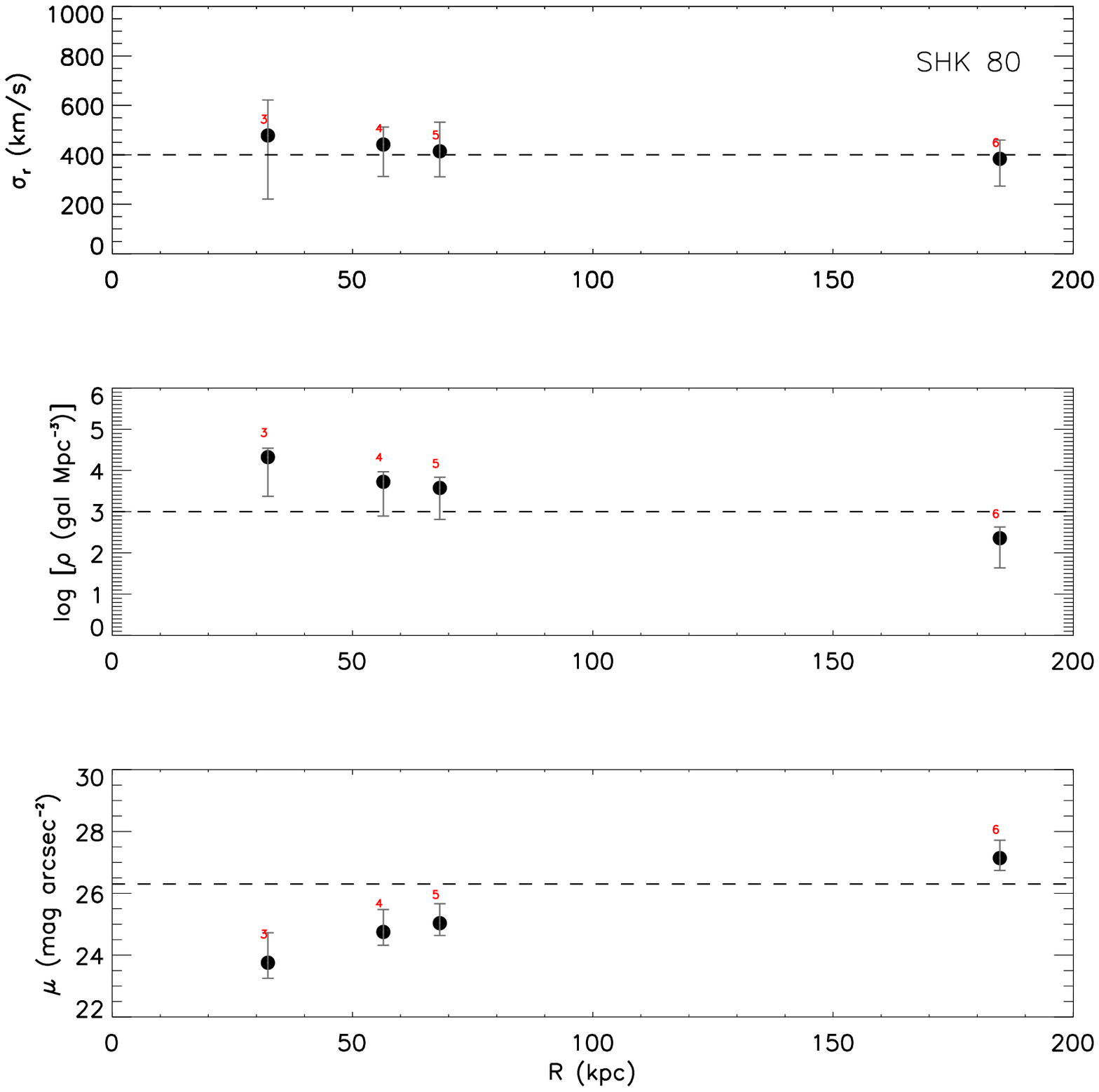}
\includegraphics[width=0.55\textwidth]{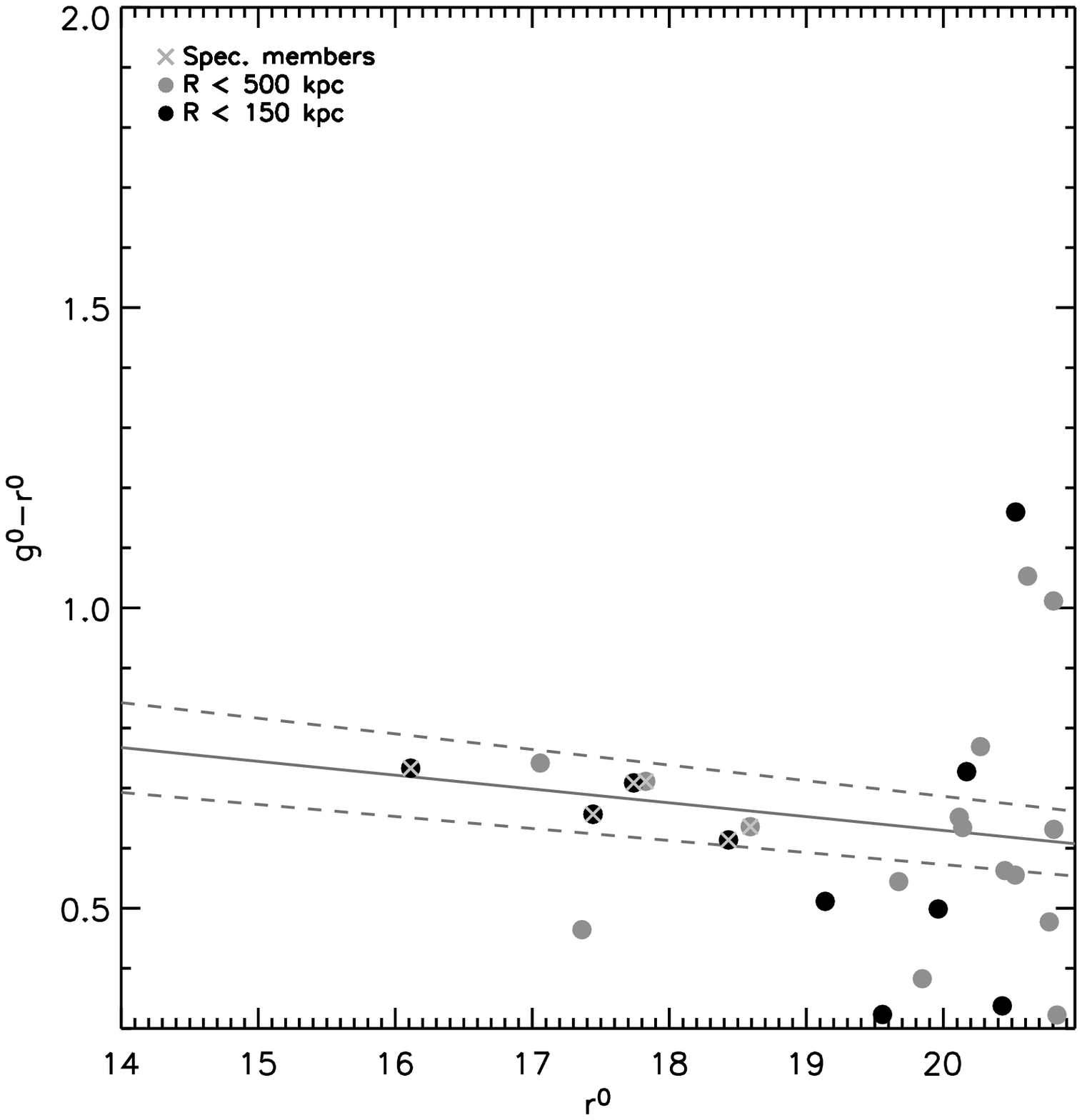}
\caption{{\bf SHK 80.} {\it Upper panel}: radial profiles of $\sigma_{\rm r}$, $\log(\rho)$ and $\mu$ (top to bottom) starting from the central galaxy triplet. Numbers on top of data points indicate the number of galaxies considered. {\it Lower Panel}: colour-magnitude diagram for all the galaxies contained within $500\ {\rm kpc}$ from group centroid and satisfying the criteria 2 and 3 defined in Section 3 of Cap09. Galaxies within the inner region ($R<150\ {\rm kpc}$) are represented by black dots. In addition, confirmed spectroscopic member galaxies are indicated with crosses. The solid line represents the colour-magnitude relation obtained by \citet{Ber2003b} at the groupÕs redshift, while the dashed ones indicate the scatter in its slope reported in the mentioned study.}
\label{fig:figA4b}
\end{center}
\end{figure*}

\begin{table*}
\begin{center}
\begin{large}
\caption{{\bf Spectroscopic membership of SHK 259}. {\it Col 1}: galaxy number; {\it col 2} \& {\it col 3}: galaxy coordinates; {\it col 4}: SDSS dereddened {\it r}-band magnitude; {\it col 5}: radial velocity; {\it col 6}: spectroscopic membership. Note that when a spectroscopic member galaxy is within both our and the SDSS spectroscopic catalogues, this galaxy is only counted once for the determination of the group spectro-photometric properties. In particular only the galaxy contained in our spectroscopic galaxy catalogue is considered. Galaxies with no velocity measurement are those whose spectra had too a low S/N ratio to produce significant CCFs (see Section \ref{sec:sec4}).}            
\label{tab:TableA5}  
\begin{threeparttable}                          
\begin{tabular}{l c c c c l}        
\hline\hline
 & & &{\bf SHK 259} & & \\
\hline                 
Galaxy\tnote{1} & R.A.          & Dec.           & {\it r}                        & v                                 & Membership \\   
             &({\rm hh:mm:ss})& ({\rm dd:mm:ss})&  ({\rm mag})           & ({\rm km/s})               &                        \\
\hline                       
01 (a) &15:39:27.09	&+37:50:59.3	&17.29  		 &$44926\pm43$    &yes\\  
02 (b) &15:39:28.98	&+37:51:09.1	&17.43  		 &$45484\pm59$    &yes\\
03 (c) &15:39:29.48	&+37:50:57.3	&17.50  		 &$44830\pm40$    &yes\\
04 (c) &15:39:29.48	&+37:50:57.3	&17.50  		 &$44309\pm60$    &yes (SDSS same as 03)\\
05     &15:39:09.21	&+37:51:27.9	&17.80  		 &$44459\pm60$    &no$\:$ (SDSS)\\
06 (d) &15:39:32.24	&+37:50:10.1	&18.03  		 &$45750\pm87$    &yes\\
07 (e) &15:39:30.39	&+37:50:31.1	&18.41  		 &$45563\pm87$    &yes\\
08 (f) &15:39:32.40	&+37:51:22.2	&19.00  		 &$45956\pm74$    &yes\\
09     &15:39:36.52	&+37:50:43.3	&19.09  		 &$50101\pm69$    &no\\
10     &15:39:29.58	&+37:51:09.4	&19.13  		 &$32250\pm71$    &no\\
11     &15:39:22.62	&+37:52:09.5	&19.17  		 &$59242\pm78$    &no \\
12     &15:39:44.64	&+37:53:26.2	&19.18  		 &$44323\pm56$    &no\\
13     &15:39:34.75	&+37:48:20.6	&19.20  		 &--		  &--\\
14     &15:39:13.94	&+37:49:45.0	&19.21  		 &--		  &--\\
15     &15:39:30.53	&+37:52:02.0	&19.32  		 &$50572\pm80$    &no \\
16     &15:39:18.93	&+37:50:03.1	&19.35  		 &--		  &--\\
17     &15:39:10.02	&+37:51:02.7	&19.39  		 &--		  &--\\
18 (g) &15:39:34.23	&+37:50:19.5	&19.49  		 &$45591\pm52$    &yes\\
19     &15:39:32.97	&+37:47:04.7	&19.72  		 &--		  &--\\
\hline                                   
\end{tabular}
\begin{tablenotes}\footnotesize 
\item[1] Letters are associated to spectroscopically confirmed member galaxies only, ordered from the brightest to the faintest. They do not indicate the original nominal scheme used by Shakhbazyan and collaborators.
\end{tablenotes}
\end{threeparttable}
\end{large}  
\end{center}
\end{table*}
\vspace{5cm} 
\clearpage

\begin{figure*}
\begin{center}
\includegraphics[width=0.62\textwidth]{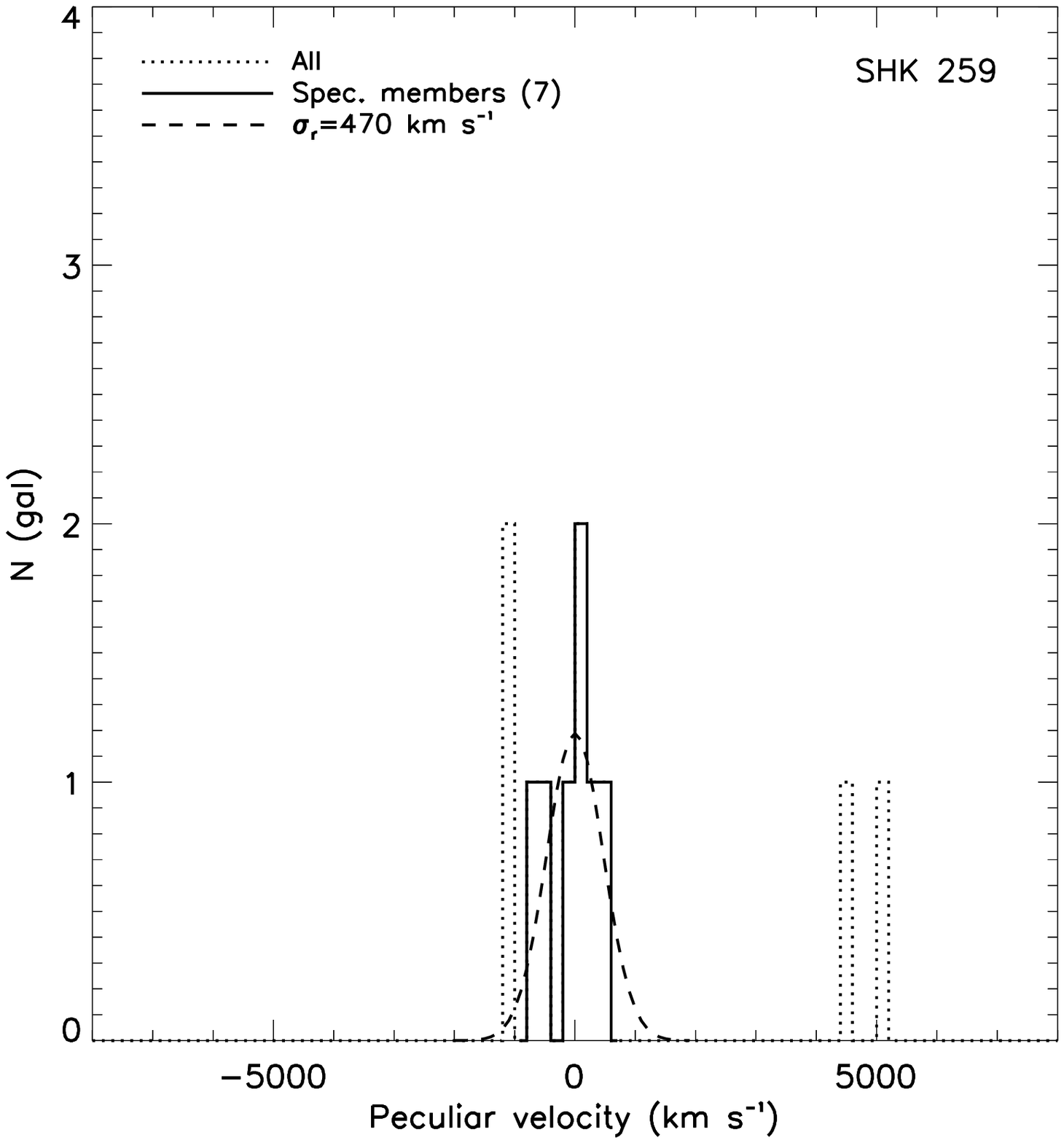}\\
\qquad \includegraphics[width=0.55\textwidth]{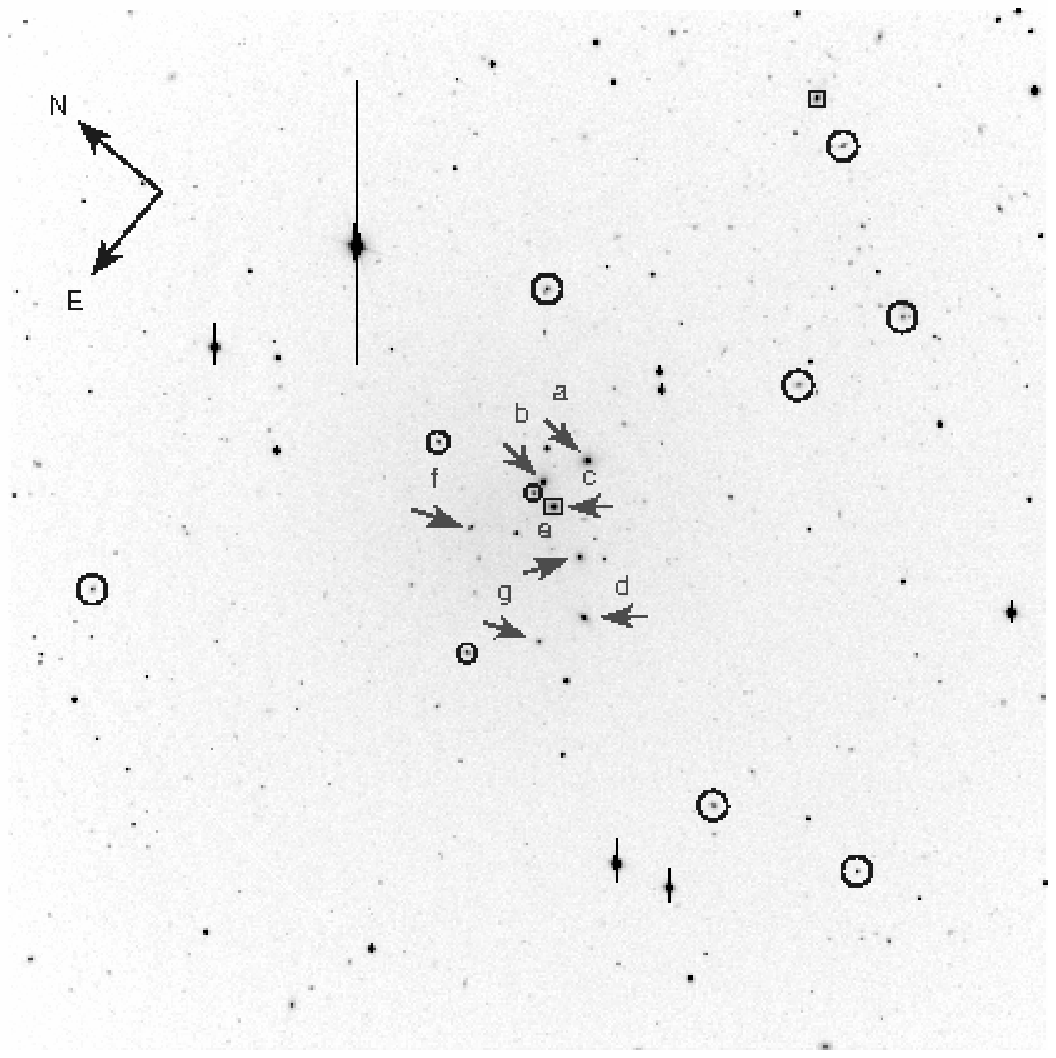}
\caption{{\bf SHK 259.} {\it Upper panel}: velocity distribution obtained by using spectroscopically confirmed member galaxies. A Gaussian distribution centred on zero and defined by the value of the estimated $\sigma_{\rm r}$ is over-plotted (dashed line). {\it Lower panel}: {\it r}-band image showing observed and SDSS spectroscopically confirmed group member galaxies (arrows), observed non-member galaxies (circles) and galaxies from the SDSS spectroscopic catalogue (squares). The size of the field is $8.6\ {\rm arcmin}\times 8.6\ {\rm arcmin}$ (DOLORES imaging mode).}
\label{fig:figA5}
\end{center}
\end{figure*}

\clearpage

\begin{figure*}
\begin{center}
\includegraphics[width=0.62\textwidth]{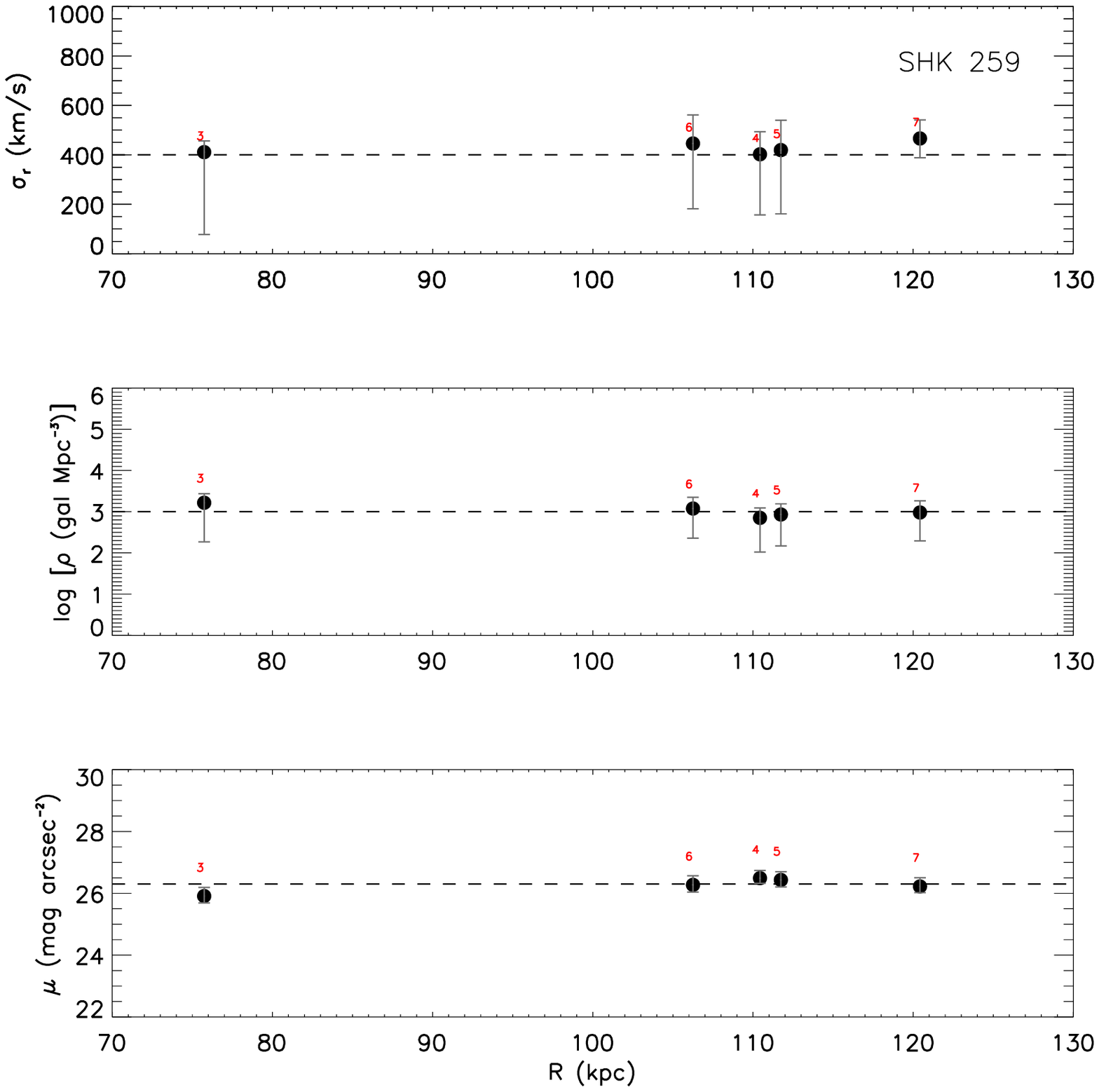}
\includegraphics[width=0.55\textwidth]{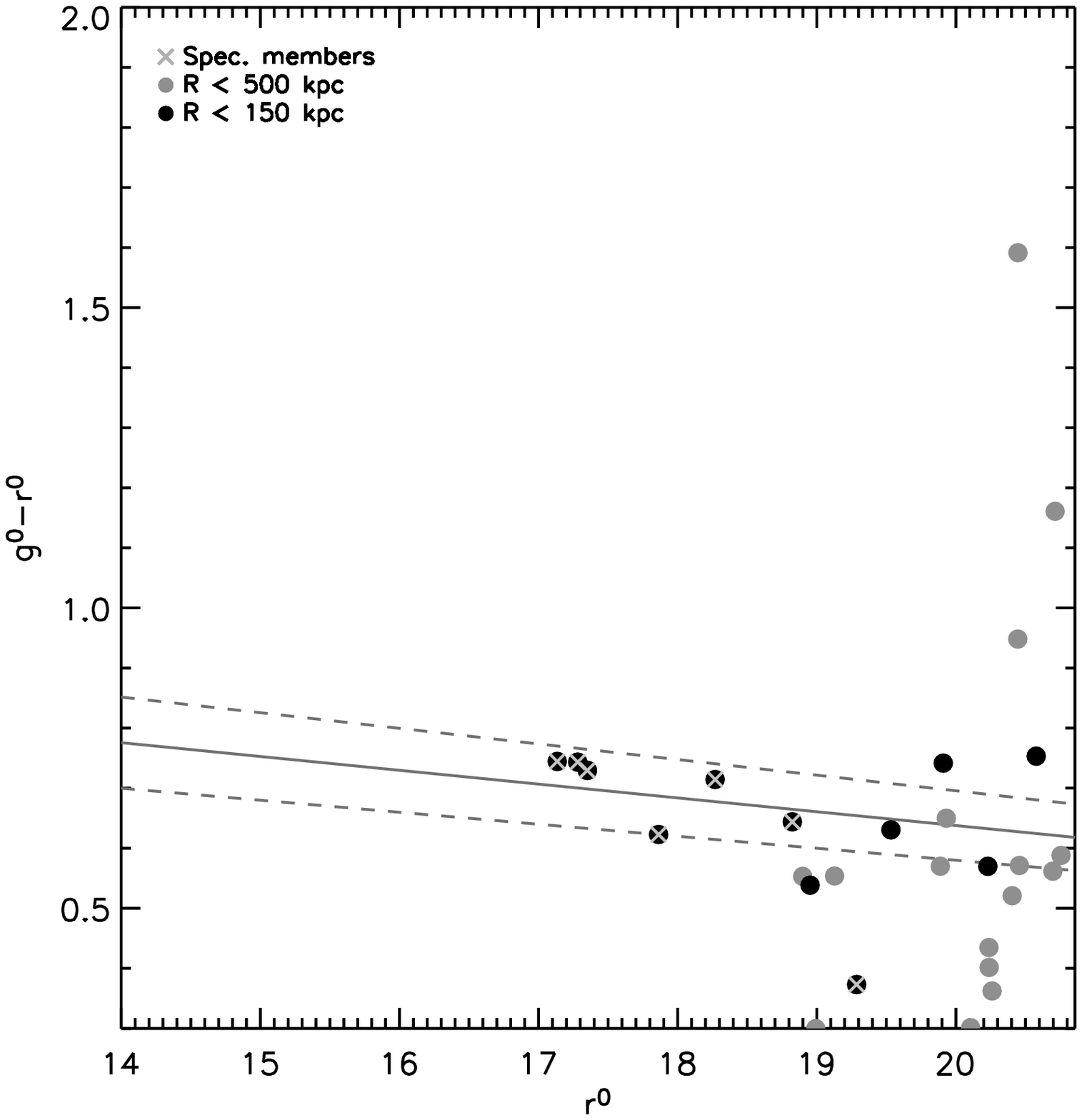}
\caption{{\bf SHK 259.} {\it Upper panel}: radial profiles of $\sigma_{\rm r}$, $\log(\rho)$ and $\mu$ (top to bottom) starting from the central galaxy triplet. Numbers on top of data points indicate the number of galaxies considered. {\it Lower Panel}: colour-magnitude diagram for all the galaxies contained within $500\ {\rm kpc}$ from group centroid and satisfying the criteria 2 and 3 defined in Section 3 of Cap09. Galaxies within the inner region ($R<150\ {\rm kpc}$) are represented by black dots. \textbf{In addition, confirmed spectroscopic member galaxies are indicated with light grey crosses.} The solid line represents the colour-magnitude relation obtained by \citet{Ber2003b} at the groupÕs redshift, while the dashed ones indicate the scatter in its slope reported in the mentioned study.}
\label{fig:figA5b}
\end{center}
\end{figure*}

\clearpage

\begin{figure*}
\begin{center}
\includegraphics[width=0.62\textwidth]{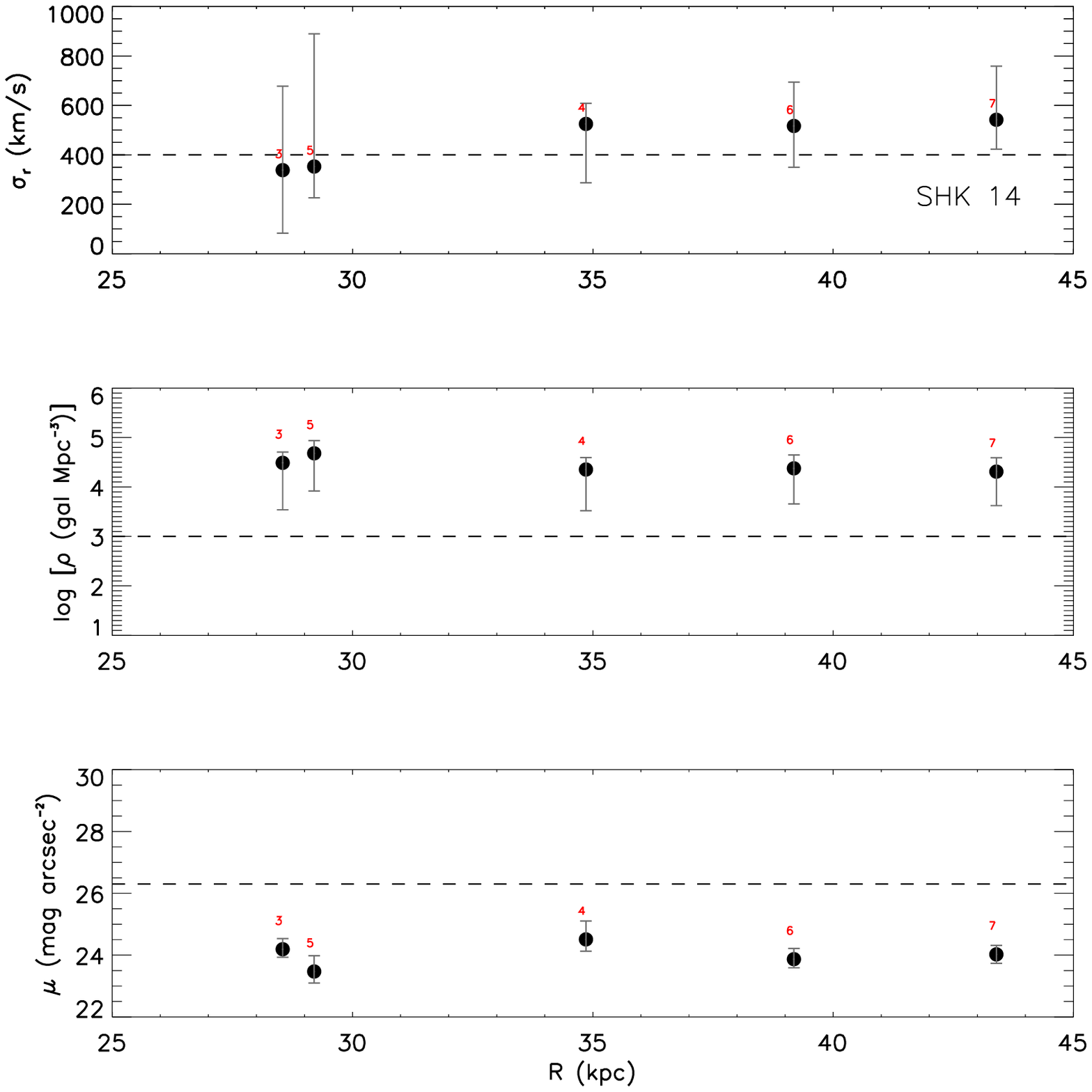}
\includegraphics[width=0.55\textwidth]{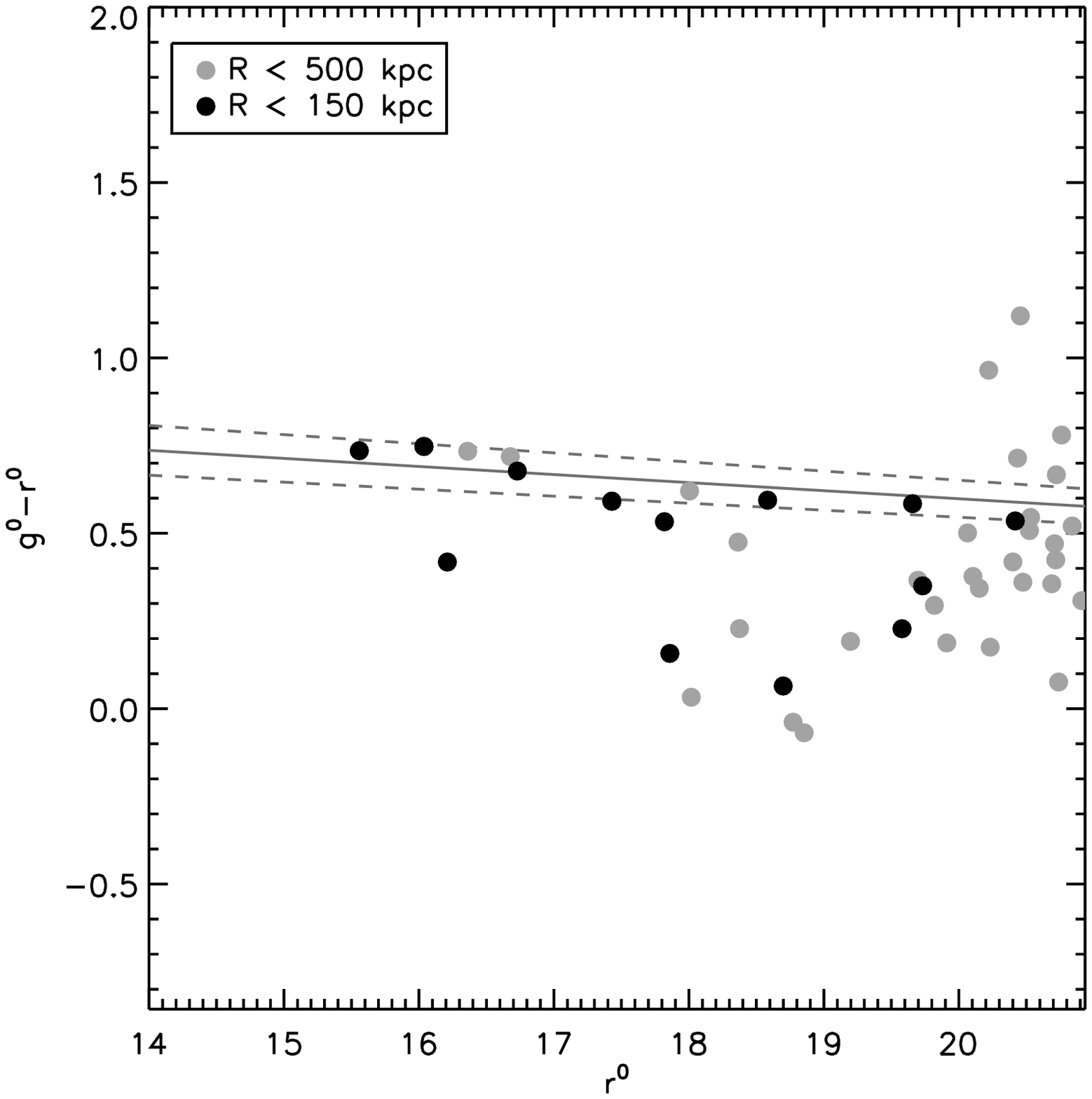}
\caption{{\bf SHK 14.} {\it Upper panel}: radial profiles of $\sigma_{\rm r}$, $\log(\rho)$ and $\mu$ (top to bottom) starting from the central galaxy triplet. Numbers on top of data points indicate the number of galaxies considered. {\it Lower Panel}: colour-magnitude diagram for all the galaxies contained within $500\ {\rm kpc}$ from group centroid and satisfying the criteria 2 and 3 defined in Section 3 of Cap09. Galaxies within the inner region ($R<150\ {\rm kpc}$) are represented by black dots. The solid line represents the colour-magnitude relation obtained by \citet{Ber2003b} at the groupÕs redshift, while the dashed ones indicate the scatter in its slope reported in the mentioned study.}
\label{fig:FigA6}
\end{center}
\end{figure*}

\begin{figure*}
\begin{center}
\includegraphics[width=0.62\textwidth]{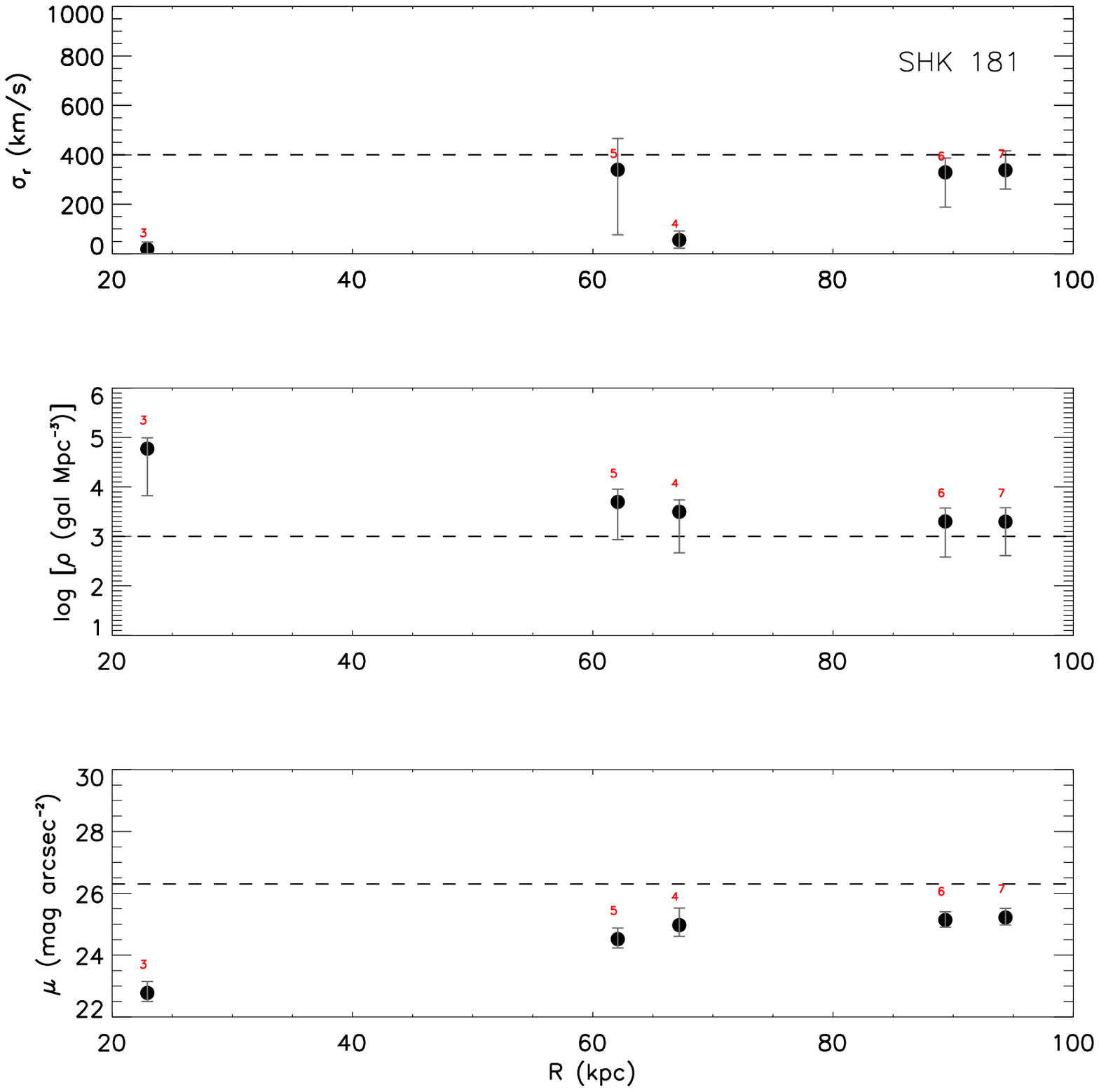}
\includegraphics[width=0.55\textwidth]{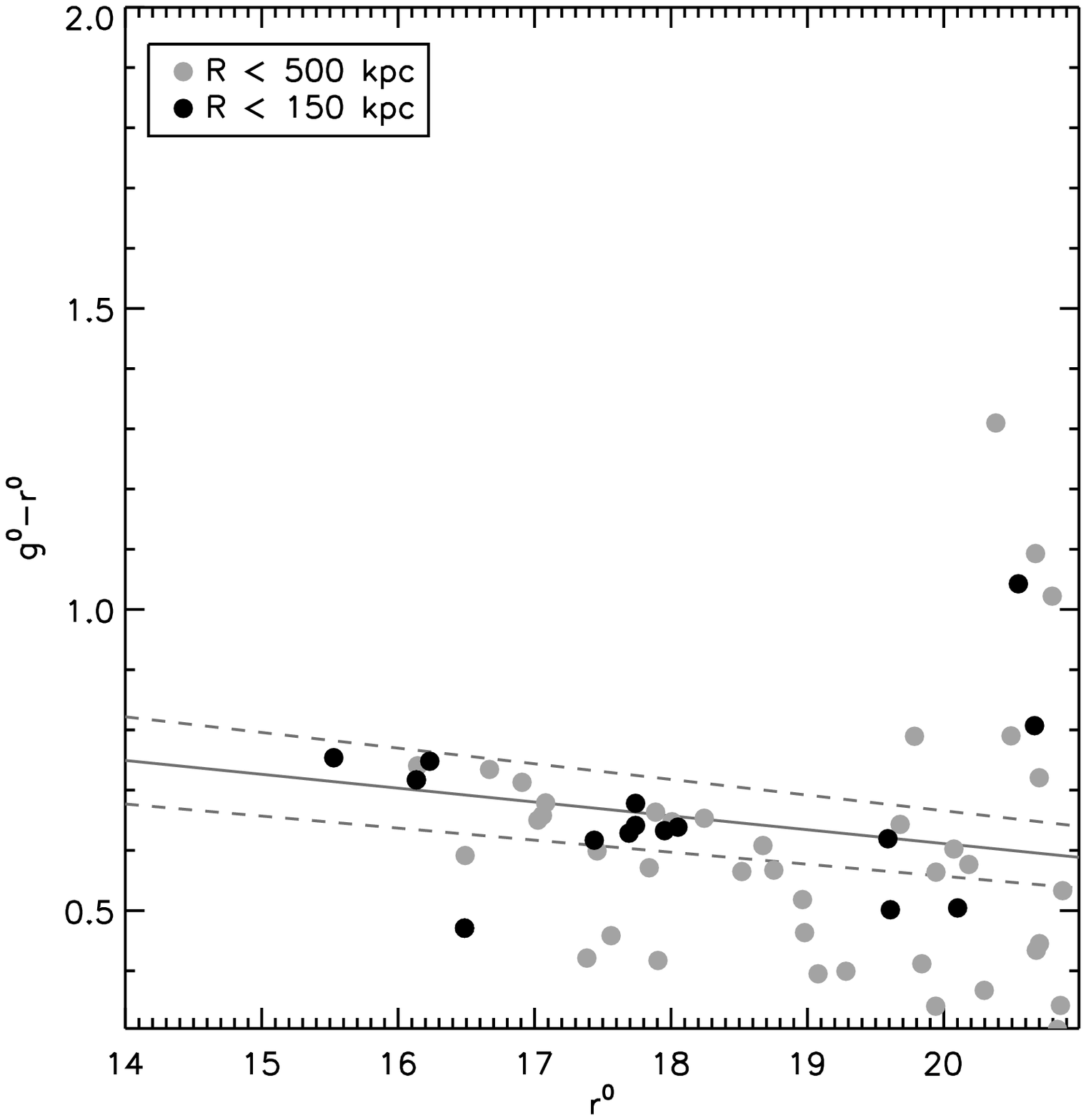}
\caption{{\bf SHK 181.} {\it Upper panel}: radial profiles of $\sigma_{\rm r}$, $\log(\rho)$ and $\mu$ (top to bottom) starting from the central galaxy triplet. Numbers on top of data points indicate the number of galaxies considered. {\it Lower Panel}: colour-magnitude diagram for all the galaxies contained within $500\ {\rm kpc}$ from group centroid and satisfying the criteria 2 and 3 defined in Section 3 of Cap09. Galaxies within the inner region ($R<150\ {\rm kpc}$) are represented by black dots. The solid line represents the colour-magnitude relation obtained by \citet{Ber2003b} at the groupÕs redshift, while the dashed ones indicate the scatter in its slope reported in the mentioned study.}
\label{fig:FigA7}
\end{center}
\end{figure*}

\end{document}